\documentclass[prd, aps,nofootinbib, preprintnumbers, showpacs,
superscriptaddress,twocolumn]{revtex4} 
\usepackage{latexsym}
\usepackage{amssymb}
\usepackage{amsfonts}
\usepackage{amsmath}
\usepackage{numprint}
\usepackage[dvips]{graphicx}\usepackage{color}

\usepackage{ulem}

\renewcommand{\emph}[1]{{\it #1}}

\newcommand{\be}{\begin{equation}}
\newcommand{\ee}{\end{equation}}
\newcommand{\beast}{\begin{equation*}}
\newcommand{\eeast}{\end{equation*}}
\newcommand{\bea}{\begin{eqnarray}}
\newcommand{\eea}{\end{eqnarray}}
\newcommand{\beqn}{\begin{eqnarray*}}
\newcommand{\eeqn}{\end{eqnarray*}}
\newcommand{\ba}{\begin{align}}
\newcommand{\ea}{\end{align}}

\newcommand{\eg}{\textit{e.g.}~}
\newcommand{\ie}{\textit{i.e.}~}
\newcommand{\etal}{\textit{et al.}~}
\newcommand{\np}{\numprint}
\newcommand{\tabsize}{\footnotesize}

\newcommand{\innprod}[2]{(#1 | #2)}

\def\Re{\mathrm{Re}}

\def\subPN{_{\mathrm{\textsc{pn}}}}

\def\subEOB{_{\mathrm{\textsc{eob}}}}
\def\superEOB{^{\mathrm{\textsc{eob}}}}
\def\subLSO{_{\mathrm{\textsc{lso}}}}
\def\subC{_{\mathrm{\textsc{c}}}}
\def\subGW{_{\mathrm{\textsc{gw}}}}
\def\superPN{^{\mathrm{\textsc{pn}}}}
\def\superVone{^{\mathrm{PhenV1}}}
\def\superVtwo{^{\mathrm{PhenV2}}}
\def\superPhen{^{\mathrm{\textsc{p}hen}}}
\def\fM{f\!M}
\def\Hz{\mathrm{Hz}}
\def\Mpc{\mathrm{Mpc}}

\def\Ms{M_{\odot}}
\def\eff{\mathrm{eff}}

\begin{document}


\title{Accuracy and effectualness of closed-form, frequency-domain waveforms for non-spinning black hole binaries} 

  \author{Thibault \surname{Damour}}
  \affiliation{Institut des Hautes Etudes Scientifiques, 91440 Bures-sur-Yvette, France}
  \affiliation{ICRANet, 65122 Pescara, Italy}

  \author{Alessandro \surname{Nagar}}
  \affiliation{Institut des Hautes Etudes Scientifiques, 91440 Bures-sur-Yvette, France}
  
  \author{Miquel \surname{Trias}}
  \affiliation{Departament de Fisica, Universitat de les Illes Balears, Carretera Valldemossa km.
7.5, 07122 Palma de Mallorca, Spain}

  \date{\today}
  
\begin{abstract}
The coalescences of binary black hole (BBH) systems, here taken to be non-spinning,
are among the most promising sources for gravitational wave (GW) ground-based detectors, 
such as LIGO and Virgo. To detect the GW signals emitted by BBHs, and measure
the parameters of the source, one needs to have in hand a bank of GW templates that are both
\emph{effectual} (for detection), and \emph{accurate} (for measurement). We study the \emph {effectualness}
and the \emph{accuracy} of the two types of parametrized banks of templates that are directly
defined in the frequency-domain by means of closed-form expressions, namely `post-Newtonian' (PN) and
`phenomenological' models. In absence of knowledge of the (continuous family of) exact waveforms, 
our study assumes as fiducial, target waveforms the ones generated by the most accurate version
of the effective-one-body (EOB) formalism,
calibrated upon a few high-accuracy numerical relativity (NR) waveforms. 
We find that,  for \emph{initial GW detectors} the use, at each point of
parameter space, of the best closed-form template (among PN and phenomenological models)
leads to an \emph{effectualness} $>97\%$ over the entire mass range and $>99\%$
in an important fraction of parameter space; however,
when considering \emph{advanced detectors}, both of the 
closed-form frequency-domain
models \emph{fail to be effectual enough} in significant domains of 
 the two-dimensional [total mass and mass ratio] parameter space.
Moreover, we find that, \emph{both for initial
and  advanced detectors}, the two
closed-form frequency-domain
models \emph{fail to satisfy the minimal required accuracy standard}
in a very large domain of the two-dimensional parameter space.
 In addition, a side
result of our study is the determination, as a function of the mass ratio, of the maximum frequency
at which a frequency-domain PN waveform can be `joined' onto a NR-calibrated
EOB waveform without
undue loss of accuracy. In the case of mass ratios larger than 4:1 this maximum frequency occurs 
well before the last stable orbit, leaving probably too many orbital cycles to be covered by current NR techniques 
if one wanted to  construct accurate enough hybrid PN-NR waveforms.
This problem will, however, be probably greatly alleviated, or even solved, by
using the EOB formalism instead of PN theory.
\end{abstract}

  
  \maketitle
  

\section{Introduction}
\label{sec:intro}

During the last $5-10$ years very significant progress has been made towards the 
detection of gravitational wave (GW) signals from binary black holes (BBH) coalescences. 
On the experimental side, current interferometric GW detectors have been operating at 
their design sensitivity for several years
\cite{Abramovici:1992ah, Abbott:2003vs, Caron:1997hu, Acernese:2006bj, Luck:2006ug, Tsubono:1997, Hough:2000}
and advanced versions
of these detectors (more sensitive by an order of magnitude \cite{Harry:2010zz}) are being planned and 
are expected to be operational in four to five years. For example, a $(10+10)\Ms$ binary [$\Ms$ denoting the mass of the Sun]
can be detected out to distances of $\sim 160\Mpc$ by initial detectors and $\sim 2200\Mpc$ 
by advanced detectors \cite{Buonanno:2009zt}. On the 
theoretical side, one has recently obtained a good understanding of all the stages in 
the dynamics of coalescing BBHs and their emitted gravitational radiation
thanks to a combination of theoretical techniques: (i) the early, adiabatic inspiral 
stage of the evolution can be described in terms of a post-Newtonian (PN) expansion up to 3.5PN order in phase and 3PN in amplitude \cite{Blanchet:2004ek,Blanchet:2002av,Kidder:2008,Blanchet:2008je}; 
(ii) the late inspiral, the merger and the ring-down 
can be computed by numerically solving  Einstein's 
equations~\cite{Pretorius:2005gq,Campanelli:2005dd,
Baker:2005vv,Gonzalez:2006md,Gonzalez:2007hi,Pollney:2007ss,
Boyle:2007ft,Boyle:2008ge,
Scheel:2008rj, Hannam:2009hh,Pollney:2009yz,Reisswig:2009us,
Hannam:2010ec,Pretorius:2007nq};
and (iii) the effective-one-body (EOB) 
formalism~\cite{Buonanno:1998gg,Buonanno:2000ef,Damour:2000we,Damour:2001tu} 
has shown its capability of incorporating information
coming both from PN theory and numerical relativity (NR) results into an accurate 
description~\cite{Buonanno:2006ui,Buonanno:2007pf,Damour:2007yf,
Damour:2007vq,Damour:2008te,Damour:2009kr,
Buonanno:2009qa,Yunes:2009ef,Pan:2009wj} 
of the whole BBHs dynamics and GW radiation,  from the early inspiral, 
right across the last stable orbit, to the ``plunge'', 
the merger and the final ringdown. 

From the data analysis point of view, the detection of the GW signals emitted by BBHs necessitates
the prior knowledge of `model waveforms', i.e. of faithful representations of the
emitted gravitational waveforms. These model waveforms are used to detect, and then to measure the
properties of, a putative BBH by employing them to match-filter~\cite{Helmstrom:1968} the 
output of the detector. 
As the expected GW signals depend on several continuous parameters (notably the 
masses of the two BHs in the non-spinning case), one needs to construct very large 
banks of model waveforms, in order to densely sample the multi-dimensional space of 
possible signals. Full NR simulations are computationally much too demanding
to provide sufficiently long waveforms, for sufficiently many parameter values. 
Faster methods, making use of our analytical knowledge of BBHs, need to be considered 
for data analysis purposes. Several possibilities can be (and are being) pursued: 
(i) one can use PN theory by itself to construct template waveforms representing 
the inspiral part of the evolution; and the PN templates can be defined
\emph{either in the time-domain, or the frequency-domain} (see~\cite{DIS:2001} 
for various possibilities and~\cite{Buonanno:2009zt} for a recent comparison of 
these possibilities); (ii) one can use EOB theory, calibrated to a small sample of 
high-accuracy NR waveforms, to construct \emph{time-domain} waveforms that cover 
the whole BBH evolution, from early inspiral to ringdown; (iii) a third possibility 
is to construct closed-form, \emph{frequency-domain} ``phenomenological'' waveform models
\cite{PhenV1,PhenV2,Santamaria:2010yb}; these models are obtained by fitting some piece-wise analytical 
formulae to the Fourier transforms of a small sample of hybrid time-domain PN-NR waveforms.

From a data analyst point of view, an attractive feature of both the frequency-domain PN models,
and the phenomenological ones, is that they are defined by closed-form expressions directly in the Fourier domain, converting the waveform generation to an extremely fast process, which is very convenient to perform extensive searches or parameter estimation studies. The EOB waveform models are obtained, in the time domain,
by integrating ordinary differential equations (ODE's) that depend on the continuous parameters of the
considered BBH. Each EOB waveform
can be obtained quite fast, both in the time domain, and then (using a Fast Fourier Transform (FFT)) in the frequency domain 
(see below for numbers). However, the exploration of variations in the
continuous parameters of the source is much more time-consuming than in the case of analytic, closed-form
frequency-domain models. It is therefore useful to compare the closed-form model
waveforms to the NR-calibrated EOB ones. If one were to find that the closed-form frequency-domain
waveforms were so close to the NR-calibrated EOB ones that they would be essentially equivalent
for performing the data analysis of GW signals, there would be no need to construct large banks of
EOB templates.

This raises the issue of having quantitative measures of the ``closeness''
 of two model waveforms that directly translate into their closeness 
in the data analysis of GW signals. We recall that, in GW data analysis, 
model waveforms can be used in two different ways, either for the mere \emph{detection} 
of the signals, and/or, after detection, for the \emph{measurement} of properties 
of the source. In both cases one uses  match-filtering~\cite{Helmstrom:1968}, 
i.e. a convolution of the measured strain in the detector with a large bank of 
model waveforms. Ideally, one can and should (for consistency) use the same 
bank of waveforms for detection and measurement. But, as the accuracy requirements 
are different for detection and measurement (and are stricter for measurement), 
it might be advantageous, from a practical point of view, to have in hand two 
different banks of template waveforms. The issue of defining, in a quantitative 
manner that is related to data analysis, the closeness of two model waveforms has 
been discussed in several works~\cite{DIS:1998,fairhurst,lindblom:2008,lindblom:2009,lindblom:2010}, 
and has already been applied to comparing  different  models in, e.g., 
Refs.~\cite{DIS:1998,DIS:2000,DIS:2001,Damour:2002vi,Buonanno:2002ft,Cokelaer:2004pw,Pan:2007nw,Buonanno:2009zt}.
Ref.~\cite{DIS:1998} introduced a special 
nomenclature for referring to two particular measures of the closeness of a 
model waveform to a supposedly exact waveform: (i) the \emph{effectualness} 
$\mathcal E$ (related to the maximum signal-to-noise ratio (SNR) associated 
to the use of a bank of model waveforms); and (ii) the \emph{faithfulness} 
$\mathcal F$ (measuring the SNR associated to a model waveform having the same 
physical parameters as the exact one). The \emph{effectualness} $\mathcal E$ 
is a useful measure for \emph{detection} purposes, and we shall use it below. 
The \emph{faithfulness} $\mathcal F$ was defined as a quantitative way 
of gauging the quality of a model waveform for \emph{measurement} purposes.
The recent literature~\cite{fairhurst,lindblom:2008}  has emphasized
that, for  \emph{measurement}, there is a better, data-analysis-relevant, 
measure of the closeness of a model waveform to a supposedly exact one, 
namely the Wiener (noise-weighted) squared distance between the two waveforms. 
Below, after discussing the motivation for the use of this measure, we shall introduce 
a certain related quantity that we shall call the \emph{inaccuracy} $\mathcal I$ 
of a model waveform (with respect to a supposedly given exact waveform). 
Then, we shall use this \emph{inaccuracy} $\mathcal I$ to gauge
the quality of the closed-form, frequency-domain model waveforms. [We shall also see that,
in most cases, the `inaccuracy' is essentially proportional to the ``unfaithfulness'' $1-\mathcal F$.]

The aim of this paper is to study, by means of the two data-analysis relevant 
measures of \emph{effectualness} and \emph{(in-)accuracy}, the two
classes of closed-form, frequency domain waveform models for non-spinning BBH systems, 
namely PN expansions and phenomenological waveforms. There remains, however, to discuss 
the issue of the choice of supposedly exact target waveforms. 
In absence of knowledge of the (continuous family of) exact waveforms, our study will 
assume as fiducial, target waveforms the ones generated by the most accurate 
version~\cite{Damour:2009kr} of the effective one body formalism, calibrated upon 
a few high-accuracy numerical relativity waveforms. We recall that 
Refs.~\cite{Damour:2009kr,Buonanno:2009qa} have shown that, after calibration of a few 
EOB ``flexibility'' parameters onto some nonperturbative information extracted from high-accuracy
NR results, the EOB waveforms exhibited an excellent agreement (within numerical errors) with 
state-of-the-art NR waveforms, for \emph{several mass ratios} (1:1, 2:1 and 3:1). 
One is therefore justified in considering the two-parameter family of NR-calibrated EOB 
waveforms as our current best approximation to the (unknown) continuous family of 
(non-spinning) exact waveforms. Independently of this argument, it is anyway 
(as already said above) interesting to discuss, in quantitative terms, the closeness between
the bank of closed-form waveforms, and the bank of EOB ones.
 Given the small computational time required for each  EOB simulation, we shall be able 
to exhaustively explore the whole parameter space, and to identify the regions where the 
closed-form  models lack in effectualness or accuracy. In the process, we shall also describe 
how to construct frequency-domain hybrid EOB waveforms by joining 
(in a smooth manner) an early-inspiral, closed-form frequency-domain PN waveform, to a 
subsequent Fourier-transformed EOB waveform.  This will lead us to discussing 
the maximum frequency  where the ``junction''  can take place without undue
loss of accuracy. 

This paper is organized as follows: in Sec.~\ref{sec:GW_DA} we briefly review the 
basic concepts and definitions used in GW data analysis and discuss the accuracy 
standards that will be used in the rest of the paper. Section~\ref{sec:models} 
summarizes the structure of the closed-form waveform models considered in our 
study, mainly with the purpose of establishing a common notation. 
In Sec.~\ref{sec:EOB-PN_merge} we define the time-domain EOB models 
we use, and describe the computation of their Fourier-transform, as well 
as a procedure for constructing hybrid PN$\cup$EOB waveforms in the Fourier 
domain. Then, in Sec.~\ref{sec:PN-SPA_range} we study what is the maximum frequency 
value where the joining  (in the frequency domain)  between  the early PN waveform 
and the subsequent EOB one can be done without undue loss of accuracy. 
 The main results of the paper, concerning quantitative measures of the closeness  
between closed-form models, and EOB ones, are presented and discussed in 
Sec.~\ref{sec:results}.  Finally,  Sec.~\ref{sec:conclusions} presents our conclusions
and Sec.~\ref{sec:future_work} the pointers for future work. 

Also, let us note that we generally use units such that $G=c=1$. In addition, when dealing
with  the EOB dynamics, and Fourier transform,  we shall often use (without warning) 
the adimensionalized time $t\subEOB = t_{\rm physical}/M$.


\section{Detecting gravitational wave signals into the noise}
\label{sec:GW_DA}

In this section, we start by providing a brief review of the basic concepts and 
definitions used in GW data analysis, mainly to establish our notation.  
Then we  give the definition of the two quantitative measures 
({\it effectualness} and {\it inaccuracy})  that we shall use 
 to quantify the accuracy of approximated waveform models in comparison with 
some (fiducial)  exact waveform. All our derivations will 
be done assuming observations from a single GW detector, although they could 
be easily generalized to multiple detectors.

\subsection{Conventions and notation}

The measured strain in an interferometric GW detector, $s(t)$, is a time series 
data consisting in the combination of the detector noise, $n(t)$, and the  
exact GW signal that we want to detect, $h_x(t)$ (here, and in the following, 
the subscript $x$ stands for ``exact'').  
By assuming additive noise, we have
\be
s(t) = n(t) + h_x(t) \; .
\ee
Commonly, the noise is modeled to be zero-mean, Gaussian and stationary, and is characterized by
its (one-sided) power spectral density (PSD), $S_n(f)$, which is defined as
\be
\overline{\tilde{n}^*(f) \tilde{n}(f')} = \frac{1}{2} S_n(f) \delta (f-f') \; .
\ee
Here,  the overline denotes the ensemble average, the superscript star  indicates 
complex conjugation, and the tilde denotes the Fourier transform,
\be \label{eq:FTdef}
\tilde{n}(f) = \int_{-\infty}^{\infty} n(t) e^{-2\pi i f t} \mathrm{d}t \; .
\ee
Note that we use here the LIGO Scientific Collaboration (LSC) sign convention 
for the Fourier transform, with a factor  $\int dt e^{-2\pi i f t} (...)$.  
This sign convention is opposite to the one used in many early GW papers
(such as Refs.~\cite{DIS:1998,DIS:2000,DIS:2001}) which used the (theoretical-physics)
convention $\int dt e^{+2\pi i f t} (...)$. This change of convention introduces several sign
changes in key formulas, that we shall indicate below. [Let us warn in advance the reader
that, because of this, there are sign errors in some of the recent GW 
literature (such as~\cite{AISS_2005}).]

\begin{figure}
\centering
\includegraphics[width = 8cm]{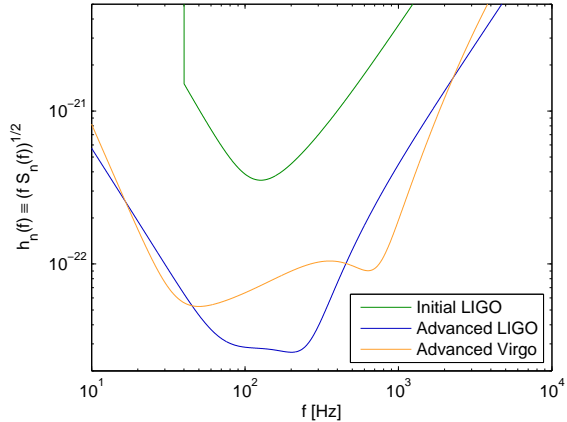}
\caption{Effective GW noise $h_n(f) \equiv (f S_n(f))^{1/2}$  of the three GW 
ground-based detectors considered in the paper.}
\label{fig:IFO_PSDs}
\end{figure}

In this paper, we shall consider three different ground-based detectors: 
Initial LIGO, Advanced LIGO and Advanced Virgo. Following Ref.~\cite{Ajith:2009fz},  
the one-sided noise PSDs of these detectors is approximated by some
simple analytic expressions of a dimensionless frequency $x=f/f_0$ 
(see Fig.~\ref{fig:IFO_PSDs}). For Initial LIGO,  $f_0 = 150$ Hz,  
and (using a low frequency cut-off, $f_{\mathrm{low}} = 40$ Hz)
\begin{eqnarray} \label{eq:iLIGOpsd}
S_n\left(x(f)\right) & = &  9 \times 10^{-46} \left [ (4.49x)^{-56} + 0.16 x^{-4.52} \right. \nonumber \\
 & & \left. + 0.52 + 0.32 x^2\right ] \, .
\end{eqnarray}
For Advanced LIGO,  $f_0 = 215$ Hz, and (using $f_{\mathrm{low}} = 10$ Hz)
\begin{equation} \label{eq:aLIGOpsd}
S_n\left(x(f)\right) = 10^{-49} \left [x^{-4.14} - 5 x^{-2} + 111 \Big(\tfrac{1 - x^2 + x^4/2}{1 
+ x^2/2}\Big)\right] \, .
\end{equation}
For Advanced Virgo, $f_0 = 720$ Hz, $f_{\mathrm{low}} = 10$ Hz and
\begin{eqnarray} \label{eq:aVIRGOpsd}
S_n\left(x(f)\right) & = & 10^{-47} \left [  2.67 \times 10^{-7} x^{-5.6} \right. \nonumber \\
 & & + 0.59 \, e^{(\ln\,x)^2\,[-3.2 - 1.08 \ln(x) - 0.13 (\ln\,x)^2]}\,x^{-4.1} \nonumber \\
 & & \left. + 0.68 \, e^{-0.73 (\ln\,x)^2}\,x^{5.34} \right] \, .
\end{eqnarray}
Let us note here that the initial LIGO noise PSD corresponds to the one in the
Science Requirements Document, which is very close to (but not the same as) the actual
noise curve obtained during LIGO's S5 and S6 science runs. The noise PSDs that we use for advanced
detectors are also slightly different to the ones currently considered by the LSC and
Virgo Collaboration (see Fig.~2 of Ref. \cite{Abadie:2010cfa}).

See Fig.~\ref{fig:IFO_PSDs} for a plot of these three PSDs in the form of the dimensionless
\emph{effective GW noise} $h_n(f) \equiv (f S_n(f))^{1/2}$.  We recall that $h_n(f)$ is
a useful measure of the noise in that, together with the corresponding
dimensionless frequency-domain \emph{effective GW signal}
$h_s(f) \equiv | f \tilde h(f)|$, it yields the squared SNR $\rho^2$
 as an integral over the logarithmic 
frequency:  $\rho^2 = \int_0^\infty (df/f) (h_s(f)/h_n(f))^2$ \cite{DIS:2000}.

The assumed Gaussianity of the noise  naturally leads to  the definition of an inner product
(Wiener scalar product) between two (real) time series, $a(t)$ and $b(t)$,  namely
\be \label{eq:def_innprod}
\innprod{a}{b} \equiv 4 ~ \Re \int_{0}^{\infty} \mathrm{d}f ~ \frac{\tilde{a}^*(f) \tilde{b}(f) }{S_n(f)}
\; .
\ee
This Wiener scalar product is real and symmetric. The associated norm, say
\be \label{eq:def_norm}
\vert a \vert^2 \equiv\innprod{a}{a} = 4 ~ \Re \int_{0}^{\infty} \mathrm{d}f
~ \frac{\tilde{a}^*(f) \tilde{a}(f) }{S_n(f)}
\; .
\ee
is positive definite, and endows the space of (real) signals with an Euclidean structure.  

Moreover, the normalization of this norm is such  that the probability of  having 
the realization of a particular noise series  $n(t)$ is 
$p(n) \propto e^{-\frac{1}{2} \innprod{n}{n}}$. From this definition, the 
Neyman-Pearson criterion leads to an optimal search statistic equivalent to the form  
$\innprod{\widehat{h}_x}{s}$ (Wiener filter), cf. Eq.~(A24) of Ref.~\cite{Cutler:1994ys}.
Here, we are using a hat to 
denote normalized templates $\widehat{h} =
\frac{h}{\innprod{h}{h}^{1/2}}$, so that $\innprod{\widehat{h}}{\widehat{h}} = 1$.
By analogy with the usual Euclidean space, the inner product between two normalized signals,
say $\widehat{h}, \widehat{g}$,
varies between $-1$ and $+1$ and can be thought of as defining the cosine of
the ``angle'' $\alpha(\widehat{h}, \widehat{g})$ between the two corresponding ``unit vectors''.
One often refers to this cosine as being the ``overlap''  between the two normalized signals, say
\be \label{eq:def_overlap}
\mathcal O (\widehat{h}, \widehat{g}) \equiv
\cos \alpha(\widehat{h}, \widehat{g}) \equiv
\innprod{\widehat{h}}{\widehat{g}} 
\; .
\ee
 As we do not know the exact signal $h_x$, but only some approximate model of it, 
say $h_m$ (where the subscript $m$ stands for `model'), one uses as
practical search statistic the quantity $\innprod{\widehat{h}_m}{s}$.  In presence of a non-zero
signal within the detector strain signal $s$, the latter quantity is the sum of a  
zero-mean random Gaussian variable, say $N \equiv \innprod{\widehat{h}_m}{n}$ (filtered noise) 
and of the non-random ``filtered signal'', $\rho_m$,  defined as,
\be
\rho_m \equiv \innprod{\widehat{h}_m}{h_x} = \frac{\innprod{h_m}{h_x}}{\innprod{h_m}{h_m}^{1/2}} \; ,
\ee
Because of our use of a normalized template $\widehat{h}_m$ it is easily checked that the 
 zero-mean random Gaussian
variable  $N \equiv \innprod{\widehat{h}_m}{n}$ has a variance equal to \emph{unity}.
Therefore, the filtered signal, $\rho_m$, represents the mean signal-to-noise ratio (SNR) 
for the detection of a GW signal within the detector output $s=h_x +n$, when 
using  $h_m$ as model template. The optimal SNR will be obtained when the template 
used $h_m$ is the exact signal   $h_x$ that we want to detect, i.e.
\be
\rho_{\mathrm{opt}} = \innprod{\widehat{h}_x}{h_x}  =
\frac{\innprod{h_x}{h_x}}{\innprod{h_x}{h_x}^{1/2}} = \innprod{h_x}{h_x}^{1/2} \; .
\ee
In general, we can represent a complex gravitational wave signal in the Fourier domain in
its polar form as
\be \label{eq:wvfF}
\tilde{h}(f) = A_h(f) e^{-i \psi_h(f)} \; ,
\ee
where the  amplitude $A_h(f)$  and the phase $\psi_h(f)$ are real quantities. 
Since we are considering here real signals, we have $\tilde{h}(f) = \tilde{h}^*(-f)$, 
which implies that the amplitude and the phase are,  respectively, even and odd 
functions of the frequency. Note that here we have explicitly introduced a 
\emph{minus sign} in the phase to ensure, in view of the 
(LSC) sign convention Eq.~(\ref{eq:FTdef}) in the definition of the Fourier transform, that
$\psi(\omega)$ (with $\omega = 2 \pi f$) is, in the stationary phase approximation (SPA),
the Legendre transform of the time-domain phase (modulo $ - \frac{\pi}{4}$): 
$ \psi(\omega) = \omega t - \phi\subGW(t)  - \frac{\pi}{4}$, with $d\phi\subGW(t)/dt = \omega$ .  This has the advantage that the so-defined frequency-domain phase $\psi(f)$
then coincides (possibly modulo$ - \frac{\pi}{4}$)  with the SPA phase $\psi(f)$ used
in many other papers (such as \cite{DIS:2000,DIS:2001}).  The reader should be wary
of possible errors in the literature due to the change in the sign convention for
the Fourier transform, as many early papers used the theoretical-physics convention.

Note, for future use, that, in terms of real amplitudes and phases, the Wiener  products 
$\innprod{h}{h}$ and $\innprod{h}{g}$ read
\bea \label{eq:innprod_explicit}
\innprod{h}{h} & = & 4 \int_0^\infty \mathrm{d}f \frac{A_h(f)^2}{S_n(f)} \\
\innprod{h}{g} & = & 4 \int_0^\infty \mathrm{d}f \frac{A_h(f) A_g(f) \cos\left( \psi_h(f) -
\psi_g(f)\right)}{S_n(f)} \nonumber \; .
\eea

Before proceeding to the definition of  the accuracy standards 
of a particular waveform model, let us spell out the notation we shall use for  the different
waveforms that will enter our analysis: \\
\begin{tabular}{p{0.75cm} p{7cm}}
$h_x$ & Exact waveform \\
$h_m$ & Model waveform having the same physical parameters as $h_x$ \\
$h_{\bar{m}}$ & Model waveform having parameters that maximize the overlap with
$h_x$ \\
$h_b$ & Model waveform of a certain  discrete template bank having a maximum overlap with
$h_x$.
\end{tabular}

Finally, let us mention that we shall be neglecting here the eventual calibration 
errors from the detectors. Taking them into account would imply a strengthening 
of the accuracy standards discussed below (see~\cite{lindblom:2008}).

\subsection{Effectualness of model waveforms}
\label{subsec:effectualness}
Any modeled waveform $h_m$ represents an approximation to the real gravitational 
wave pattern $h_x$ that is emitted in nature, so that the practical search 
statistic  $\innprod{\widehat{h}_m}{s}$ will necessarily be less efficient
in detecting a signal than the optimal Wiener filter  $\innprod{\widehat{h}_x}{s}$.
 In GW searches, one says that
$h_m$ is an  \emph{effectual}  model of the exact signal if it allows for a successfull 
 \emph{detection}  of the presence of a GW signal in the detector output, even if the values of
parameters entering $h_m$ are not close to the real values of the parameters
entering the exact signal $h_x$. [One sometimes even allow the parameters of $h_m$ to take
physically unrealizable values.] 
Following~\cite{DIS:1998}, we shall quantify the  \emph{effectualness}, say $\mathcal E$,
of a continuous (multi-parameter) bank of model waveforms by considering the maximum value 
of the ratio between the SNR,  $\rho_m$, obtained by using $h_m$ as filter, and the optimal SNR, 
$\rho_{\mathrm{opt}} = \rho_x$, obtained by using the exact Wiener 
filter  $\innprod{\widehat{h}_x}{s}$, i.e.
\be \label{eq:def_effectualness}
\mathrm{effectualness} \equiv {\mathcal E} \equiv \innprod{\widehat{h}_{\bar{m}}}{\widehat{h}_x} =
\frac{\rho_{\bar{m}}}{\rho_{\mathrm{opt}}} \; ,
\ee
where we recall that $\widehat{h}_{\bar{m}}$ stands for the normalized waveform
in the continuous bank of models that maximizes the overlap with the exact (normalized) signal.
 In other words,
if we represent the set of parameters that characterize our model by the vector $\vec{\lambda}$;
$h_{\bar{m}}$ is such that
\be \label{eq:maximization_effectualness}
\innprod{\widehat{h}_x}{\widehat{h}_{\bar{m}}} = \max_{\vec\lambda}~
\innprod{\widehat{h}_x}{\widehat{h}_{m(\vec\lambda)}} \; .
\ee
In geometrical language, this means that the cosine of the angle (in Wiener's signal space) 
between the unit vector $\widehat{h}_x$ and any unit vector in the continuous bank $\widehat{h}_{m(\vec\lambda)}$ is maximized\footnote{Rigorously speaking we should
be talking about a supremum rather than a maximum, as there is no mathematical
guarantee, for a continuous bank, that the supremum is reached for some model. Anyway,
in practice one always uses discrete banks where the maximum will be attained by some model.}
by the unit vector $\widehat{h}_{\bar{m}}$.  This maximum cosine, corresponding to a
\emph{minimum angle}, will necessarily range between $-1$ and $+1$, and in practice
will be close to, but strictly smaller than, $+1$.

In this paper, we will restrict the source parameters to take values with real 
physical interpretation. In particular, the symmetric mass ratio, 
$\nu = m_1 m_2 / (m_1+m_2)^2$, will be constrained to be $\nu \leq 0.25$. [Otherwise,
we would be considering systems with imaginary individual masses, $m_1$ and $m_2$ !]

It is also convenient to work with the \emph{ineffectualness}, i.e. the quantity complementary
(with respect to 1)  of the effectualness,  say $\bar  {\mathcal E} \equiv 1- \mathcal E$.
In other words, the ineffectualness of a   \emph{continuous} bank of models $h_m$ in
detecting a certain exact signal $h_x$ is given by
\be \label{eq:def_dxm}
\bar  {\mathcal E}_{x\bar{m}} = 1 - \innprod{\widehat{h}_x}{\widehat{h}_{\bar{m}}} \; .
\ee
In choosing an upper bound on the value of the ineffectualness one must, however, remember
that, in practice, one approximates the continuous bank of model templates by some
finite, discrete template bank.  The closeness of a discrete bank to its continuous
version is measured by the so-called \emph{minimal match} (MM) \cite{Owen:1995tm}, i.e. the
minimum overlap between any model $h_m$ in the continuous bank, and any 
model $h_b$ in the discrete bank. One can replace the minimal match by its complementary,
i.e. the \emph{maximal mismatch} $\epsilon_{MM} \equiv 1-MM$.  It is easily
seen (by viewing the complementaries of overlaps, $1- \cos \alpha \simeq \alpha^2/2$,
as squared distances between nearby points on the Wiener sphere of unit signals) 
that, when setting thresholds on the ineffectualness of the discretized
version of a template bank (say, $b$), one should replace the
continuous ineffectualness $\bar  {\mathcal E}_{x\bar{m}}$ by the ``effective'' ineffectualness
\cite{lindblom:2008}
\be \label{eq:pythagoras_d}
\bar  {\mathcal E}_{xb} =\bar  {\mathcal E}_{x\bar{m}} + \epsilon_{MM} \; .
\ee
Since in any GW observation, the distance to the source is inversely proportional to the observed
SNR and the number of events is proportional to the observable volume, we can express the
fractional loss of potential events in terms of the effective ineffectualness of a given 
discrete template bank,
\be \label{eq:missed_events}
\frac{\mathrm{\# \, events_{~b}}}{\mathrm{\# \, events_{~total}}} = \left( 1 - \bar  {\mathcal E}_{xb} \right)^3 \;
.
\ee
A threshold on the allowed fraction of missed events will set up the effectualness condition for
any approximated template bank. Typically, one works with template banks with a maximal mismatch
\footnote{The choice of this value depends on the computational cost to generate the resulting total number of templates
and the maximum false alarm rate allowed for the search.},
$\epsilon_{MM}$, of $0.005-0.030$ \cite{CBC_search_1, CBC_search_2, CBC_search_3, CBC_search_4}
and allow for a fraction of missed events 
smaller than $10\%$ $\left(
 \bar  {\mathcal E}_{xb} < 0.0345 \right)$, which making use of Eq.~(\ref{eq:pythagoras_d}), translates to the following
effectualness condition,
\be \label{eq:effectualness}
\mathcal{E}_{x\bar{m}} = \innprod{\widehat{h}_x}{\widehat{h}_{\bar{m}}} > 0.9705-0.9955 \; .
\ee
In view of the fact that it is much more difficult, and expensive, to improve the
sensitivity of the detectors than the quality of the template banks, one should
aim at imposing  on $\bar  {\mathcal E}_{xb}$ a bound significantly stricter
than that corresponding to a loss of $10\%$ of the hard-won potential events.
Then, if we assume the reasonable condition that the two contributions
on the r.h.s. of  Eq.~(\ref{eq:pythagoras_d}) are comparable, i.e.
 $\epsilon_{MM} \sim  \bar  {\mathcal E}_{x\bar{m}}$, we see that we should
really aim at imposing bounds on the effectualness which are on the
stricter side of the condition above.

Let us end this section by a brief discussion of the procedure we used to
compute the effectualness, i.e. to maximize the overlap 
$\innprod{\widehat{h}_m}{\widehat{h}_x}$ over all the parameters entering the model
waveform $h_m$.  There are two types of parameters we should maximize upon:
(i)  the physical parameters of the source (which, in the present case, are the
two masses $m_1, m_2$, or equivalently the total mass $M \equiv m_1+m_2$ and the
symmetric mass ratio $\nu \equiv m_1 m_2/M^2$); and (ii) some kinematical parameters.
In the present case, the relevant kinematical parameters of $h_m$ are some time origin $t_m^0$,
and some initial phase $\phi_m^0$. 

Among these parameters, the only one upon which we can maximize analytically
is the arbitrary phase $\phi_m^0$. Note first that, in the following, we shall work with a given
target waveform $h_x$ (including a quasi-random initial phase), so that our problem here
is only to maximize over the phase of the template waveform $h_m$. [See 
appendix B of~\cite{DIS:1998} for the case where one varies the initial phases of both the
template and the target.]
To do this we start by noting a simplifying feature of the frequency domain, with respect to
the time domain. In the frequency domain, shifting the phase of
  any waveform $\tilde{h}(f) = A_h(f) e^{-i \psi_h(f)}$
simply corresponds to adding an arbitrary constant, say $\phi_0$, to the phase $ \psi_h(f)$.
A look at the explicit form  Eq.~(\ref{eq:innprod_explicit}) of the Wiener scalar product
shows that two waveforms whose frequency-domain phases  $ \psi_h(f)$,  $ \psi_g(f)$
only differ by $\pi/2$ are orthogonal to each other. [This property does not hold 
in general when shifting time-domain phases by $\pi/2$; except when: (i) one considers just
one harmonic, and (ii) the signal is such that the stationary phase approximation is accurate,
see Sec.~\ref{subsec:models_PN-SPA}.] In addition, two waveforms having the same
frequency-domain amplitude $A_h(f)$ have the same Wiener squared length $\innprod{h}{h}$.
This shows that the two model waveforms  $\{ h^{\phi_0 = 0},h^{\phi_0 = \pi/2} \}$
define an equal-length, orthogonal basis of the two-plane spanned by all the
phase-shifted models  $ h^{\phi_0 }$. Moreover, a generic phase-shifted model can
be written as $ h^{\phi_0 }=  \cos \phi_0 h^{\phi_0 = 0} + \sin \phi_0 h^{\phi_0 = \pi/2} $.
These properties allow one to compute the analytical dependence upon $\phi_0$ of
the scalar product of  $ h^{\phi_0 }$ with any given waveform $g$:
\be
\innprod{g}{h^{\phi_0}} =  \cos \phi_0  \innprod{g}{h^0} + \sin \phi_0 \innprod{g}{h^{\pi/2}}
\ee
With this, one can maximize $\innprod{g}{h^{\phi_0}}$  analytically over $ \phi_0$, with
the result
\be 
\max_{\phi_0} \innprod{g}{h^{\phi_0}} = \left[ \innprod{g}{h^0}^2 + \innprod{g}{h^{\pi/2}}^2
\right] ^{1/2} \; .
\ee

On the other hand,  the maximization over the other parameters 
must be done numerically. There is, however, a simplification when
dealing with the maximization with respect to the time origin $t_m^0$.
Indeed, the dependence on the time origin is known analytically,
so that there is, in particular, no need   to recompute the
full waveform every time we need to shift a time origin. 
Indeed, we have for the phase difference between any two
waveforms
\be
\psi_h(f,t_{0,\,h}) - \psi_g(f,t_{0,\,g}) = \psi_h(f,0) - \psi_g(f,0) + 2\pi f \Delta t_0 \; .
\ee
where $\Delta t_0 = t_{0,\,h}-t_{0,\,g}$.
Then, it is just a matter of numerically maximizing the integral of $\innprod{h}{g}$ given by
Eq.~(\ref{eq:innprod_explicit}) over $\Delta t_0$.
%

\subsection{ Accuracy standards for a model waveform}
\label{subsec:faithfulness}

Gravitational wave observations will open a new observational window on the Universe.
In order to do actual
astrophysics with GW, we need not only to detect signals, but also to extract reliable physical
information out of them.  In other words, we need to have at hand waveform models that
are close enough to the exact signals  to allow for an accurate enough measurement of
the physical properties of the GW source. [As mentioned above, here we could be talking 
either about the same template bank used for detection, or about a complementary
template bank used for analyzing in detail signals having passed a first detection threshold.]
It has been recently emphasized~\cite{fairhurst,lindblom:2008,lindblom:2009} that a simple 
accuracy standard ensuring that a model waveform $h_m$
can be used for extracting most of
the  scientific content of the GW signal $h_x$ (buried within the detector noisy output
$s= h_x+n$) is that the Wiener squared distance between $h_m$ and $h_x$ 
be smaller than one:
\be \label{eq:faith_cond}
\vert h_m - h_x \vert^2  <  \overline{\vert n \vert^2}  \equiv 1 \; ,
\ee
where the overline denotes the ensemble average.
Here it is understood that $h_x$ and $h_m$ share the same physical
parameters. [But, one is allowed to vary purely kinematical parameters such as an 
initial time $t_0$ and an initial phase $\phi_0$.]

There are several ways of thinking about the meaning of the inequality Eq.~(\ref{eq:faith_cond}).
Ref.~\cite{lindblom:2008} motivates it by considering the accuracy with which one can
measure the parameter $\xi$ gauging the affine distance along a straight line
connecting $h_x$ to $h_m$ in Euclidean Wiener space:
$h(\xi) = (1-\xi) h_x + \xi h_m$. Assuming Gaussian noise,
the variance of the measurement of $\xi$ is found to be given by
\be 
\sigma_{\xi}^{2} = \vert h_m - h_x \vert^{-2} \; .
\ee
Therefore,  the inequality~(\ref{eq:faith_cond}) implies that the error in $\xi$ 
is larger than its variation when it interpolates  between the two models, $\sigma_\xi >
1$. 

Another way of motivating Eq.~(\ref{eq:faith_cond}) is to think in terms of the
Wiener geometry of signal space.  If the detector were noiseless,
a measurement of the detector output $s$ would give us the exact location, 
in signal space, of the incoming signal $h_x$, namely   $h_x=s$.
However, the presence of noise in the detector is introducing a certain uncertainty
in our knowledge of $h_x$. More precisely, under our assumption of Gaussian noise,
the probability distribution, in signal space, of the measured signal $s$ is
 $p(s) \propto e^{-\frac{1}{2} \vert s-h_x \vert^2}$. Therefore, at, say, the $68$\% 
probability level,  our knowledge is only that the exact signal $h_x$  lies
somewhere within a ``unit ball'' in Wiener space, centered around the observed
signal $s$:  $\vert h_x -s\vert^2 =1$.  The maximum knowledge
we can extract from the GW data about the parameters $\vec{\lambda}$
of the source is therefore that $ \vert h_x(\vec{\lambda}) - s  \vert = \vert n \vert$, with
$\overline{\vert n \vert^2}  \equiv 1$.
However, we do not know the exact form of the functional dependence
of the signal upon the source parameters , i.e. the function  $h_x(\vec{\lambda})$.
We only know an approximation of this functional dependence, namely the
one given by the waveform model  $h_m(\vec{\lambda})$. By using the \emph{triangular
inequality in (Euclidean) Wiener space}, we can write the inequality
\begin{align} \label{triangular}
 \vert h_m(\vec{\lambda}) - s \vert  & \leq  \vert h_m(\vec{\lambda}) 
- h_x(\vec{\lambda}) \vert +  \vert h_x(\vec{\lambda}) - s \vert \nonumber \\
     & =  \vert h_m(\vec{\lambda}) - h_x(\vec{\lambda}) \vert +  \vert n \vert\; .
\end{align}
This inequality shows that the condition Eq.~(\ref{eq:faith_cond}) is such that, 
in the worst case\footnote{As we are discussing the case where we have only \emph{one}
observed datum $s$, one cannot use the fact that ``on average'' $s$ is centered around $h_x$
to try to get a better general inequality on the distance between $h_m$ and $s$.}, 
it degrades our knowledge on the values of the source parameters 
by being equivalent to \emph{doubling the noise level of the detector}.
Let us recall that our only knowledge
about the source parameters is derived from the fact that
the measured signal $s$ satisfies  $ \vert h_x(\vec{\lambda}) - s  \vert = \vert n \vert$,
so that it is the noise level in the detector which determines the
statistical errors in parameter estimation.
Therefore, a doubling (in the worst case) of the noise will add 
systematic biases to measured parameters of the same order
as the statistical errors. This is
clearly the minimal requirement we should put on the difference between
$h_m$ and $h_x$.  However,  as in the discussion above of
the threshold for effectualness, we should remember that
 it is much more difficult, and expensive, to improve the
sensitivity of the detectors than the quality of the template banks. Therefore,
we should not allow  the lack of accuracy of waveforms to lead to
any significant (effective) increase of the detector noise level.  In other words,
we should  strengthen the condition (\ref{eq:faith_cond}) into
a condition of the form
\be \label{eq:accuracy}
\vert h_m(\vec{\lambda}) - h_x(\vec{\lambda}) \vert^2  <  \epsilon^2 \overline{\vert n \vert^2}  \equiv  \epsilon^2 \; ,
\ee
where $\epsilon<1$ must be chosen so as to limit the
 effective increase of the noise level on the r.h.s. of the
triangular inequality Eq.~(\ref{triangular}) to an acceptable level.
  Note indeed that the r.h.s. of  Eq.~(\ref{triangular})
now reads $\leq (1+\epsilon) \vert n \vert$.  One should  take
at least $\epsilon \sim 1/2$, and probably   $\epsilon \sim 1/3$ or more\footnote{Note that,
according to Ref.~\cite{lindblom:2009b}, the consideration of calibration errors in GW detectors
also lead to suggesting a strengthening of the original condition Eq.~(\ref{eq:accuracy}) by a factor
$\epsilon_\mathrm{calib}$ smaller than one. But such a `calibration factor' $\epsilon_\mathrm{calib}<1$ 
should be taken \emph{on top of} the one of Eq.~(\ref{eq:accuracy}), 
which has a different motivation.
In other words, one should use a ``safety factor''   
$\epsilon_\mathrm{total} = \epsilon \cdot \epsilon_\mathrm{calib}$.}.
Only with such small values of $\epsilon$ can we really consider $h_m$
as being effectively indistinguishable from $h_x$.

\subsection{ Definition of an \emph{inaccuracy} functional}

It is useful to rewrite the accuracy condition Eq.~(\ref{eq:accuracy})  in a different form,
which highlights its connection with the  \emph{faithfulness}  functional, $\mathcal F$,
introduced in \cite{DIS:1998}. The l.h.s. of Eq.~(\ref{eq:accuracy}) depends  on the absolute
magnitudes of the signals (and therefore on the distance to the source). It also depends
on the normalization of the Wiener metric,  i.e. on the overall normalization
of the PSD of the noise.  Let us instead define the following functional of
$h_m$ and $h_x$:
\be \label{eq:inaccuracy1}
\mathcal{I}[h_m;h_x] \equiv \frac{\vert h_m(\vec{\lambda}) - h_x(\vec{\lambda}) \vert^2}{ \vert h_x(\vec{\lambda}) \vert^2}  \; .
\ee
We shall call $\mathcal{I}[h_m;h_x] $ the \emph{inaccuracy} of the model waveform
$h_m$ with respect to $h_x$\footnote{Note that this inaccuracy functional is not
a symmetric function of its two arguments. One could also consider
defining the symmetric functional obtained by replacing in 
Eq.~(\ref{eq:inaccuracy1}) the denominator $\vert h_x(\vec{\lambda}) \vert^2$ by $| h_m(\vec{\lambda})| | h_x(\vec{\lambda})|$. 
However, this would lead to an upper limit on this symmetric functional
which differs from the inverse squared SNR by a factor involving the ratio $|h_m|/|h_x|$.}.  
The functional $\mathcal{I}[h_m;h_x] $ remains invariant
both when one multiplies $h_m$ and $h_x$ by a common factor, and when one 
multiplies the noise density $S_n(f)$ by an arbitrary factor. In other words,
it depends only on the shape of the noise PSD, and on the relative shapes
of the two waveforms $h_m$ and $h_x$. 
As explicitly indicated, both waveforms refer to the
same values of the physical parameters of the source. However, we can still 
maximize over the kinematical parameters of the waveform model $h_m$,
namely some arbitrary time origin $t_m^0$, as well as some initial phase $\phi_m^0$.

Using the identity
\be
\frac{|\vec a - \vec b|^2}{|\vec b|^2}  \equiv \left( \frac{|\vec a|}{|\vec b|} -1 \right) ^2 + 2 ~\frac{|\vec a|}{|\vec b|} \left( 1-\cos\alpha \right)  \; ,
\ee
valid in any Euclidean space (with $\alpha$ denoting the angle between the two vectors), 
we can write more explicitly the inaccuracy functional in
the following expanded form
\be \label{eq:inaccuracy2}
\mathcal{I}[h_m;h_x] =\left[\frac{\innprod{h_m}{h_m}^{1/2}}{\innprod{h_x}{h_x}^{1/2}} -1\right]^2 + 
2 \frac{\innprod{h_m}{h_m}^{1/2}}{\innprod{h_x}{h_x}^{1/2}}
\left[1-\innprod{\widehat{h}_x}{\widehat{h}_m}\right]  \; .
\ee
With this notation, the accuracy condition  Eq.~(\ref{eq:accuracy}) can now written in
the following form
\be 
\label{eq:faith_cond_expanded}
\mathcal{I}[h_m;h_x]  < \epsilon^2 \frac{\innprod{n}{n}}{\innprod{h_x}{h_x}} = \frac{\epsilon^2}{\rho^2} \; .
\ee
where $\rho^2 \equiv |h_x|^2 \equiv |h_x|^2/|n|^2$ denotes the squared SNR.
This reformulation of the accuracy condition has several useful features:
\begin{itemize}

\item  All the dependence on the
absolute magnitudes of both the signals and the noise is rejected to the r.h.s.,
and in a very simple form involving the (squared) SNR. This allows one to compute
the `inaccuracy' on the l.h.s. from the shapes of the signals and the shape of the noise PSD
independently from the various accuracy thresholds set by the r.h.s.

\item The inaccuracy  $\mathcal{I}[h_m;h_x] $,   Eq.~(\ref{eq:inaccuracy2}),  is seen to
consist of the sum of two positive-definite contributions: 
(i) the first one depends only on the ratio
of the Wiener squares of the two signals (and vanishes when $|h_m|^2=|h_x|^2$);
 while (ii) the second one is proportional (and nearly equal) to twice the 
`unfaithfulness', $ 2  \bar{ \mathcal{F}}_{mx} = 2 (1- \mathcal{F}_{mx})$, where
$\mathcal{F}_{mx} \equiv \innprod{\widehat{h}_x}{\widehat{h}_m} $ \cite{DIS:1998}.
Note that, using the explicit frequency-domain 
expression of the inner product in terms of the amplitudes and
phases of the waveforms, Eq.~(\ref{eq:innprod_explicit}),
the first contribution to the inaccuracy only depends on the frequency-domain amplitudes,
$A_m(f)$ and $A_x(f)$, and vanishes when they coincide; 
whereas the second contribution, proportional to $2  \bar{ \mathcal{F}}_{mx}$, 
vanishes when both the amplitudes, and  the phases $\psi_m(f)$ and $\psi_x(f)$ coincide.

\item Note in passing that if, instead of evaluating $\mathcal{I}[h_m;h_x]$ for the same
values of the physical parameters, one were considering the analog of the `effectualness' 
$\mathcal E$, i.e. if one were minimizing $\mathcal{I}[h_m;h_x]$ over the physical parameters
of the model waveform, the minimization over the effective distance of $h_m$ would
require the condition $|h_m|^2=|h_x|^2$ (because of the positive-definite character
of the first contribution (i)), so that $\mathcal{I}[h_{\bar m};h_x]$ would reduce
to the `ineffectualness' $ 2 (1- \mathcal E) =2(   1 - \innprod{\widehat{h}_x}{\widehat{h}_{\bar{m}}})$.
\item In our comparisons below, our experience has been that the inaccuracy
contribution from the amplitude differences was typically several orders of magnitude smaller than the one from the phase errors. This means that the accuracy condition  Eq.~(\ref{eq:accuracy}) is practically equivalent to the following faithfulness condition
\be
1- \mathcal{F}=1-\innprod{\widehat{h}_x}{\widehat{h}_m} < \frac{\epsilon^2}{2\rho^2} \; ,
\ee

\item  If  (as in \cite{lindblom:2008,lindblom:2009}) we expand the inaccuracy to
leading order in the amplitude and phase differences, $\delta A(f) \equiv A_m(f) - A_x(f)$,
$\delta \psi(f) \equiv\psi_m(f) - \psi_x(f)$, it becomes
\be
\mathcal{I}[h_m;h_x]=\langle \left( \frac{\delta A}{A_x} \right)^2  \rangle + \langle \delta\psi^2 \rangle \;,
\ee
where the angular brackets denote an average with respect to the normalized measure
entering the squared SNR, i.e.  $dw(f) = 4 N df A_x(f)^2/ S_n(f)$, with the
normalization factor $N \equiv 1/\rho_{\mathrm{opt}}^2$
chosen so that $\int_0^\infty dw(f)=1$.  This expanded expression is too coarse to be
numerically useful, but it is useful in indicating that the accuracy condition 
Eq.~(\ref{eq:accuracy}) essentially means that
\be
\langle \left( \frac{\delta A}{A_x} \right)^2  \rangle + \langle \delta\psi^2 \rangle< \frac{\epsilon^2}{\rho^2} \;,
\ee
i.e. that  the noise-weighted quadrature sum of the fractional
amplitude discrepancy and of the phase discrepancy (between $h_m(f)$ and $h_x(f)$)
must be smaller than $\epsilon/\rho$. [As we shall further discuss below, this motivates
one to considering $\rho_{\rm eff} \equiv \rho/\epsilon$ as an \emph{effective SNR}.]
 If, for instance, we consider a SNR of $10$ and
take $\epsilon=1/2$ (corresponding to an effective SNR $\rho_{\rm eff}=20$),
 we see that the accuracy condition is restricting the 
 noise-weighted quadrature sum of the fractional
amplitude discrepancy and of the phase discrepancy to be smaller than $0.05$ (i.e.
 5 \%  or 0.05 radians). Such a constraint is more difficult to satisfy for the phase
(for which any error in its dynamical description is accumulated over many GW cycles).

\end{itemize}

In summary, we shall require of  any model waveform that it satisfies the following conditions:
(i) for \emph{detection} the effectualness condition Eq.~(\ref{eq:effectualness}); and
(ii) for \emph{measurement} the accuracy condition  Eq.~(\ref{eq:faith_cond_expanded}), where
the \emph{inaccuracy} functional is defined by  Eq.~(\ref{eq:inaccuracy1}), or,
in more explicit form, by Eq.~(\ref{eq:inaccuracy2}). From its r.h.s. we  see that the higher 
the observed SNR, $\rho$, is, the more restrictive  the accuracy condition becomes. In addition,
one must choose some $\epsilon <1$ to guarantee that the inaccuracy of the
model waveform only increases the detector noise by a factor $1+\epsilon$  remaining
close to 1.


\section{Closed-form, frequency-domain waveform models considered in our study}
\label{sec:models}

In this paper we consider GW observations from a single interferometer, assuming the
detected signals to be short enough so that the orientation 
of the detector (due to Earth's motion) can be considered constant during the time it
is being observed. [For simplicity, we
shall also not explicitly write down the (Doppler) effect on the GW signal coming from
the relative motion (and cosmological redshift) between
the detector and the source.]
These approximations are commonly used 
for first and second generations of ground-based detectors, and will allow us to reduce
all the dependence on the sky location, luminosity 
distance, polarization and inclination angles to an effective distance parameter,
$D_\eff$ (see Eq.~(\ref{eq:Cdef}) below), and to
an additional phase shift $\varphi_0$ (see Eq.~(\ref{eq:phi0def}) below). Moreover, we
shall only consider the dominant $(2,2)$ mode from 
the spin-weighted spherical harmonics decomposition of a signal from a non-spinning coalescing binary system.

In a suitable frame adapted to the quasi-circular orbit of the considered compact
binary system, the two orthogonal polarizations of the emitted time-domain quadrupolar GW signal read
\begin{eqnarray}
h_+(t)  & = &  \left[\dfrac{1+\cos^2\iota}{2}\right] \, a\subGW(t) \cos\phi\subGW(t) \, , \nonumber \\
h_\times(t)& = & [\cos\iota] \, a\subGW(t) \sin\phi\subGW(t) \, .
\end{eqnarray}
The relative weight between the polarizations depends on the inclination angle, $\iota$, i.e.
the angle between the angular momentum of the system and the propagation direction. 
The time series   $a\subGW(t)$ and $\phi\subGW(t)$ define the (time-domain) amplitude
and phase of the quadrupolar waveform. [The time-domain phase  $\phi\subGW(t)$ is
conventionally defined so that it increases with time.]

The corresponding GW signal seen in the detector will be given by a single time series strain, $h(t)$, resulting from the projection of the two GW polarizations with the beam antenna patterns, $F_+$ and $F_\times$, which depend on the sky location of the source (i.e. the relative orientation of the line of sight and the normal direction of the detector's plane) and the polarization angle \cite{Thorne:300yr}. Using basic trigonometric relations, the detected strain can be written as a single cosine function, with amplitude $a(t)$ and phase $\phi(t)$, as follows:
\begin{eqnarray}
h(t) &     =    & F_+ h_+ + F_\times h_\times = \mathcal{C} a\subGW(t) \cos(\phi\subGW(t) + \varphi_0) \nonumber \\
      & \equiv & a(t) \cos\phi(t) \, ,
\end{eqnarray}
where we defined
\begin{eqnarray} \label{eq:Cdef}
\mathcal{C} & \equiv & \frac{1}{2} \sqrt{(1+\cos^2\iota)^2 F_+^2 + 4 \cos^2\iota F_\times^2} \\
 \label{eq:phi0def}
\varphi_0 & \equiv & \arctan\left[ \frac{-2 F_\times \cos\iota}{F_+ (1+\cos^2\iota)} \right] \, .
\end{eqnarray}
If we were considering observations from multiple detectors, the amplitude and
phase differences between the various detected strains would involve different
$\mathcal{C}$ and $\varphi_0$ values through the antenna 
patterns of each detector, and thereby we could infer the sky
location and polarization angle of the source. 
Also if we were observing a GW signal for a long time period, the time dependence
of these three parameters could be used to estimate 
the position of the source. In this paper, we are considering short-lived signals
detected by a single interferometer and therefore
$\mathcal{C}$ and $\varphi_0$ will be numerical constants and absorbed,
respectively, into an effective distance and the initial phase of the signal.

In the Fourier domain, the measured GW signal can be represented generically
in the polar form Eq.~(\ref{eq:wvfF}). In what follows, we shall describe the
different waveform models considered in our study in terms of the frequency-domain
amplitude and phase, $A(f)$ and $\psi(f)$.

\subsection{Frequency-domain Post-Newtonian model: PN(f)}
\label{subsec:models_PN-SPA}

In the limit where the two coalescing compact objects are still far away from the merger, the 
post-Newtonian approximation to  Einstein's theory provides an accurate representation of the dynamics and radiation of the system. For non-spinning black holes binaries, the time-domain phase, $\phi\subPN(t)$,  is known up 3.5PN order  and the amplitude, $a\subPN(t)$, up to 3PN \cite{Blanchet:2004ek,Blanchet:2002av,Kidder:2008,Blanchet:2008je}.

A closed-form expression for the PN waveform in the Fourier domain can be obtained by making use of the Stationary Phase Approximation (SPA). Using this approximation is not an extra
assumption beyond the usual ones underlying the PN method. Indeed, the derivation of the
time-domain PN phasing already assumes that the inspiral is \emph{adiabatic}, which is the
basic assumption needed for using the SPA.  More precisely the SPA is valid
 when the amplitude and frequency of the signal evolve on time scales much longer than the
orbital period:  $\epsilon_a=\dot{a}(t) /( a(t) \dot{\phi}(t)) \ll 1$ and $\epsilon_f=\ddot{\phi}(t)/\dot{\phi}^2(t) \ll 1$.
For a coalescing binary system the time scale of evolution of the GW amplitude and frequency
is the radiation-reaction time scale so that $\epsilon_a \sim \epsilon_f \sim \nu v^5$
[$v$ being the PN expansion parameter $v \equiv (\pi f M)^{1/3}$]
stays much smaller than one essentially up to the end of the inspiral. See 
\cite{Droz:1999qx, DIS:2000}
for a more detailed discussion of the applicability of the SPA, and for the calculation
of the first correction to the leading-order SPA. Here we shall use the leading-order SPA
which has the following expression in terms of the time-domain GW amplitude $a(t)$,
the GW phase $\phi(t)$, and the derivative of the GW frequency $F(t)= \dot{\phi}(t)/(2 \pi)$:
\begin{equation} \label{eq:SPAraw}
\tilde{h}(f) = \frac{a(t_{\!f})}{2\sqrt{\dot{F}(t_{\!f})}} e^{-i [2\pi f t_{\!f} - \frac{\pi}{4} - \phi(t_{\!f}) ]} \, .
\end{equation}
Here $t_{\!f}$ is the saddle point defined as the instant when the GW frequency
$F(t)$ is equal to the Fourier-domain argument $f$. It is obtained by solving the
equation $\dot{\phi}(t) = 2\pi f$ in $t$.

By inserting in the general SPA expression (\ref{eq:SPAraw}) the time-domain PN expressions
one can obtain a closed-form expression for the PN-approximated frequency-domain
inspiral waveform.  As discussed in \cite{DIS:2001} there are actually several
different ways of using the PN information to define $\tilde{h}(f)$. Here, we shall
consider the most straightforward way of  doing so (called `TaylorF2' in \cite{DIS:2001}),
consisting in expanding both the frequency-domain amplitude and phase in
powers of the PN expansion parameter $v(f) \equiv (\pi f M)^{1/3}$. We recall that
$M = m_1 + m_2$ denotes the total mass of the system ($m_1$ and $m_2$ being the individual masses),  and $\nu = m_1 m_2/M^2$  the symmetric mass ratio.  In the following, we
shall denote $v(f)$ simply as $v$, but one must remember that it is a function of the
Fourier variable $f$. For brevity, we shall refer in the following to this PN-expanded
frequency-domain waveform as `PN(f)'. We can write it as
\begin{equation} \label{eq:PNf}
\tilde{h}\subPN(f) \equiv A\subPN(f) e^{-i\psi\subPN(f)}  \theta(v\subLSO -v)\; ,
\end{equation}
where we introduced a frequency-domain cut-off at the 
Schwarzschild-like Last Stable Orbit (LSO) frequency $v(f\subLSO) \equiv 6^{-1/2}$
($\theta(x)$ denoting Heaviside's step function).

The angular argument $\psi\subPN(f) $ in Eq.~(\ref{eq:PNf}) is given by the PN series
\begin{equation} \label{eq:psiPN}
\psi\subPN(f) = 2\pi f t\subC - \phi\subC + \frac{3}{128 \nu v^5} \sum_{k=0}^{7} \alpha_k(\nu, \ln v) v^k \; ,
\end{equation}
where $t\subC$ and $\phi\subC$ represent some arbitrary time and phase shifts. [We absorbed
the SPA phase shift $- \frac{\pi}{4}$ into  $\phi\subC$.] The explicit expressions of the
PN expansion coefficients $\alpha_k$ (up to 3.5PN) can be identified from Eq.~(3.18) of \cite{Buonanno:2009zt}. [Notice that Ref.~\cite{Buonanno:2009zt} uses the opposite sign convention for the Fourier transform, which translates into a global sign in the frequency-domain phase $\psi(f)$, usually written explicitly in the definition of $\tilde{h}(f)$ so that $\psi(f)$ has the same expression with both sign conventions; compare, for instance, Eq.~(\ref{eq:wvfF}) with Eq.~(3.17) of \cite{Buonanno:2009zt}.]

In many early GW papers, the GW amplitude was taken only at leading order in the
PN expansion, i.e. at the `Newtonian'  (frequency-domain) approximation:
\begin{equation} \label{eq:AN}
A_{\textsc{n}}(f) = -\sqrt{\frac{5\pi}{24}} \frac{\nu^{1/2} M^2}{D_\eff} v^{-7/2} \, .
\end{equation}

Here we have kept an explicit negative sign coming from the second derivative of a complex exponential function,  and we have defined the effective distance $D_\eff \equiv D_L / \mathcal{C}$ which absorbs all the sky position, polarization angle and orientation dependencies,  see Eq.~(\ref{eq:Cdef}) above.  The 3PN correction to the (quadrupolar)
frequency-domain amplitude is then obtained by combining the 3PN-accurate expression
of  $h^{22}(t)$ \cite{Kidder:2008} with the 3PN expansion of  $\dot{F}$  (given, \eg , in \cite{Buonanno:2009zt}). Re-expanding the result as a single PN series leads to
\begin{widetext}
\begin{eqnarray} \label{eq:APN}
A\subPN (f) & = & A_{\textsc{n}}(f)
\left[ 1 +
\left( \frac{451 \nu }{168} -\frac{323}{224} \right) v^2 +
\left( \frac{\np{105271} \nu ^2}{\np{24192}} -\frac{\np{1975055} \nu }{\np{338688}} - \frac{\np{27312085}}{\np{8128512}} \right) v^4 +
\left( \frac{85 \pi }{64} (4\nu - 1) -24 i \nu \right) v^5 \right . \nonumber \\ 
 & & \left. +
\left( \frac{\np{34473079} \nu ^3}{\np{6386688}} - \frac{\np{3248849057} \nu ^2}{\np{178827264}}
+\frac{\np{545384828789} \nu }{\np{5007163392}} - \frac{205 \pi ^2 \nu }{48}
-\frac{\np{177520268561}}{\np{8583708672}}+\frac{428 i \pi }{105} \right) v^6 \right] \, .
\end{eqnarray}
\end{widetext}
Note that,  contrary to the convention generally assumed in the rest of the paper,  this `3PN amplitude' $A\subPN (f)$
contains some (small)  imaginary contributions (as  $h^{22}(t)$ in \cite{Kidder:2008}). 
 When, in the
following, we shall talk about the amplitude of the PN(f) waveform we shall mean the
modulus $|A\subPN (f)|$. Correlatively, when we shall talk about the phase of the
PN(f) waveform we shall mean (minus) the argument of the complex number $\tilde{h}\subPN(f)$,
defined in  Eq.~(\ref{eq:PNf}) (i.e. $\psi\subPN(f)$ minus the argument of $A\subPN (f)$).

\begin{figure}
\centering
\includegraphics[width = 8cm]{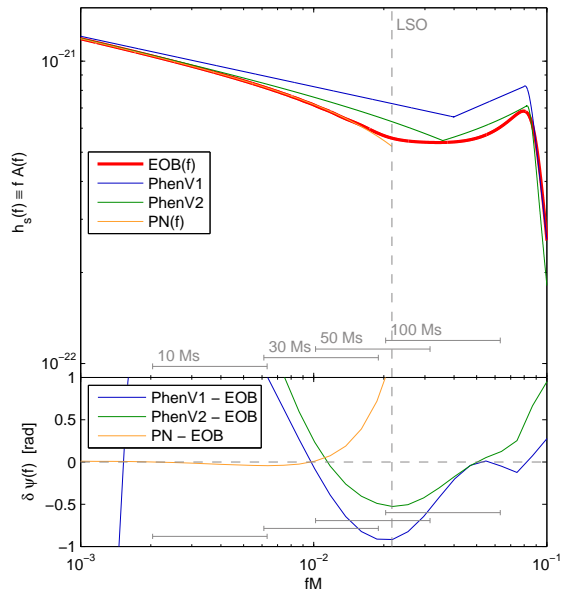}
\caption{Upper panel: effective GW amplitudes $h_s(f) \equiv | f \tilde h (f)|$ (see text)  of the various
frequency-domain  waveform
models considered in this paper for an equal-mass system [assuming a fiducial distance of $1$ Gpc and a total mass of $M=75\Ms$, in order to get actual amplitude values].  Lower panel: differences between the phases of the
closed-form frequency-domain waveforms, and the phase of the frequency-domain EOB waveform. 
One has subtracted a linear term $a f+b$ corresponding to an arbitrary 
initial time, and an arbitrary initial phase. 
The vertical dashed line indicates the GW frequency of the ``Schwarzschild-like'' LSO.
The horizontal segments near the bottom of the two panels indicate the shortest 
frequency intervals over which $70 \%$ of the SNR for an inspiral signal
is accumulated when observed with Adv. LIGO (see text).}
\label{fig:waveforms}
\end{figure}

Contrary to the  time-domain amplitude  $|h(t)|$, which is dimensionless, the
 amplitude of a Fourier-transformed GW signal $A(f)= |\tilde h (f)|$ has the dimension
of a time.
By contrast, the effective GW signal amplitude
$h_s(f) \equiv | f | A(f) \equiv  | f  \tilde h (f)|$  is dimensionless. As dimensionless
quantities, the effective amplitudes of the signal, $h_s(f)$, and of the noise, $h_n(f)$,
are not affected when one rescales the frequency, by means of the total mass $M$,
into the dimensionless variable $\fM$. As mentioned above, the SNR$^2$ is given by
the logarithmic frequency integral:  $\rho^2 = \int_0^\infty (df/f) (h_s(f)/h_n(f))^2$.
The effective GW amplitude of the PN(f) waveform is represented 
(for the equal-mass case $\nu=1/4$) in Fig.~\ref{fig:waveforms}. This figure also
plots the effective GW amplitudes of the other waveforms we shall use in the
present study: the EOB one, and the two phenomenological ones that we shall
introduce next. Fig.~\ref{fig:waveforms} represents also the \emph{difference}
between the phase  of the  PN(f) signal, and that of the corresponding EOB(f) signal.
This figure  indicates that both the amplitude and the phase of the PN(f) waveform stay close
to that of the EOB signal up to a frequency $\fM\sim 0.015$, after which they start
to deviate quite strongly.  For reference, note that the  GW frequency at
the `Schwarzschild-like' LSO is equal to $M f\subLSO = (6\pi \sqrt{6})^{-1} \approx 0.02166$, 
while the GW frequency at the adiabatic-EOB LSO is, for $\nu=1/4$, approximately $42 \%$ larger, being equal  to $M f\subLSO\superEOB \approx 0.0308$.  In other words,  the PN(f) model starts to strongly deviate
from the EOB(f) model at a frequency  smaller by a factor two than the adiabatic-EOB LSO 
frequency. Later we shall study in more detail after which frequency the difference
between PN(f) and EOB(f)  starts to introduce an unacceptable level of `inaccuracy'
(in the technical sense defined above).

Finally, note that the horizontal segments near the bottom of
the two panels of  Fig.~\ref{fig:waveforms} indicate (for each quoted total mass $M$)
the shortest frequency intervals  $[f_1M,f_2M]$ upon which  70 \% of the SNR
for an inspiral signal ($h_s(f) \propto f^{-1/6}$) is accumulated. Here we used the Adv. LIGO PSD, and found that $f_1= 41$ Hz and $f_2=128$ Hz span the \emph{shortest} interval such that
$\int_{f_1}^{f_2} (df/f) (h_s(f)/h_n(f))^2= 0.7^2 \int_0^\infty (df/f) (h_s(f)/h_n(f))^2$,
with $h_s(f) \propto f^{-1/6}$.

\subsection{Non-spinning phenomenological waveform model (PhenV1)}
\label{subsec:models_PhenV1}

In this paper we shall consider two different phenomenological models. [See,
however, the concluding section for additional results
concerning a recently published, third phenomenological model.]
The first model was built by Ajith \etal \cite{PhenV1} by starting from $\sim 30$ \emph{time-domain, hybrid} non-spinning waveforms in the mass-ratio range, $0.16 \leq \nu \leq 0.25$. 
These time-domain hybrid waveforms were constructed by joining together, in the time domain,
some early 3.5 PN TaylorT1 restricted waveforms with relatively short ($\sim 4$ inspiral cycles, plus merger) subsequent numerical waveforms generated by the Jena group using their \textsc{bam} code \cite{BAM1,BAM2} with fourth-order finite differencing. Afterwards, the resulting phenomenological model was tested against some other hybrid waveforms built from longer and more accurate numerical simulations produced using the \textsc{ccatie} \cite{Pollney:2007ss} and the same \textsc{bam} code employing sixth-order finite differencing.  These time-domain
hybrid waveforms were then Fourier-transformed, and the resulting sample of frequency-domain
waveforms $h_{\rm hybrid}(f)$ was then fitted to some piecewise-continuous multi-parameter analytic expressions, which define the closed-form `version-1 phenomenological waveform', say
\begin{equation} \label{eq:PhenV1}
\tilde{h}_{\mathrm{PhenV1}}(f) \equiv A_{\mathrm{PhenV1}}(f) e^{-i\psi_{\mathrm{PhenV1}}(f)}  \; .
\end{equation}
The phenomenological waveforms are defined directly in the Fourier domain: the amplitude, $A_{\mathrm{PhenV1}}(f)$, is defined as a piecewise function distinguishing inspiral, merger and ring-down stages, while the phase\footnote{Notice the explicit change of sign of the phenomenological phase defined in \cite{PhenV1} compared with our definition here, Eq.~(\ref{eq:wvfF}).}, $\psi_{\mathrm{PhenV1}}(f)$, is defined as a PN-type series with phenomenological coefficients. The whole waveform is characterized by $10$ phenomenological parameters ($4$ for the amplitude and $6$ for the phase), each of them fitted to a quadratic polynomial of the symmetric mass ratio, $\nu$. We refer the reader to Sec.~IV.C of Ref.~\cite{PhenV1} for all the information about the actual definition of this phenomenological model.

Since the hybrid waveforms were built using a \emph{restricted} PN model, i.e. with a
time-domain Newtonian amplitude, we expect (in view of the fact that the time-domain
PN correction is smaller than one during the early inspiral) that the amplitudes of the
corresponding Fourier-domain waveforms will be larger than the ones fully incorporating PN 
effects;  and indeed one can see on Fig.~\ref{fig:waveforms} that 
$A_{\mathrm{PhenV1}}(f)$ is roughly $6\%$ larger than the
EOB waveform amplitude.  Concerning the phase, note that, though full PN information (up to 3.5PN order) was used to construct the time-domain hybrid signals, after fitting their
Fourier transform to closed-form expressions, the final frequency-domain phase $\psi_{\mathrm{PhenV1}}(f)$ exhibits significant  differences 
with respect to the EOB phase.  These \emph{global} discrepancies cannot be absorbed in
the arbitrary term $a f+b$ present in the phase, and will be quantified below
in terms of effectualness and inaccuracy.

\subsection{Spinning, non-precessing phenomenological waveform model (PhenV2)}
\label{subsec:models_PhenV2}

Recently, Ajith \etal have extended the first phenomenological model \cite{PhenV1} to black hole binary systems with non-precessing spins (i.e. spins parallel/anti-parallel to the orbital angular momentum) \cite{PhenV2}. Moreover, some other improvements have been considered; for instance, they are using longer numerical waveforms, covering at least eight cycles before merger and also, they are matching (in the time domain) NR simulations to more accurate PN TaylorT1 waveforms that include  3.5PN phase accuracy, 3PN amplitude corrections to the dominant quadrupole mode, and  2.5PN spin-dependent corrections.  The improvement in the
 amplitude modeling due to this new set of phenomenological waveforms is visible on
 Fig.~\ref{fig:waveforms}.

For the purposes of our study, we shall only consider non-spinning waveforms, and therefore we shall be using just a particular case ($\chi = 0$) of the phenomenological waveforms defined in  \cite{PhenV2}. The definition of the closed-form, frequency-domain phenomenological amplitude and phase, $\{ A_{\mathrm{PhenV2}}(f) , \psi_{\mathrm{PhenV2}}(f) \}$, is very similar to what has been explained in the previous section, and is given in Eqs.~(1)-(3) and Table I of Ref.~\cite{PhenV2}.

In what follows, we will refer to this `version-2' phenomenological model as PhenV2, in contrast with the former PhenV1. The effective amplitude and phase  PhenV2 are compared to EOB,
and the other models, in Fig.~\ref{fig:waveforms}.


\section{Fiducial target waveform used in this study: NR-calibrated EOB waveform}
\label{sec:EOB-PN_merge}

\subsection{Effective-One-Body (EOB) formalism }
\label{subsec:models_EOB}

 The EOB formalism \cite{Buonanno:1998gg,Buonanno:2000ef,Damour:2000we,Damour:2001tu} 
has the unique feature of being able to incorporate  information
coming both from PN theory and numerical relativity (NR) simulations into an  accurate 
description of  the full dynamics, and GW radiation, of a compact binary coalescence, 
from the early inspiral up to the end of  the merger. In addition to incorporating the
full PN information available, the EOB formalism  goes beyond PN theory, even during
the inspiral, in several ways: (i) it replaces the PN-expanded results by \emph{resummed} expressions that have been shown to be more accurate in various cases; (ii) it includes 
more physical effects, notably those associated with the
\emph{non adiabatic} aspects of the inspiral. We recall
that all the various  `PN models'  presently considered in the literature make the
approximation that the GW phase can be computed in an adiabatic manner, by
using simply the balance between the GW energy flux and the adiabatic
loss of binding energy of the binary system. On the other hand, it has been found in 
\cite{Buonanno:2000ef} that the number of GW cycles before the LSO over which
non-adiabatic effects are important is of order $\sim (4\nu)^{-1/5}$. 
[See Equations (4.47)-(4.49) and Figure 7 in \cite{Buonanno:2000ef}; and also
Ref.~\cite{Ori:2000zn} for a similar result in the $\nu\to 0$ limit.]  
This indicates that the PN phasing will
accumulate, because of non-adiabatic effects, a dephasing with respect to the EOB one
which becomes $\sim (4\nu)^{-1/5}$ radians when reaching the LSO (see Figure 11 of
\cite{Buonanno:2000ef} for the $\nu=1/4$ case).  The accumulated dephasing due
to the non-resummed/resummed difference between PN and EOB can easily be larger than
this non-adiabatic effect.

Over the last years, it has become possible to improve the EOB waveform by
comparing it to the results of accurate NR simulations and by using the  natural `flexibility'
of the EOB formalism to `calibrate'  some EOB parameters representing either higher-order
perturbative effects that have not yet been analytically calculated, or non-perturbative effects
that can only be accessed to by NR simulations
\cite{Buonanno:2007pf,Damour:2007yf,Damour:2007vq,Damour:2008te,Damour:2009kr,Buonanno:2009qa}.
In addition, several theoretical improvements have been brought in the EOB formalism, 
notably concerning new ways of resumming the GW waveform \cite{Damour:2007yf,DIN},
and the GW radiation reaction \cite{Damour:2009kr}.

The currently most accurate version of the EOB waveform \cite{Damour:2009kr} agrees, 
within the numerical error bars, with the currently most accurate NR waveform
\cite{Damour:2008te,Boyle:2007ft,Scheel:2008rj}.  This means dephasing levels
of order of $\pm 0.02$ during the entire inspiral and plunge.  In view of the
expressions given for the `inaccuracy' functional above, this roughly corresponds
to an inaccuracy level $\mathcal I \sim (0.02)^2 = 4 \times 10^{-4}$. 
As we shall find below that the inaccuracy, with respect to EOB, of the closed-form
models presented in the previous section is much larger than this level, we can
meaningfully use the continuous bank of EOB waveforms as  `fiducially exact'
target waveforms in order to gauge the effectualness, and the accuracy of the closed-form
models.  More precisely, we shall use the EOB model
defined in Ref. \cite{Damour:2009kr}  with EOB parameters $a_5= -6.37$ and $a_6= 50$
and with a Pad\'e (3,2) resummation of the function $f_{22}(x)$.

\subsection{Time-domain EOB waveform}

From the practical point of view,  the EOB waveform is defined in the \emph{time domain}
by integrating some ODE's that cover the inspiral down to the `EOB merger',  and then
by attaching a ring-down signal, in a smooth manner, at the  instant of the `EOB merger'.
[Note that the time variable used in the EOB model is the dimensionless variable
$\hat{t} \equiv t/M$. The only parameter entering the integrated ODE's
is the symmetric mass ratio $\nu$.]
The EOB model is therefore `semi-analytical' in the sense that it needs to use
some numerical integration of the basic analytical ODE's defining the EOB dynamics.
We have used a Matlab implementation of the EOB dynamics that has not been optimized
for speed and/or efficiency. The integration time is typically less than $0.15$~s 
per orbital cycle [value obtained running on a single core of a 2.80 GHz quad-core Intel 
Xeon desktop machine with 4 Gb of RAM], so that we are able to
 simulate several thousands of orbital cycles in less than half an hour 
by running on a single CPU of a desktop machine\footnote{By comparison, note that 
a typical NR simulation takes about \emph{two days} per orbital cycle on a computer 
cluster with several hundreds of cores, which is a factor  $\sim 10^8$ longer 
[in actual CPU time] than EOB.}. 
Given its accuracy and speed, the EOB model is a very good candidate to be used 
as a full (inspiral-merger-ringdown) waveform generator in actual searches of GW, especially as
it is clear that some optimization would allow to significantly reduce the current integration
time per orbital cycle.

Though this procedure defines extremely long time-domain EOB waveforms, it
delivers them with an abrupt beginning, say at  the (dimensionless EOB) time $\hat{t}_b \, (=0)$.
This abrupt beginning at $\hat{t}_b $ is a necessary consequence of having started 
the EOB evolution at some finite initial radius $r_0$. [Note in passing that, though 
we start the EOB waveform with some specified initial phase at  $\hat{t}\subEOB=0$,
 the fact that we then continuously vary parameters such as $r_0$ and $\nu$ 
effectively implies that the phase of the generated EOB waveform around 
coalescence is essentially random.]

A direct Fourier transform of the   (time-windowed)  EOB waveform would introduce,
because of edge effects, some  significant oscillations in the frequency domain 
that would ``spoil'' the beginning of the frequency-domain 
waveform. [See, \eg , \cite{DIS:2000} for a detailed discussion of  the oscillations 
in $\tilde h(f)$ induced by a sharp edge in $h(t)$.]  In order to lose as little 
information as possible  from the original EOB evolution, we have 
``smoothed''  the initial edge of the time-domain waveform coming out
of the numerical integration by extending it to earlier times, $\hat{t}<\hat{t}_b$.
Namely:
\begin{enumerate}
\item Since the times $\hat{t}<\hat{t}_b$ are deep into the adiabatic inspiral, 
the signal there would be analytically well described by some post-Newtonian 
chirping time-domain model. However, as our aim is purely technical, namely to 
avoid the oscillations due to edge effects to spill over frequencies $f$ larger 
than the initial time-domain frequency $F(\hat{t}_b)$, we found sufficient to 
move the abrupt starting edge towards earlier times by attaching, say for 
$\hat{t}_a<\hat{t}<\hat{t}_b$, a simple phenomenological Newtonian time-domain
evolution of the form
\begin{equation*} \begin{array}{rcl}
A(t) & = & \alpha_2 (\alpha_1 - \hat{t})^{-1/4} \\
\phi(t) & = & \beta_3 + \beta_2 (\beta_1 - \hat{t})^{5/8}  \end{array} 
\end{equation*}
We determined  the $5$ phenomenological parameters $\alpha_n, \beta_n$
by imposing continuity, at $\hat{t}=\hat{t}_b$, of $A$, $\dot{A}$, $\phi$, 
$\dot{\phi}$ and $\ddot{\phi}$ with the EOB. Given the one-to-one relation 
between frequency and time in a chirping signal, this oversimplified  left 
extension is enough to displace the edge effects due to the abrupt starting 
of the waveform to a controllable  frequency-domain region localized around 
$F(\hat{t}_a)$. We choose $\hat{t}_a$ so that $F(\hat{t}_a)$ is sufficiently 
smaller than $F(\hat{t}_b)$, thereby having a smooth frequency-domain waveform 
for any $f \geq F(\hat{t}_b)$. 
\item Finally,  to further suppress any edge effect related to the sharp start of the
waveform at the left-ward displaced time $\hat{t}_a$,  we multiply this Newtonian-like
time-domain waveform by a  (time-domain) semi-Tukey filter in the interval
$\hat{t}_a\leq \hat{t} \leq \hat{t}_a+\frac{\hat{t}_b-\hat{t}_a}{3}$. 
This filter replaces the sharp start at $\hat{t}=\hat{t}_a$
by a progressive build up of the waveform during the first third of the Newtonian extension.
\end{enumerate}

At the end of this procedure, we have a very long, smooth  time-domain waveform that
vanishes for  $\hat{t}<\hat{t}_a$,  smoothly starts building up at $\hat{t}=\hat{t}_a$, and ends up decaying
exponentially [deep in the ring-down stage] when $\hat{t}\to + \infty$.
This time-domain waveform is now ready to be Fourier transformed. [A similar procedure has
been used by authors who needed to compute the Fourier transform of
short, time-windowed time-domain NR waveforms, see \eg \cite{Buonanno:2006ui}.] 

\subsection{Frequency-domain  EOB waveform}

Starting from the extended time-domain EOB waveform defined in the previous subsection, 
we now compute its FFT.  However, we must discard from the 
 resulting frequency-domain waveform the low-frequency contribution coming from the Newtonian extension, i.e. anything from $\fM<f_b M$, where $f_b M = \dot{\phi}\subEOB(\hat{t}_b) / (2\pi)$. This means that, so far, we have constructed only a 
version of the full frequency-domain EOB waveform which is cut-off on the low-frequency side:   $\tilde{h}\subEOB(f) \theta(f-f_b)$.
We wish now to \emph{extend to lower frequencies}, $f<f_b$, this frequency-truncated
EOB waveform in order to have the complete, frequency-domain EOB waveform.
We can do this low-frequency extension by \emph{joining together, in the frequency domain},
an early frequency-domain PN waveform PN(f) (as defined above) to the 
frequency-truncated  EOB waveform $\tilde{h}\subEOB(f) \theta(f-f_b)$ we
just constructed. We will do this  \emph{joining} of PN(f) with EOB(f) in a smooth and
progressive manner, over a certain finite frequency interval $[f_0, f_1]$,  entirely located
within the range of frequencies where we have, so far, computed $\tilde{h}\subEOB(f)$. 
In other words, we impose the inequalities $f_b \leq f_0<f_1$.  The choice $f_0=f_b$
will define the best approximation we can construct of the complete frequency-domain
EOB waveform EOB(f).

Technically, we accomplished this  progressive `joining'  between a low-frequency PN(f) 
and a higher-frequency EOB(f)  by  constructing
\begin{align} \label{eq:EOB(f)}
\tilde{h}(\fM)& = (1-\Theta_{f_0M,f_1M}(\fM)) \; \tilde{h}\subPN^{t\subC,\phi\subC}(\fM)\nonumber\\
              & + \Theta_{f_0M,f_1M}(\fM) \; \tilde{h}\subEOB(\fM)
\end{align}
where $ \Theta_{f_0M,f_1M}(x)$ is an infinitely smooth, $C^{\infty}$, step-like transition function that goes from $0$ to $1$ in a finite\footnote{Note that the often used smooth transition functions
based on hyperbolic tangent functions are also $C^{\infty}$, but interpolate between $0$ to $1$ only over
the infinite real line.} interval $[ f_0M, f_1M ]$.  Specifically, we have used
\begin{equation}
\Theta_{x_0,x_1}(x) \equiv \left\{ \begin{array}{ll}
0                                                                                  & x \leq  x_0 \\
\left[ 1+e^{\frac{1}{x-x_0}+\frac{1}{x-x_1}} \right]^{-1} &  x_0 < x < x_1 \\
1                                                                                  & x \geq x_1
\end{array} \right .
\end{equation}
In addition, we have chosen the extrinsic parameters $t\subC, \phi\subC$ entering the
PN(f) waveform $\tilde{h}\subPN(f)$ (see Eq.~(\ref{eq:psiPN})) so as to minimize
the phase difference between $\psi\subPN(f)$ and $\psi\subEOB(f)$ in the junction interval.
Note that the PN(f) waveform, $\tilde{h}\subPN(f)$ is generated 
with the same physical parameters $\{ D_\eff , M , \nu \}$ that were used for the EOB simulation.

\begin{figure}
\centering
\includegraphics[width = 8cm]{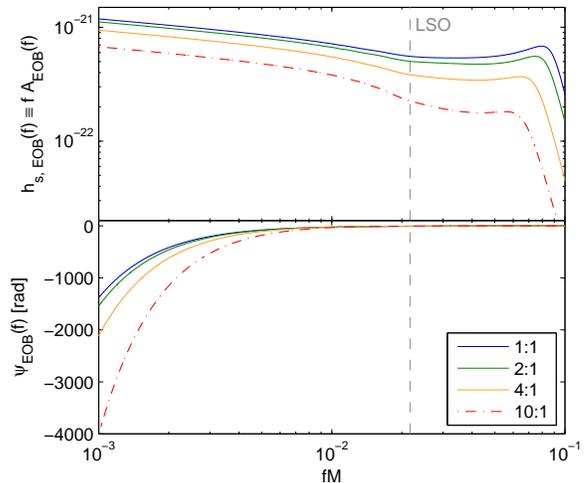}
\caption{EOB(f):  frequency-domain gravitational wave effective amplitudes, 
$f |\tilde h(f)|$ (top panel), and phases, $\psi(f)$  (bottom panel) , 
obtained from EOB simulations of BBHs coalescences with different mass ratios. 
[As in Fig.~\ref{fig:waveforms}, the actual values of the amplitude have been 
obtained assuming a total mass of $75\Ms$ at $1$ Gpc.]}
\label{fig:EOBwaveforms_different_nu}
\end{figure}

Finally,  for the practical purpose of this study, we  shall \emph{define} the 
complete frequency-domain EOB waveform EOB(f) as the 
`smooth frequency-domain PN$\cup_{f_0,f_1}$EOB hybrid'  constructed in
 Eq. (\ref{eq:EOB(f)}), with the choices $f_0=f_b$ and $f_1= 1.05 f_0$, where $f_b$ is
the GW frequency corresponding to the beginning of the longest EOB simulations
we generated. More precisely, we have explored a fine grid of mass ratios $q=m_2/m_1 \geq 1$
between $q=1$ and $q=10$. Given our current implementation and running our simulations on a 
single desktop computer with a 2.80 GHz quad-core and 4 Gb of RAM, we could, when 
$1\leq q\leq 4$, start the simulations at $r_0 =R_0/M = 90$, which corresponds to an initial
GW frequency $M f_b=1/(\pi r_0^{3/2}) \approx 3.73 \times 10^{-4}$. For larger mass ratios,
up to $q=10$,  we started at $r_0=76$, which corresponds to  $M f_b \approx 4.80 \times 10^{-4}$. 
The effective amplitude and phase of this complete EOB(f) waveform is represented
in Fig.~\ref{fig:EOBwaveforms_different_nu} for the values $q = {1,2,4,10}$ of the
 mass ratio. Note that this represents simulating $1553$, $1744$, $2414$ and $3042$ orbital cycles, 
respectively, that took between $4$ and $10$ minutes of computational time on a single CPU.

The starting point of these simulations is in the very deep inspiral phase of the evolution,
 where both SPA and PN approximations are perfectly valid. Thus, we shall consider these 
waveforms as the ``correct'' ones, using them as the target when computing the 
inaccuracy of PN-SPA.

\section{Accuracy upper limit on the junction frequency $f_0$  of hybrid 
PN$(f)\cup_{f_0} \rm{EOB}(f)$  waveforms}
\label{sec:PN-SPA_range}

In this section we shall address a question that is a side-issue with respect to the main
aim of this paper, but which is important in the general context of constructing accurate,
frequency-domain waveform models. This question concerns the \emph{upper limit} on the 
frequency $f_0$ parametrizing\footnote{For simplicity and concreteness,
we shall henceforth use, for
the smooth joining of PN(f) with EOB(f), Eq. (\ref{eq:EOB(f)}), a frequency interval $[f_0,f_1]$ with a
\emph{fixed} ratio $f_1/f_0=1.05$. This leaves the single parameter $f_0$ to characterize
the `junction frequency' between the two waveforms.}  the joining between a lower-frequency
 PN(f) waveform to a higher-frequency EOB(f) waveform which does not lead to an undue loss 
of accuracy for the resulting hybrid waveform PN(f)$\cup_{f_0}$EOB(f).  We shall measure the loss
of accuracy of the (frequency-domain) hybrid  PN(f)$\cup_{f_0}$EOB(f) by means of the
`inaccuracy functional'  $ \mathcal{I}[h_m[f_0];h_x]$, Eq. (\ref{eq:inaccuracy2}), where 
$h_m[f_0]$ denotes the hybrid PN(f)$\cup_{f_0}$EOB(f), and where $h_x$ denotes our
best approximation to the full, exact EOB(f) waveform, i.e., as explained above, the
hybrid PN(f)$\cup_{f_b}$EOB(f) generated with the very low frequency $f_b$ at which
we started the EOB simulation.  We recall that $M f_b \sim 4 \times 10^{-4}$ 
corresponds to such a non-relativistic and adiabatic part of the inspiral (thousands of
orbital cycles before the merger of the BBH) that we can safely consider the lower-frequency PN(f) 
piece in the hybrid PN(f)$\cup_{f_b}$EOB(f) as being indistinguishable from the low-frequency part
of the exact EOB(f).  On the other hand, in view of Fig.~\ref{fig:waveforms}, we expect that
PN(f) will start to significantly deviate from EOB(f) for frequencies somewhere between
$\fM \sim 10^{-3}$ and $\fM \sim 10^{-2}$, thereby leading to an increasing inaccuracy
$ \mathcal{I}[h_m[f_0];h_x]$ of the hybrid PN(f)$\cup_{f_0}$EOB(f) as $f_0 M$ reaches such values.

There are two motivations for this study. First, though we have been able, for this work, 
to easily generate many, very long time-domain EOB waveforms, containing thousands of 
orbital cycles, and then use an automated version of the procedure explained above to 
generate the corresponding frequency-domain EOB waveforms, it might be useful, when one 
will need to construct dense banks of accurate frequency-domain EOB waveforms, to cut 
down to a strict minimum the integration time of each EOB simulation.  
Determining the maximum $f_0$ leading to some given inaccuracy will be useful in 
allowing one to start the EOB simulation at a higher GW frequency $f_b = f_0$. 
A second motivation, is that our result will clearly also tell us  the maximum frequency 
at which we must start a NR simulation (\ie the minimum number of  inspiral orbital 
cycles it must include) if we wish to be able to extend (to lower frequencies) 
its Fourier-transformed GW signal by using a simple closed-form PN(f) waveform.

As said above, in order to study the frequency upper limit for the accuracy of 
PN(f) waveforms, we generate frequency-domain hybrid waveforms  $h_m[f_0](f)={\rm PN}(f)\cup_{f_0} {\rm EOB} (f)$,
 and compute the relative inaccuracy  $ \mathcal{I}[h_m[f_0];h_m[f_b]]$ as a function
of the joining frequency $Mf_0$, or, equivalently, $M\omega_0 \equiv 2\pi Mf_0$.
The inaccuracy functional will also depend on the particular symmetric mass ratio
$\nu$ one is considering, and on the value of the total mass $M$ (in seconds), since
the detector noise curve that enters the definition Eq. (\ref{eq:inaccuracy2}) depends on
the physical frequency $f$ (in Hz) and not on the dimensionless scaled frequency $\fM$.
In particular, we shall consider four different values of $\nu$ corresponding to the
mass ratios $q=1, 2, 4, 10$ and we shall maximize $\mathcal{I}(Mf_0, \nu, M)$, for
each value of $Mf_0$, and $\nu$, over $M$, \ie we shall consider the worst case
scenario for the loss of accuracy. Since we are considering ground-based detectors,
the mass range will be considered to be $M\in[3,500]\Ms$. This yields, for each
value of $\nu$,  the worst-case inaccuracy as a function of $Mf_0$. 
We have done this study for the three different detector noise curves discussed above, 
and represented in Fig.~\ref{fig:IFO_PSDs}: initial LIGO, advanced LIGO, and advanced 
Virgo. Note that the inaccuracy functional only depends on the \emph{shape} of 
these noise curves, and not on their absolute magnitudes.

\begin{figure*}[t]
\begin{center}
\begin{tabular}{p{10cm}p{7.5cm}}
\centering
\includegraphics[width = 10cm]{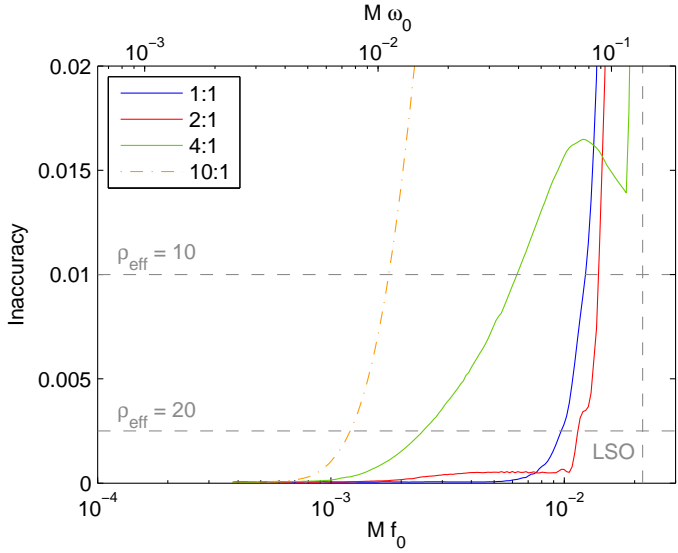}

&
\vspace{-7.5cm}
\begin{tabular}{c|cccccc}
\multicolumn{7}{c}{ {\bf Initial LIGO} ~;~  $\rho_{\rm eff}=10$ ($\rho_{\rm eff}=20$)} \\[0.05cm]
                  & ~ & $f_0M \times 10^{2}$ & ~ &    $R_0/M$       & ~ & orbits pre-merger \\ \hline \\[-0.3cm]
\bf{1:1}      & ~ &    $1.28$ ($1.02$)       & ~ &    $8.51$ ($9.89$)    & ~ &   $3$ $(5)$        \\
\bf{2:1}      & ~ &    $1.42$ ($1.15$)       & ~ &    $7.96$ ($9.16$)    & ~ &   $3$ $(4)$        \\
\bf{4:1}      & ~ &    $0.81$ ($0.31$)       & ~ &    $11.6$ ($22.1$)    & ~ &   $10$ $(65)$      \\
\bf{10:1}    & ~ &    $0.23$ ($0.17$)       & ~ &    $26.9$ ($33.0$)    & ~ &   $206$ $(359)$    \\
\multicolumn{7}{c}{} \\
\multicolumn{7}{c}{{\bf Adv. LIGO} ~;~  $\rho_{\rm eff}=10$ ($\rho_{\rm eff}=20$)} \\[0.05cm]
                  & ~ & $f_0M \times 10^{2}$ & ~ &   $R_0/M$     & ~ & orbits pre-merger \\ \hline \\[-0.3cm]
\bf{1:1}      & ~ &    $1.24$ ($0.95$)       & ~ &     $8.70$ ($10.4$)      & ~ &    $3$ $(6)$      \\
\bf{2:1}      & ~ &    $1.40$ ($1.14$)       & ~ &     $8.03$ ($9.21$)      & ~ &   $3$ $(4)$       \\
\bf{4:1}      & ~ &    $0.63$ ($0.25$)       & ~ &     $13.7$ ($25.4$)      & ~ &   $17$ $(95)$     \\
\bf{10:1}    & ~ &    $0.18$ ($0.12$)       & ~ &     $31.9$ ($41.1$)      & ~ &   $327$ $(634)$   \\
\multicolumn{7}{c}{} \\
\multicolumn{7}{c}{{\bf Adv. Virgo} ~;~  $\rho_{\rm eff}=10$ ($\rho_{\rm eff}=20$)} \\[0.05cm]
                  & ~ & $f_0M \times 10^{2}$ & ~ &    $R_0/M$       & ~ & orbits pre-merger \\ \hline \\[-0.3cm]
\bf{1:1}      & ~ &    $1.11$ ($0.90$)       & ~ &    $9.38$ ($10.8$)    & ~ &    $4$ $(6)$    \\
\bf{2:1}      & ~ &    $1.35$ ($1.11$)       & ~ &    $8.21$ ($9.34$)    & ~ &    $3$ $(4)$     \\
\bf{4:1}      & ~ &    $0.36$ ($0.18$)       & ~ &    $19.7$ ($31.8$)    & ~ &    $47$ $(171)$   \\
\bf{10:1}    & ~ &    $0.14$ ($0.10$)       & ~ &    $38.1$ ($47.8$)    & ~ &   $521$ $(943)$   \\
\end{tabular}

\end{tabular}
\caption{\emph{(Plot)} ``Inaccuracy''   $ \mathcal{I}[h_m[f_0];h_m[f_b]]$ 
[defined by Eq.~(\ref{eq:inaccuracy1})
or  Eq.~(\ref{eq:inaccuracy2})] of frequency-domain 
hybrid PN(f)$\cup_{f_0}$EOB(f) waveforms as we move their
junction frequency, $M f_0$ ($\equiv M\omega_0/2 \pi $), to higher values; taking as \emph{reference}
waveform an  hybrid PN(f)$\cup_{f_b}$EOB(f) where the junction frequency is 
extremely low ($M f_b \sim 4 \times 10^{-4}$). We have computed  $ \mathcal{I}$ for $4$ different mass ratios, 
using the Adv. LIGO noise PSD and maximizing the inaccuracy over $M \in [3,500]\Ms$. The vertical dashed 
line represents the frequency of the Schwarzschild LSO and the two horizontal dashed lines indicate the 
accuracy thresholds corresponding to the indicated effective SNRs  $\rho_{\rm eff} \equiv \rho/\epsilon$,  
according to Eq.~(\ref{eq:faith_cond_expanded}). \emph{(Table)} Upper limits (for three different 
detectors noise PSDs, and four mass ratios)
on the value of the junction frequency $M f_0$
 above which the inaccuracy of the hybrid PN(f)$\cup_{f_0}$EOB(f)
becomes larger than the threshold corresponding to an effective SNR equal to
 $10$ and $20$. We have also converted them into a minimum distance between the BHs, $R_0$, and 
a minimum number of orbital cycles before merger (see text).}
\label{fig:PN-SPA_range}
\end{center}
\end{figure*}

We represent in Fig.~\ref{fig:PN-SPA_range}  the inaccuracy, $ \mathcal{I}[h_m[f_0];h_m[f_b]]$,
 maximized over $M$, as a function of  the junction frequency  $Mf_0$,  for $4$ different mass 
ratios, and for the Advanced LIGO noise PSD. According to Eq.~\eqref{eq:faith_cond_expanded},  
the maximum inaccuracy that we can tolerate depends both on the signal-to-noise ratio, 
$\rho$, and on the choice of an additional ``safety factor'' $\epsilon$. 
More precisely, what matters is only what we can call the
\emph{effective SNR}, namely $\rho_{\rm eff} \equiv \rho/\epsilon$. 
Two models will be indistinguishable if their relative inaccuracy is smaller than
$1/\rho_{\rm eff}^2$. In the present section we shall consider the effective SNRs
$\rho_{\rm eff}=10$ or $\rho_{\rm eff}=20$. Note that, as we expect typical
observations to have a real SNR of $10$, the inaccuracy limit $1/\rho_{\rm eff}^2$ with
$\rho_{\rm eff}=10$ corresponds to taking $\epsilon =1$, \ie to allowing the inaccuracy 
of the  model waveform to \emph{double} the effect of the detector noise. 
It would seem more reasonable to take at least $\epsilon=1/2$, \ie to consider 
an effective SNR of $\rho_{\rm eff}=20$.

The most interesting feature of Fig.~\ref{fig:PN-SPA_range} is the notable
difference between the case of mass ratios near unity (namely $q=1$ and $q=2$),
and the case of large mass ratios: $q=4$ and $q=10$.  When $q=1, 2$ 
Fig.~\ref{fig:PN-SPA_range} shows that the accuracy loss remains quite small up to
frequencies rather close to the Schwarzschild-like LSO frequency 
$M f\subLSO=(6\pi \sqrt{6})^{-1} \approx 0.02166$. [One should, however, 
recall that
the real (EOB-predicted) LSO frequency is  $42 \%$ larger, being close 
to $M f\subLSO^{\mathrm EOB} \approx 0.0308$.]  In other words, the PN(f) model
produces accurate waveforms up to near the LSO when $q=1, 2$.
By contrast, when the mass ratio increases, namely $q=4, 10$, the inaccuracy
gets significant for  frequencies which are much smaller than the LSO frequency
(especially if one considers the threshold $\rho_{\rm eff}=20$).
As the number of orbital cycles between $f_0$ and $f_1$ is approximatively 
given by the ``Newtonian'' estimate (see, \eg , \cite{DIS:2000})
\be \label{eq:Norbits}
N^{\rm Newt}[f_0,f_1] = [(\pi Mf_0)^{-5/3} - (\pi Mf_1)^{-5/3}]/(64 \pi \nu)
\ee
such a smaller $M f_0$ means a much larger number of orbital cycles left
between $f_0$ and, say, $f_1 \sim f\subLSO$ (both because of the power-law
dependence on $M f_0$, and of the explicit proportionality to $1/\nu$).
Even if we consider the somewhat lax threshold $\rho_{\rm eff}=10$, we
see on  Fig.~\ref{fig:PN-SPA_range} that the PN(f) waveform can be used:
up to $M f \approx 0.014$ when $q=1$ or $2$; only up to a twice smaller frequency 
$M f \approx 0.008$ when $q=4$; and only up to a seven times smaller frequency
 $M f \approx 0.002$  when $q=10$.

The Table of Fig.~\ref{fig:PN-SPA_range}  gives more quantitative information
about this accuracy upper limit on the junction frequency to PN(f). The table
considers all three detectors noise PSD, and the two thresholds  
$\rho_{\rm eff}=10$ and $\rho_{\rm eff}=20$ (the quantitative data pertaining to the
$\rho_{\rm eff}=20$ threshold are given within parentheses).
In particular, the last two columns
indicate the  conversion of $M f_0$ in separation $r_0 = R_0/M$
between the BHs, and in number of orbital cycles left before reaching the merger.
[The conversion between $M f_0$ and $r_0 = R_0/M$ was formally done using the 
approximate formula $r=(\pi f M)^{-2/3}$, while the number of orbital cycles was
computed by using the underlying EOB dynamics; defining the ``merger'' 
as the instant where the (metric)  waveform reaches its maximum amplitude.]

Notice the steep rise of the number of orbital cycles left as the mass ratio gets larger than 4.
For a given accuracy threshold $1/\rho_{\rm eff}^2 $ the  Advanced Virgo  detector 
is the most demanding one for  the PN(f) model because the \emph{shape} of its effective noise curve
peaks at a lower frequency, but note that, for a given signal amplitude the SNR of 
Advanced Virgo would be smaller than that of Advanced LIGO. 

Leaving to  future work~\cite{DNT2} a detailed discussion of the precise meaning
of these results  (notably in terms of the intrinsic accuracy of the PN(f) waveform),
let us mention that our results suggest that the ``inaccuracy''  of 
 a frequency-domain hybrid, say PN(f)$\cup_{f_0}$NR(f), between a
lower-frequency PN(f) and a higher-frequency NR(f) waveform will be similar
to the one plotted in Fig.~\ref{fig:PN-SPA_range}.  In particular, we think that,
if one requires the same type of threshold on the inaccuracy, 
the number of orbital cycles before merger   will have to be similar
to the numbers indicated in the last columns of the Table.
On the other hand, let us recall that
numerical simulations are computationally very expensive, and the current longest 
waveforms being produced span at most the last $10-15$ orbital cycles of 
the evolution~\cite{Hannam:2009rd, Hinder:2010vn}. Recently, a joint effort 
between several NR groups and scientists producing analytical models~\cite{NRAR_col} 
has begun to make use of TeraGrid resources in order to produce longer and more 
accurate simulations. However, one does not currently foresee the possibility of producing
more than $15-20$ orbital cycles. Comparing these current NR capabilities with the 
numbers we are obtaining here for the upper-frequency-limit of  frequency-domain hybrids
using PN(f) for $f<f_0$, we observe that, for mass ratios $q \geq 4$ it will be impossible
to ``join''   a PN(f) waveform to a NR(f) one without incurring an unacceptably
large accuracy loss.  In intuitive terms, one can say that we are here facing the probable 
existence of a \emph{frequency gap}  where neither PN theory nor current NR
simulations  are able to provide accurate enough information. We think, however, 
that the effective-one-body approach, aided by suitably designed
NR calibrations, may be  able to bridge this dangerous frequency gap, and 
then to construct some accurate-enough complete waveform models.

\section{Validity ranges of PN$(f)$ and of Phenomenological models}
\label{sec:results}

This section will present the central results of our work: namely the quantitative
study of the ineffectualness and of the inaccuracy of the closed-form, frequency-domain
waveforms presented above, with respect to the fiducial target waveform defined by the 
NR-calibrated EOB waveform (computed in the frequency-domain, as explained above). 
[For brevity, we shall sometimes denote the NR-calibrated EOB
waveform as  EOB$_{\rm NR}$.]  Our aim is to assess the 
\emph{regions of the parameter space of (non-spinning) BBHs}, \ie $M$ and $\nu$, \emph{where}:

\begin{itemize}

\item  either one of the closed-form models (PN(f), PhenV1 or PhenV2) 
provides an \emph{effectual} representation
of   EOB$_{\rm NR}$; as measured by the value of the \emph{effectualness}, see  
Eq.~\eqref{eq:effectualness}; or,

\item either one of  the closed-form models  (PN(f), PhenV1 or PhenV2)
provides an \emph{accurate} representation
of   EOB$_{\rm NR}$; as measured by the value of the \emph{inaccuracy}, see
Eq.~\eqref{eq:faith_cond_expanded}.

\end{itemize}

We did not investigate whether a suitable PN(f)$\cup_{f_0}$PhenV2(f) \emph{hybrid} might
be closer to EOB$_{\rm NR}$(f). Indeed, Phen waveforms are already supposed to have incorporated
a  PN-type early behavior, with an NR-type later waveform.

Before presenting our quantitative results,  let us anticipate some general aspects of our results
by noting that one a priori expects that neither PN(f) nor phenomenological models
will be able to accurately describe all stages of the evolution of BBHs. Through the
detector noise PSD, this lack of uniform physical validity will then
 translate into validity ranges in the space of physical parameters, $M$ and $\nu$.
In particular, we expect the PN(f) waveforms to only provide a valid description of the 
early inspiral stage of the evolution up to a certain frequency [$\sim f_0(M, \nu)$, see Sec.~\ref{sec:PN-SPA_range} above], 
but to start failing as the two BHs get close 
and the dynamics becomes more relativistic and less adiabatic.
Given the inversely proportional relation between the total mass and the frequency, 
inspiral-only models, such as PN(f) (or, for that matter, any other PN waveform), 
will be adequate only for searches of relatively low mass systems, $M\in[2,20]\Ms$, 
in initial ground-based detectors 
(for such systems  the merger occurs at higher frequencies than
the detectors sensitivity window; see the horizontal segments near the
bottom of Fig.~\ref{fig:waveforms}).
 On the other hand, the phenomenological models are built to effectively describe the merger phase of 
the evolution. They  can therefore be expected to
provide an adequate method to search for high mass systems (say $ M> 50 \Ms$), where 
the merger phase is the main contribution to the SNR (see  Fig.~\ref{fig:waveforms}). 
This preliminary discussion suggests 
that there might exist an intermediate range of masses, around 
$20 \Ms \lesssim M \lesssim 50 \Ms$,  where none of the closed-form models
can provide an accurate description. We shall see that this is indeed the case.

\begin{figure*}[t]
\begin{center}
\begin{tabular}{ccc}
\includegraphics[width = 6cm]{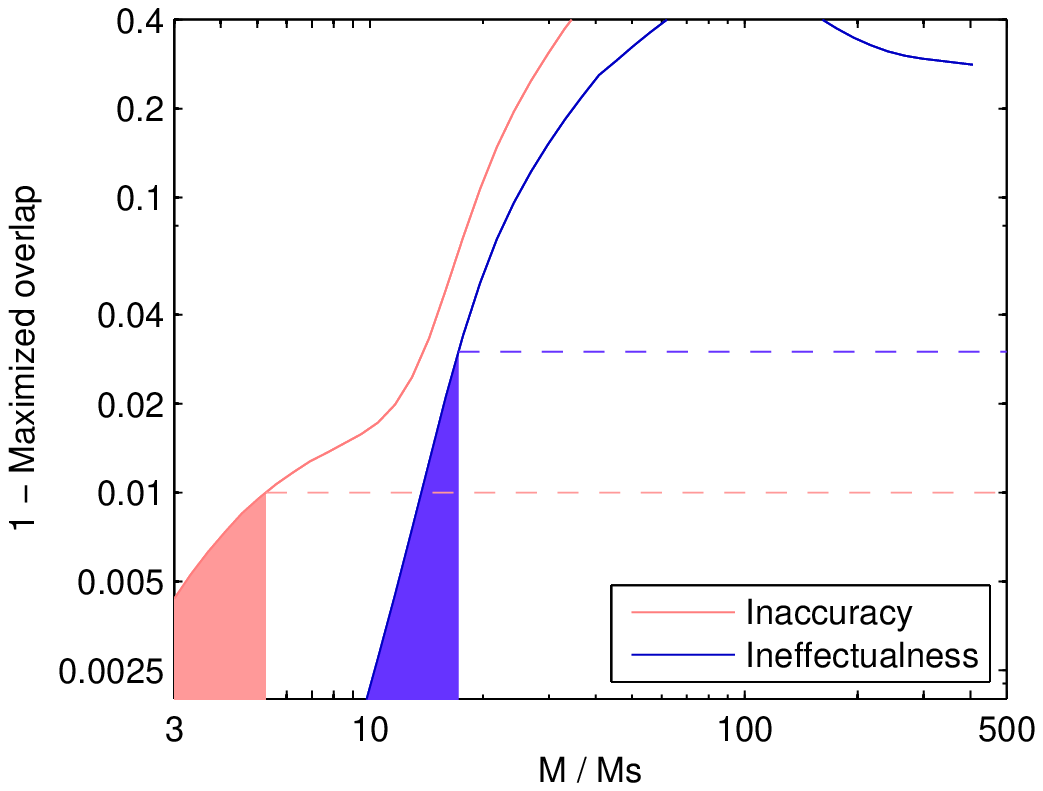} & 
\includegraphics[width = 6cm]{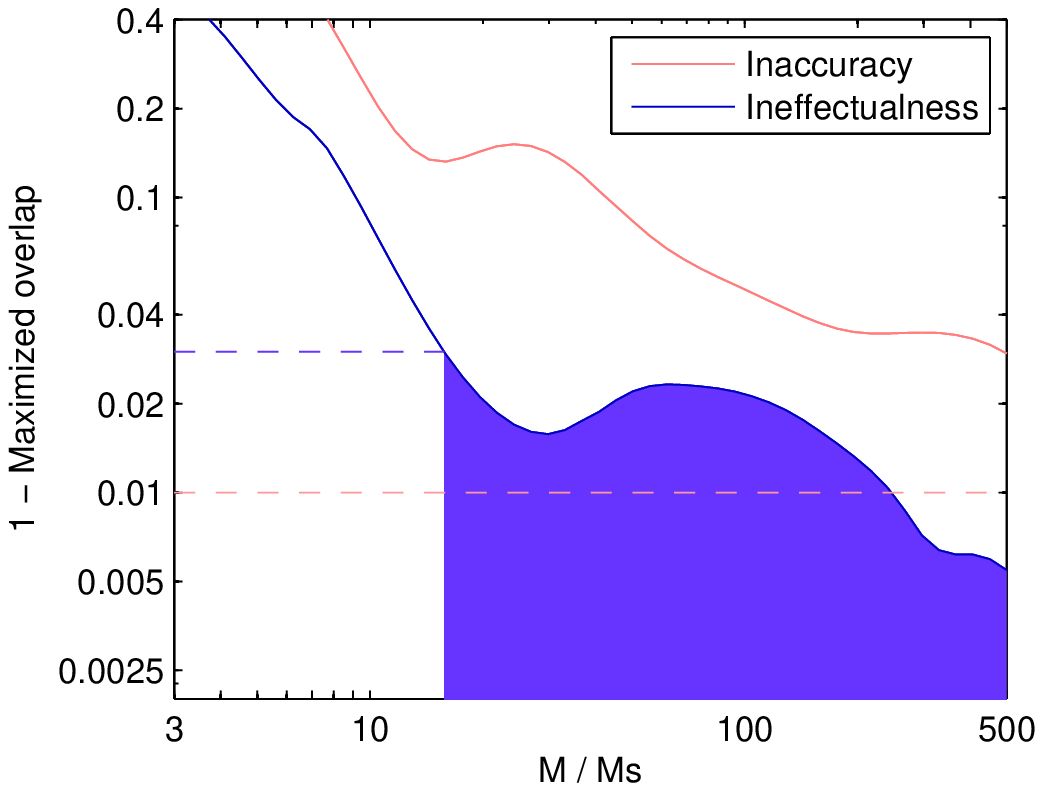} & 
\includegraphics[width = 6cm]{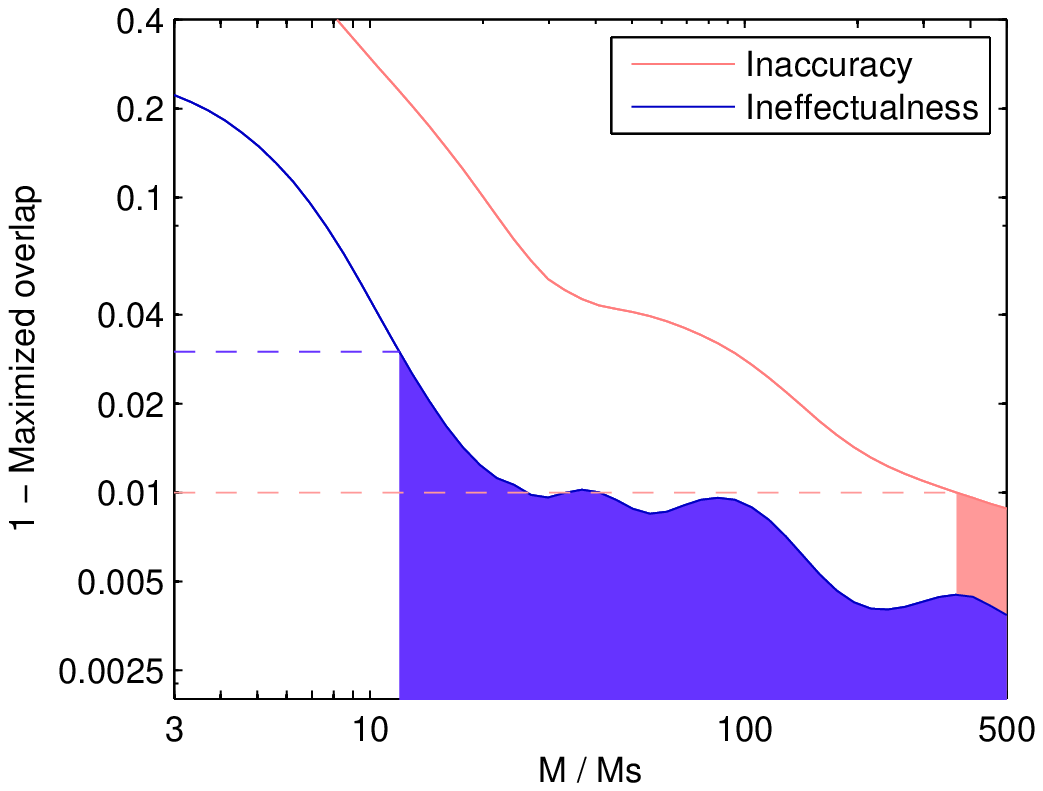} \\
\includegraphics[width = 6cm]{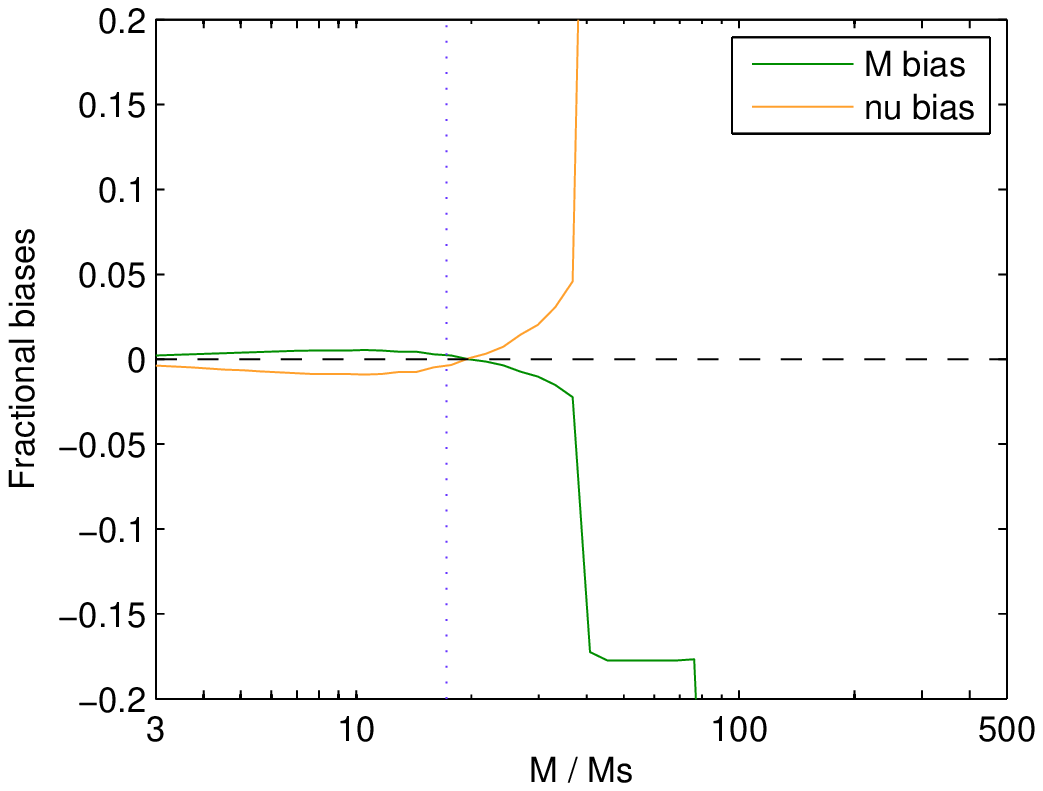} & 
\includegraphics[width = 6cm]{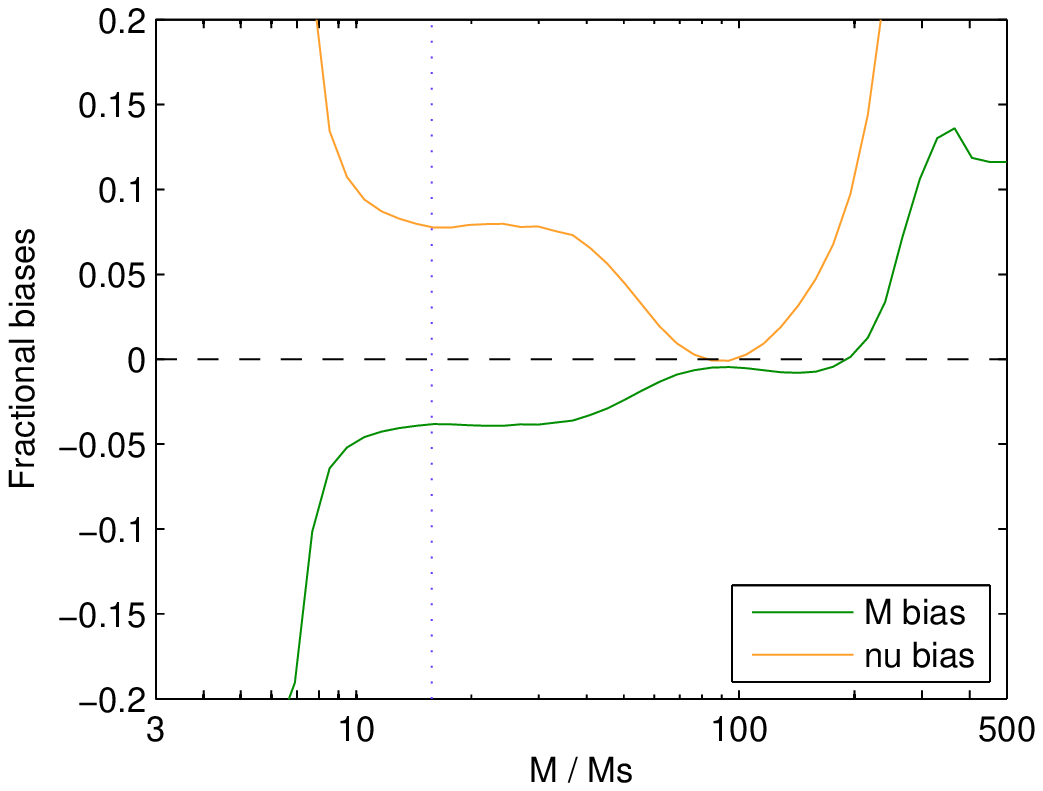} & 
\includegraphics[width = 6cm]{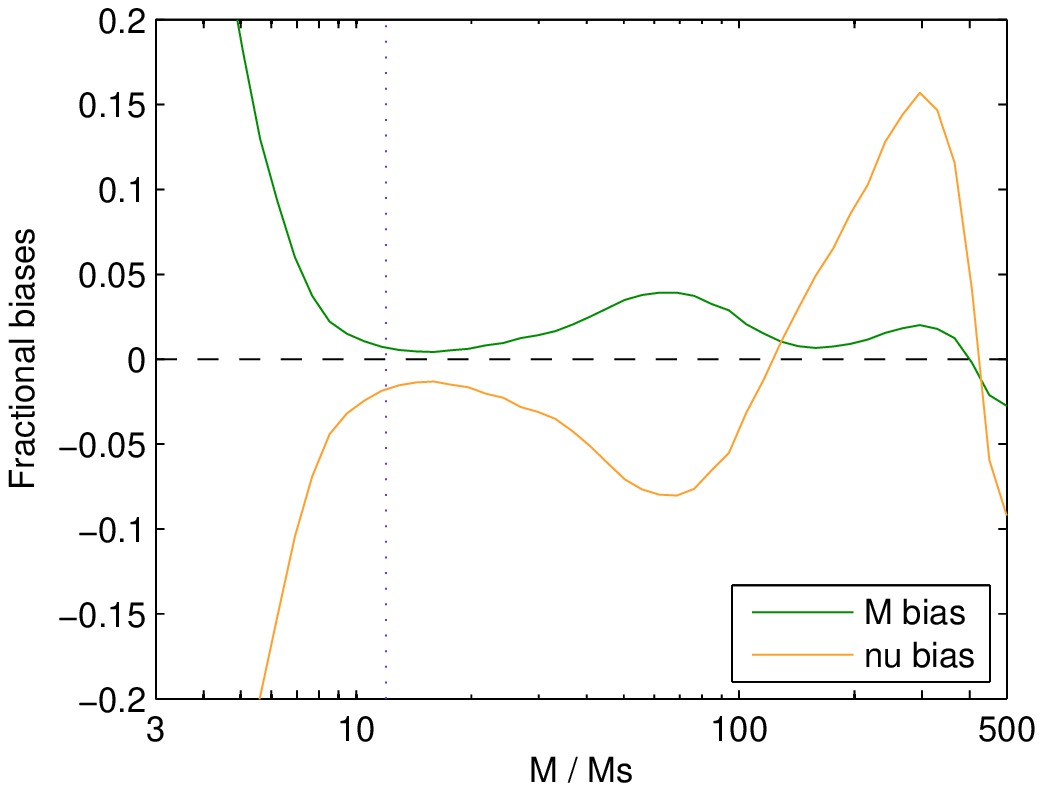} \\
(a) PN & (b) PhenV1 & (c) PhenV2
\end{tabular}
\end{center}
\caption{Top panels: Ineffectualness $\bar{\mathcal E} = 1- \mathcal E$
[dark gray (blue) solid line; see Eq.~(\ref{eq:def_effectualness})] and inaccuracy $\mathcal I$ [light gray (red) solid line; see Eq.~(\ref{eq:inaccuracy1})  or Eq.~(\ref{eq:inaccuracy2})] of the different closed-form models compared to an EOB(f) waveform, as functions of the total mass $M$ of the system (for a 4:1 mass ratio, and using Adv. LIGO noise PSD). The horizontal dashed lines represent rather \emph{minimal thresholds} on these quantities: $3\%$ for the ineffectualness, and $1\%$ for the inaccuracy (see text).
 Bottom panels: Fractional biases on the $M$ and $\nu$ parameters after maximizing the overlap between the model and target waveforms when computing the effectualness (see Eq.~(\ref{eq:maximization_effectualness})). The horizontal dashed line represents the zero biases and the vertical one denotes the $M$ value where the ineffectualness reaches $3\%$  in the top panels.}
\label{fig:overlaps}
\end{figure*}

\begin{figure*}[t]
\begin{center}
\begin{tabular}{p{7pt}cccccc}
 & & & \multicolumn{2}{c}{PN(f) -- PhenV2} & & \\
 \vspace{-2.4cm} $\bar{\mathcal{E}}$:  &
\multicolumn{2}{c}{\includegraphics[width = 5.7cm]{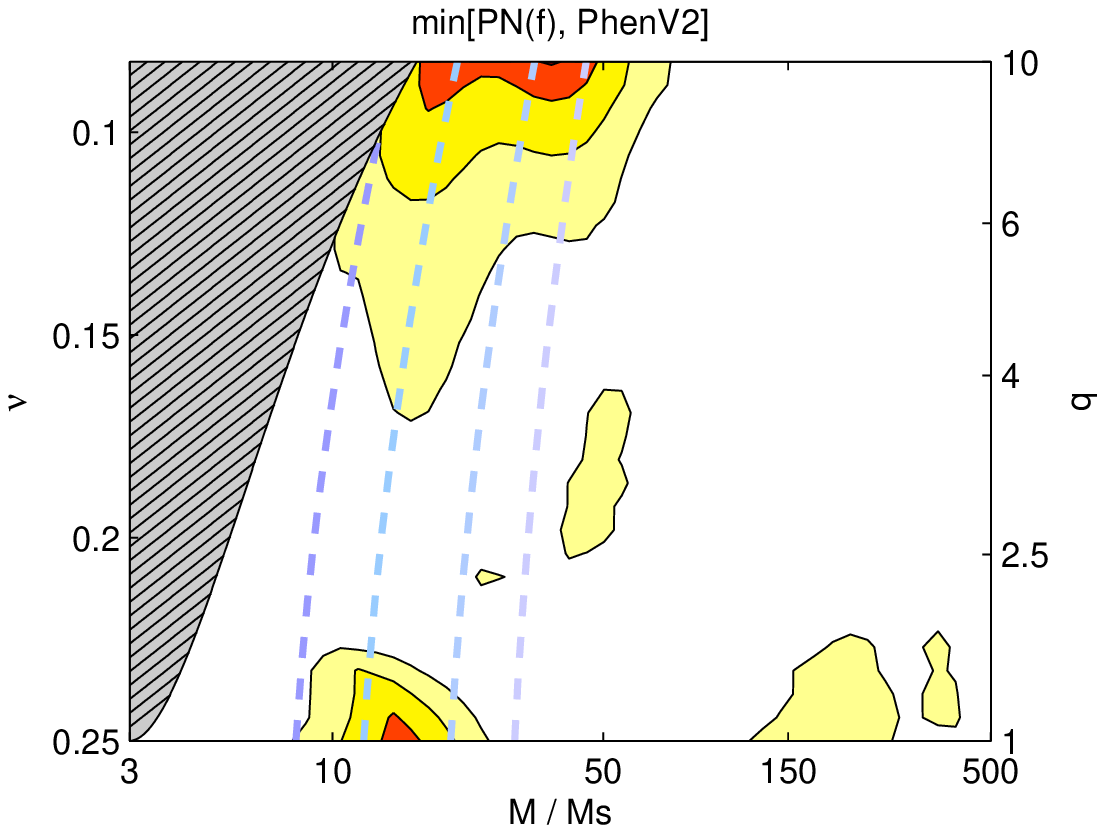}} & 
\multicolumn{2}{c}{\includegraphics[width = 5.7cm]{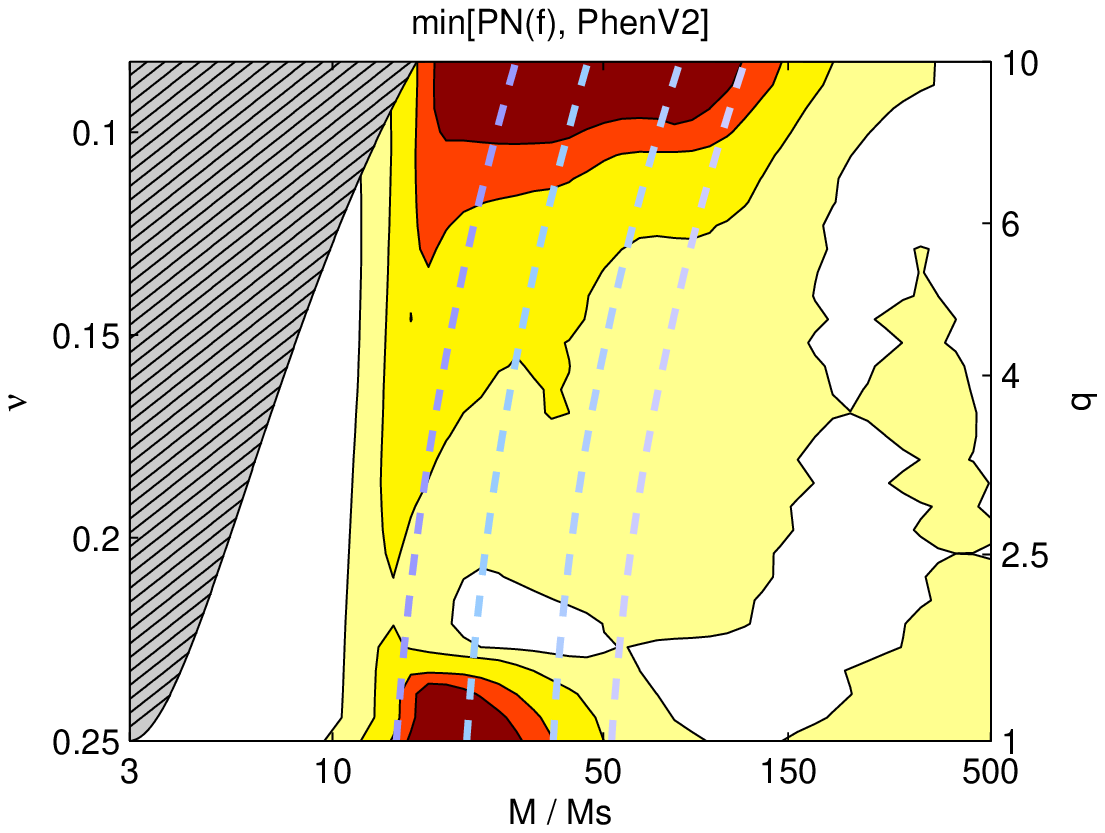}} & 
\multicolumn{2}{c}{\includegraphics[width = 5.7cm]{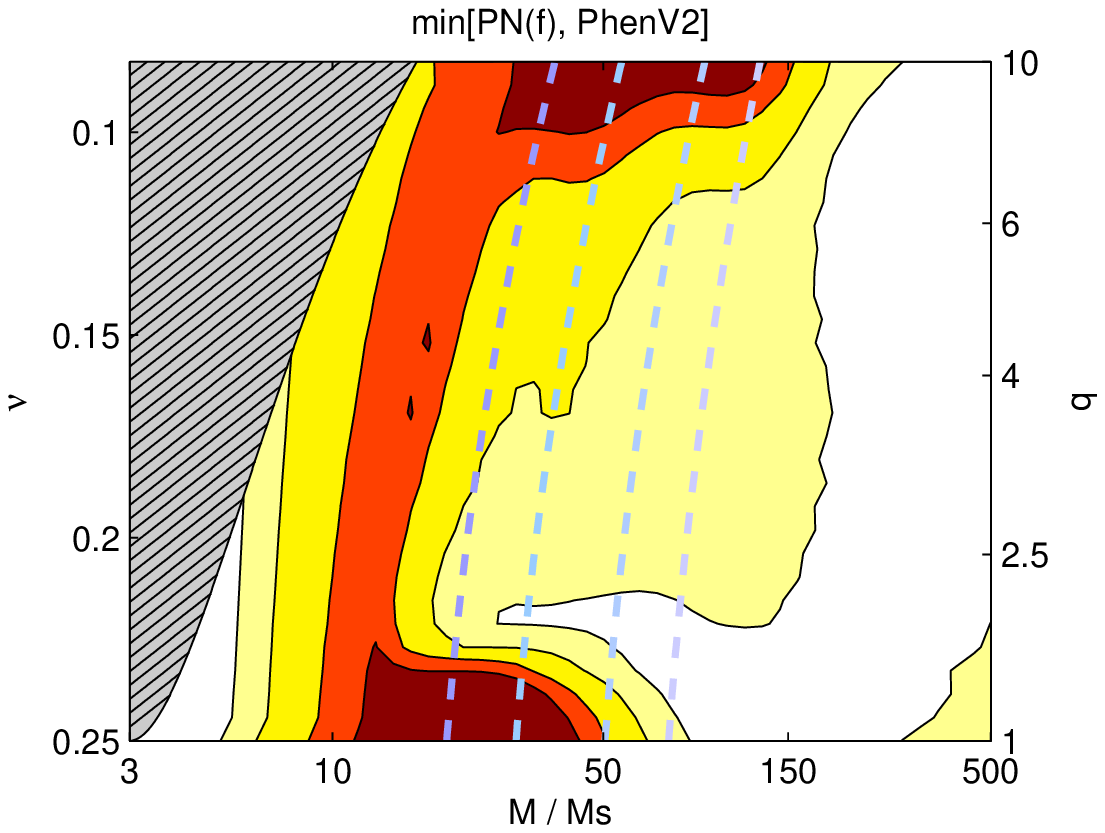}} \\
 &
\includegraphics[width = 2.8cm]{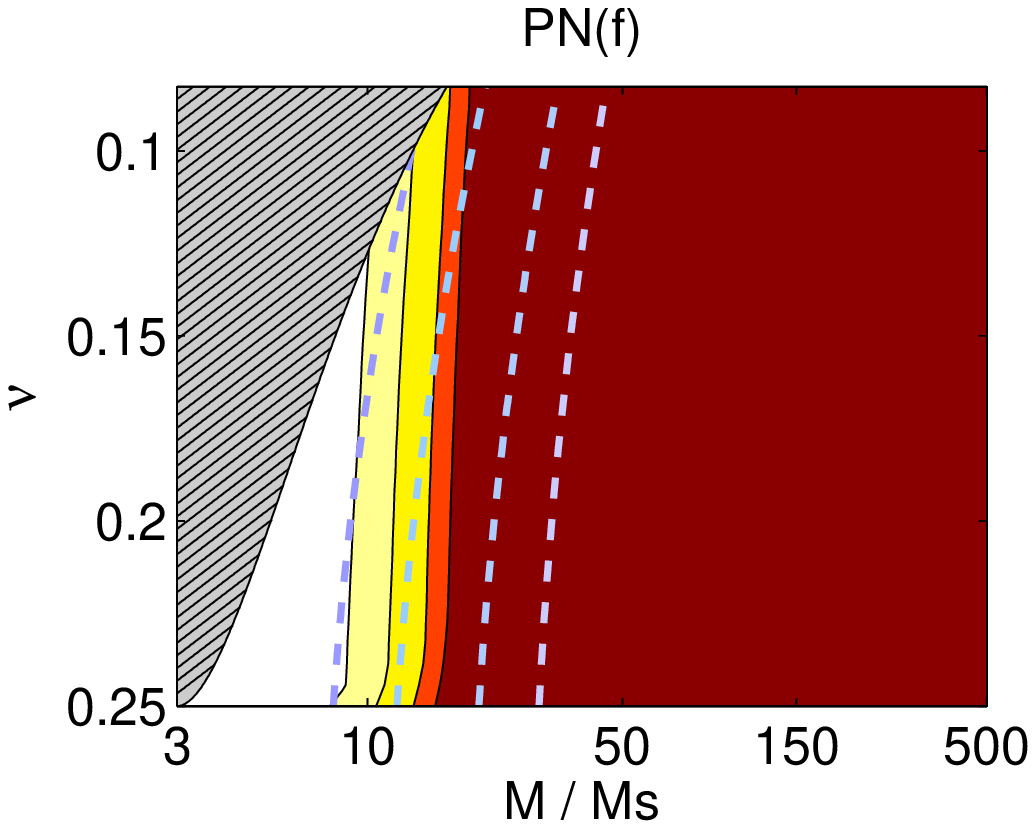} & 
\includegraphics[width = 2.8cm]{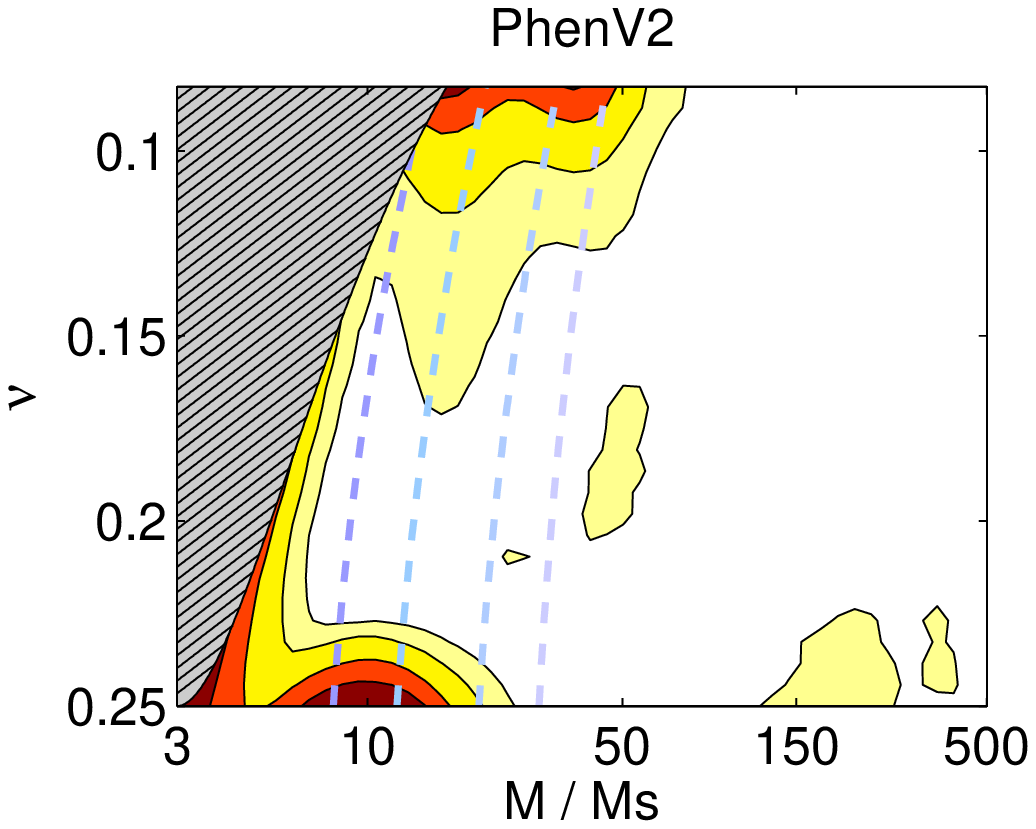} & 
\includegraphics[width = 2.8cm]{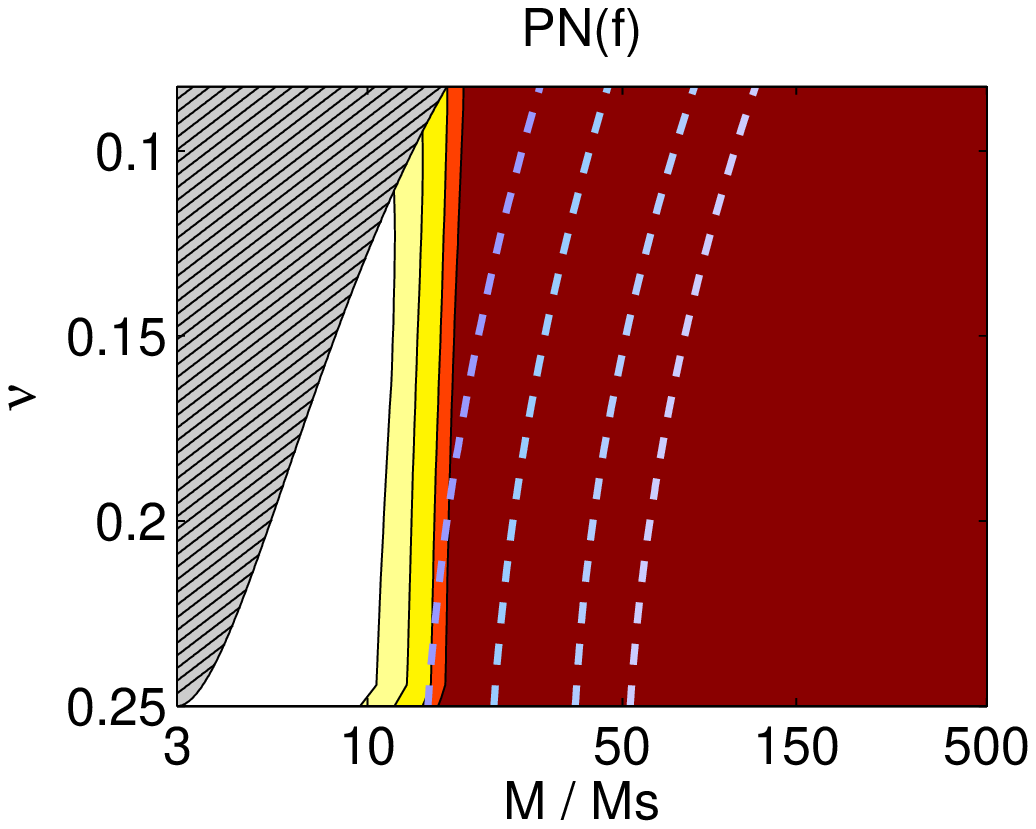} &
\includegraphics[width = 2.8cm]{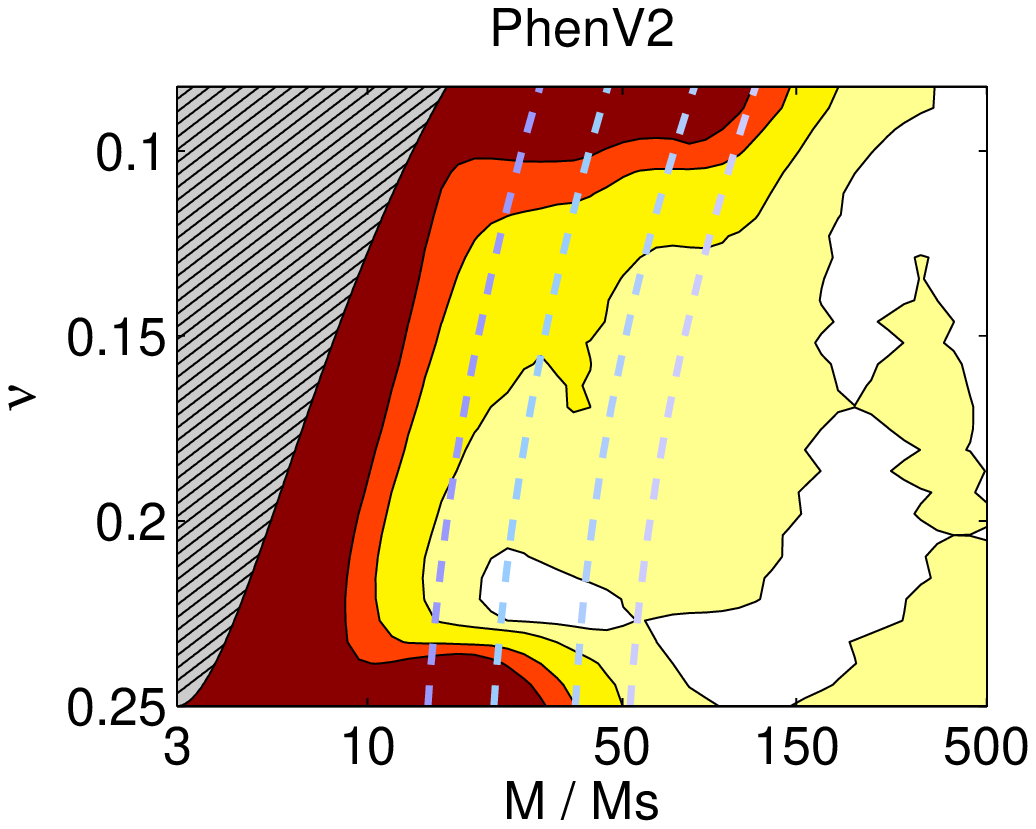} & 
\includegraphics[width = 2.8cm]{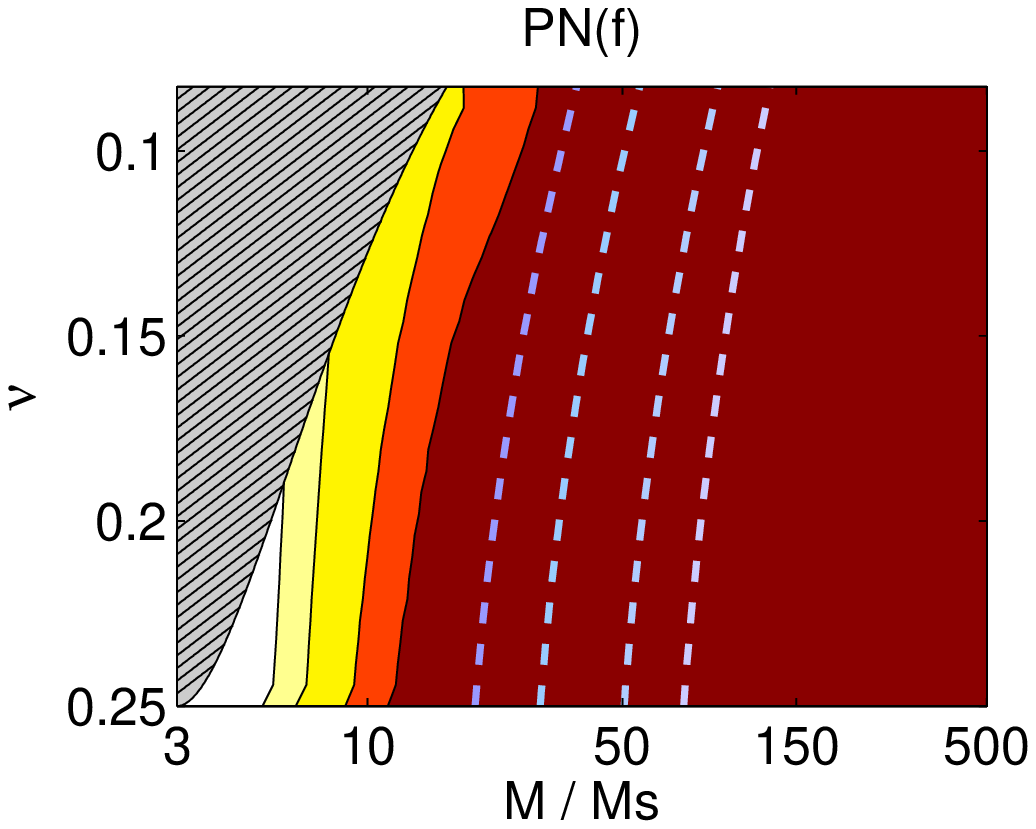} & 
\includegraphics[width = 2.8cm]{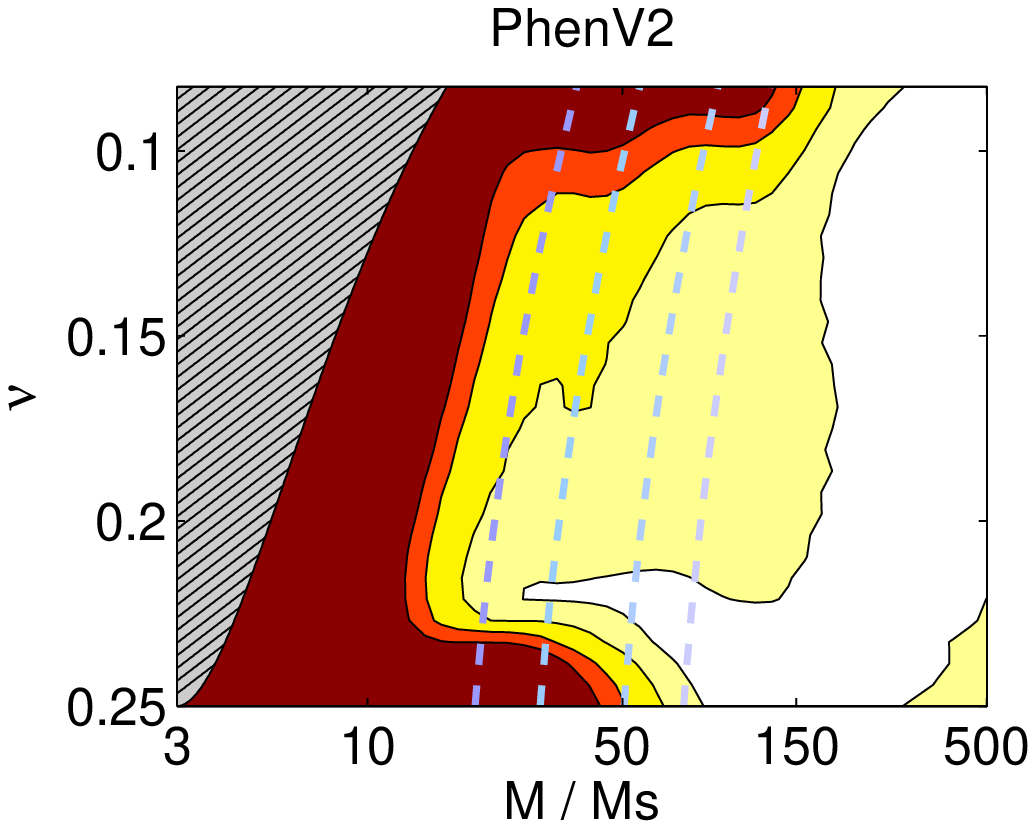} \\[1cm]
 \vspace{-2.4cm} $\mathcal{I}$:  &
\multicolumn{2}{c}{\includegraphics[width = 5.7cm]{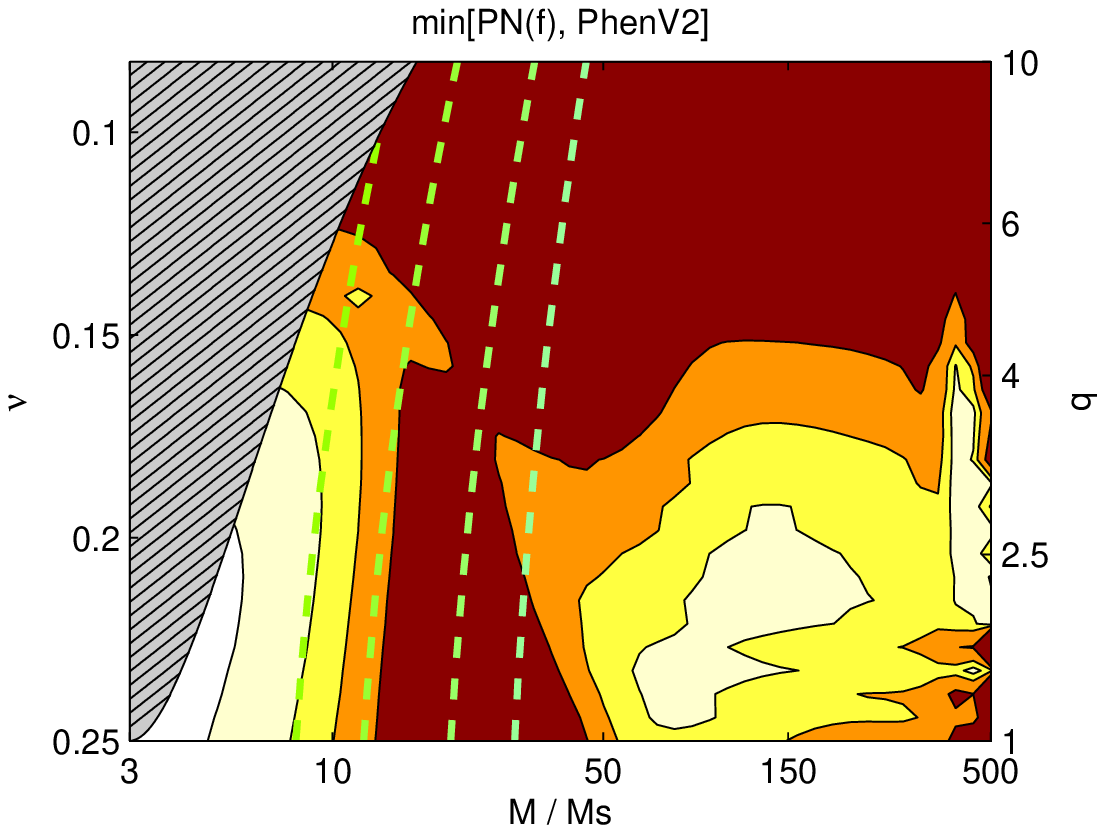}} & 
\multicolumn{2}{c}{\includegraphics[width = 5.7cm]{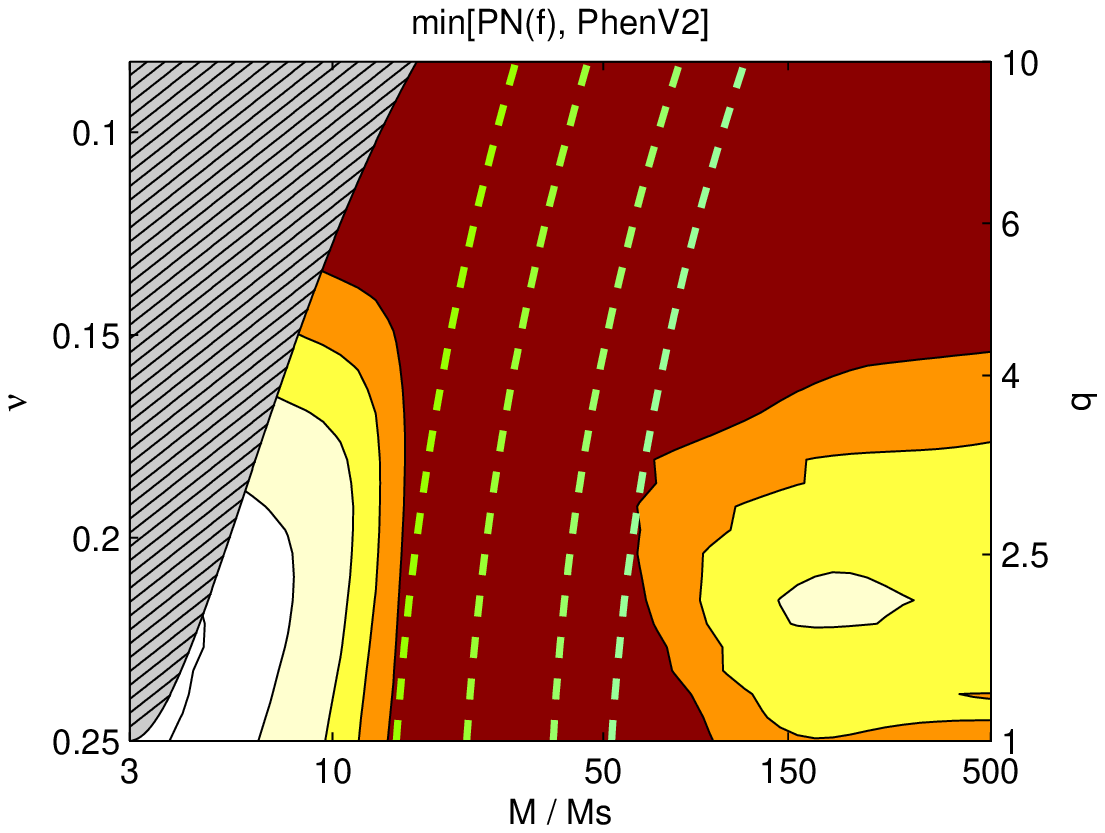}} & 
\multicolumn{2}{c}{\includegraphics[width = 5.7cm]{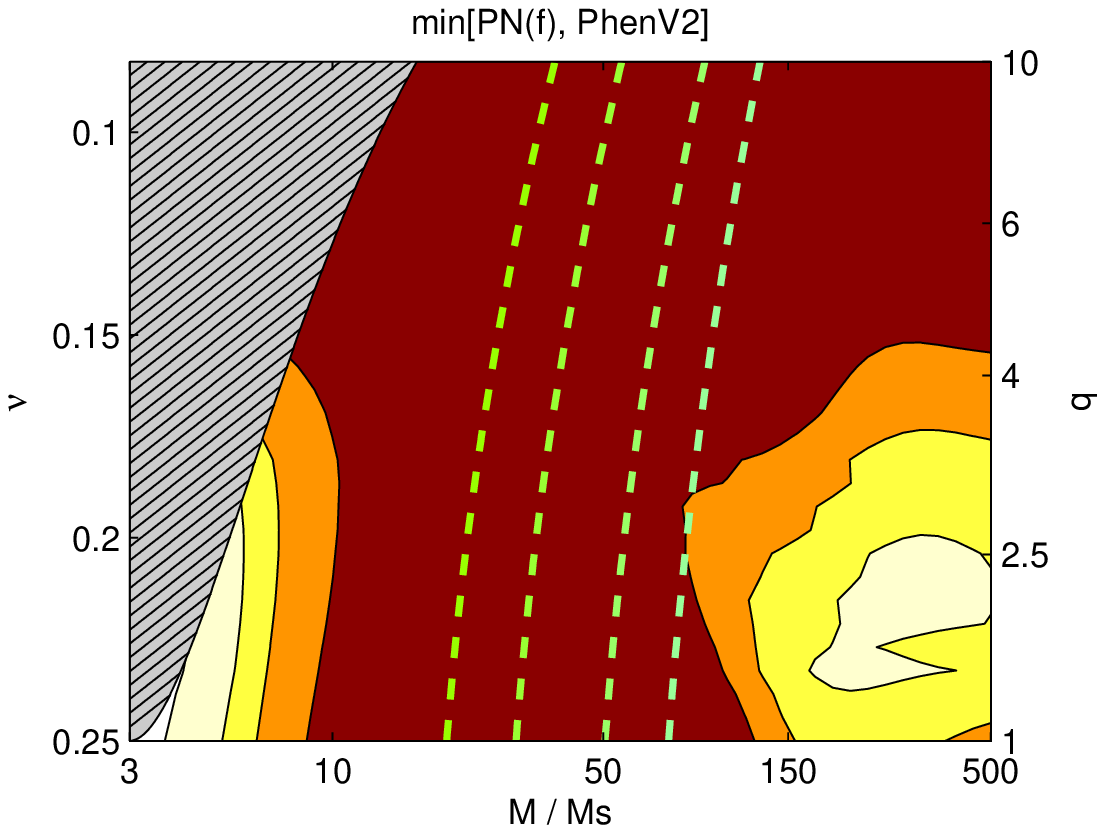}} \\
 &
\includegraphics[width = 2.8cm]{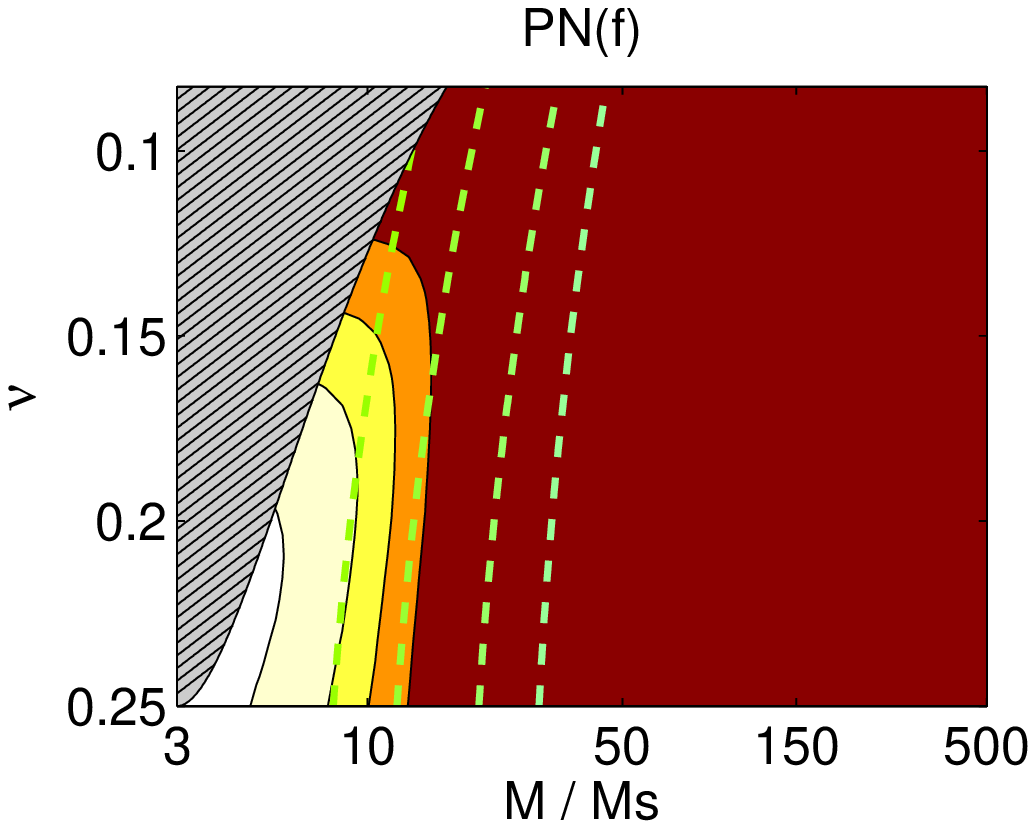} & 
\includegraphics[width = 2.8cm]{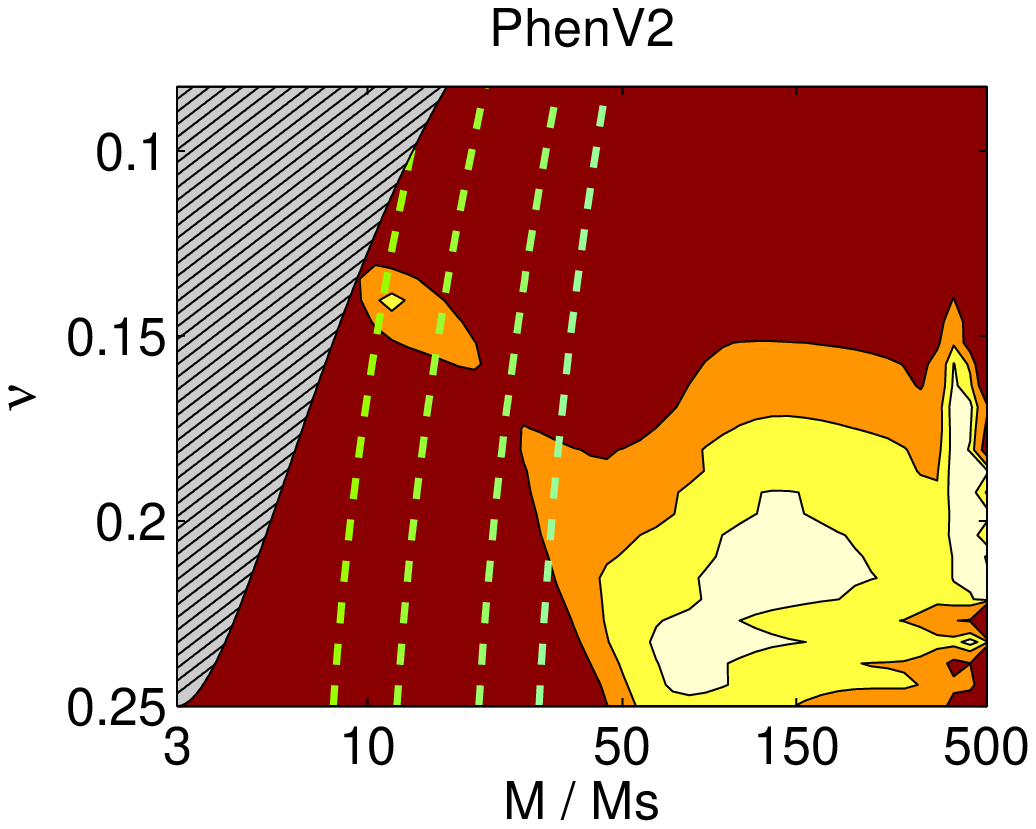} & 
\includegraphics[width = 2.8cm]{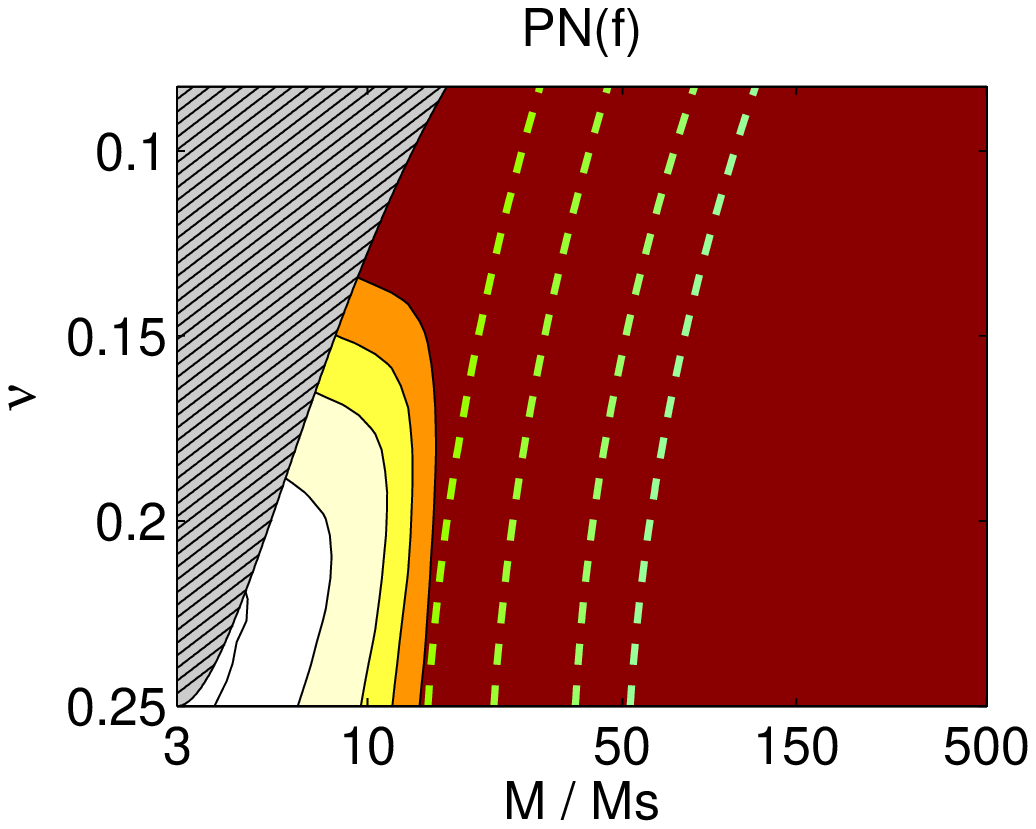} &
\includegraphics[width = 2.8cm]{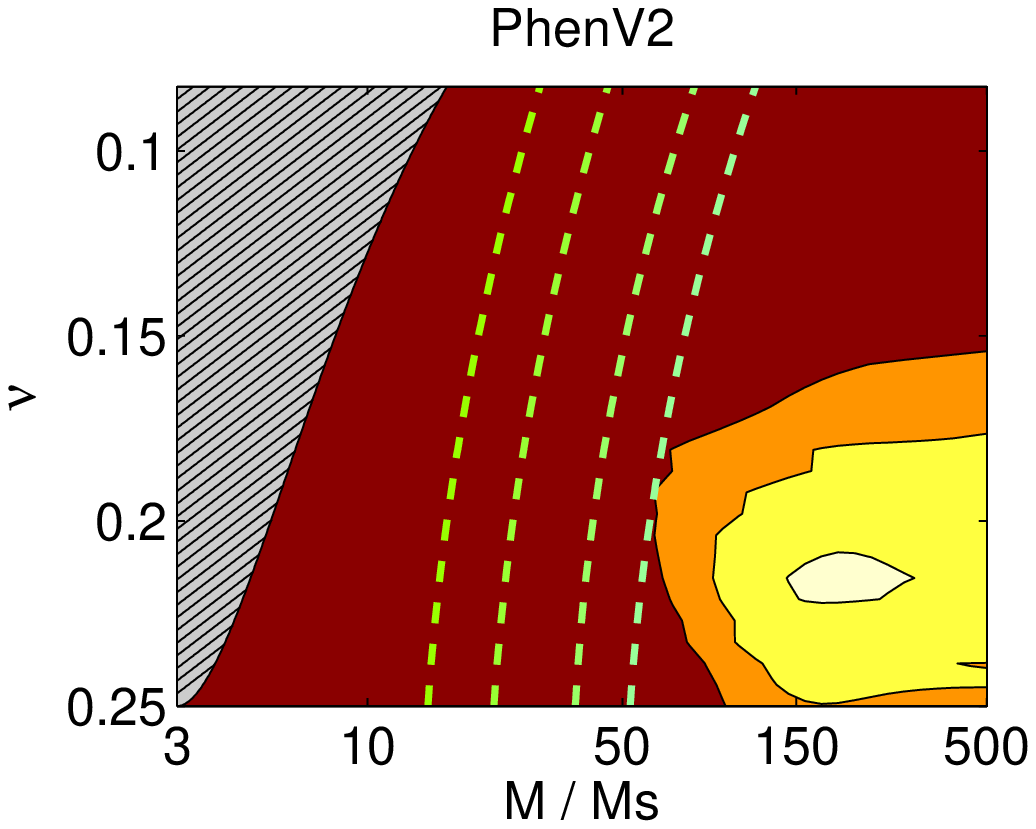} & 
\includegraphics[width = 2.8cm]{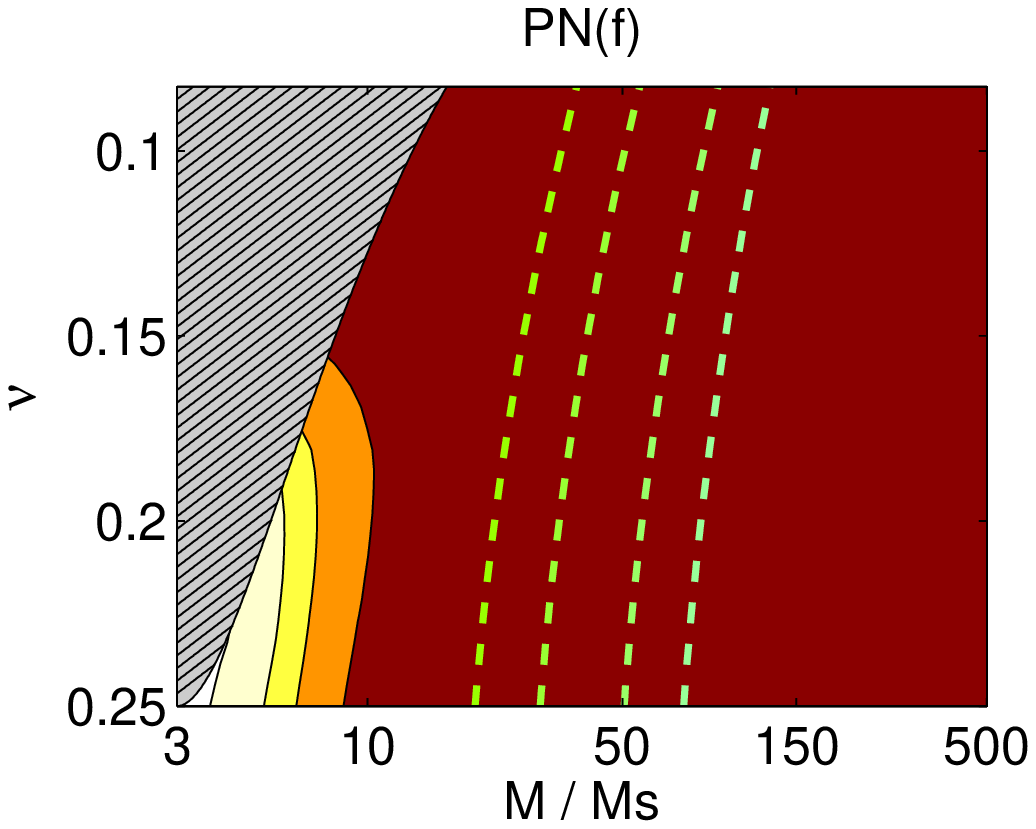} & 
\includegraphics[width = 2.8cm]{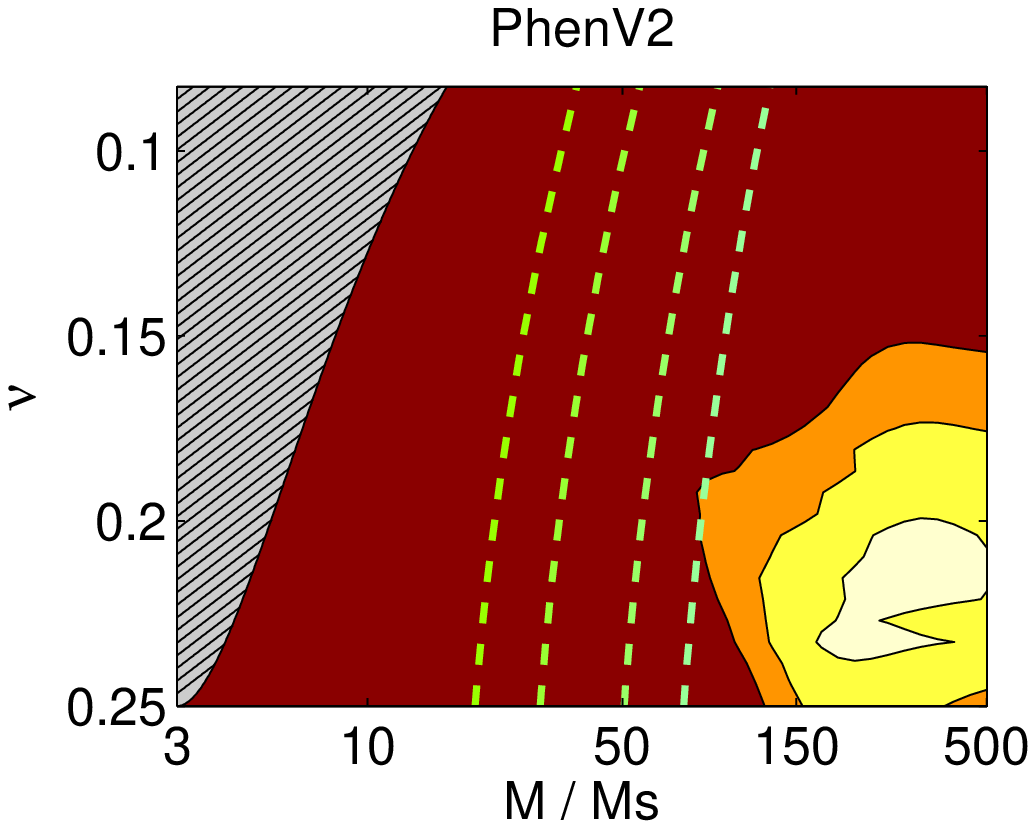} \\
 & \multicolumn{2}{c}{(a) iLIGO}
 & \multicolumn{2}{c}{(b) aLIGO}
 & \multicolumn{2}{c}{(c) aVirgo}
\end{tabular}
\end{center}
\caption{Contour plots of the same mass ranges as in Table~\ref{tab:val_ranges}, 
but plotted with a finer mass ratio grid between 1:1 and 10:1 and also considering 
the three detector noise PSDs. The top panels plot the regions of the parameter 
space where the infimum of the ineffectualnesses of the  PN(f) and PhenV2 
models is larger than $\{ 3\%, 2\%, 1\%, 0.45\% \}$ going from the darkest to the 
lightest areas; whereas the bottom panels show the parts of the parameter space 
where the infimum of the inaccuracies of the PN(f) and PhenV2 models is 
larger than $1/\rho_{\rm eff}^2$, where the effective SNR
is $\rho_{\rm eff} =\{5, 7, 10, 20\}$, when going from the darkest to the lightest 
region. The small sub-panels located below the larger ones represent the
results obtained when considering each closed-form model \emph{separately} (with
PN(f) on the left and PhenV2 on the right). Here  we only consider 
BBH systems with individual components heavier than $1.5\Ms$; this excludes the
dashed area in the lowest mass range of each plot. The four parallel dashed lines represent the 
locus of the systems where (at least)  $N_{0.99} = \{100, 50, 20, 10\}$ orbital cycles (going from left to right)  contribute   $99\%$ of the SNR (see text).}
\label{fig:scan_param_space}
\end{figure*}

Our quantitative results on the validity range of these two models, both from the effectualness
(detection) and accuracy (measurement) points of view  (using  EOB$_{\rm NR}$ waveforms 
as fiducial targets), are presented in Figs.~\ref{fig:overlaps} and \ref{fig:scan_param_space}, 
and in Table~\ref{tab:val_ranges}. We now explain them  in  detail.


\begin{table*}[t]
\begin{center}
\begin{tabular}{ p{9cm} p{9cm} }
\begin{center}
{\bf Initial LIGO}
\begin{tabular}{c|cccc}
  &   ~~~\, $>3\%$ ~~~\,   &  ~~~\,  $>2\%$   ~~~\, & ~~~\, $>1\%$  ~~~\,  &  ~~~\, $>0.45\%$  ~~~\,  \\ \hline
{\bf 1:1}    & \tabsize $[]$                     &   \tabsize $(13,17)\Ms$      &   \tabsize $(11,21)\Ms$      &   \tabsize $^*(8,278)\Ms$  \\ 
{\bf 2:1}    & \tabsize $[]$                     &   \tabsize $[]$                     &   \tabsize $[]$                     &   \tabsize $[]$                   \\ 
{\bf 4:1}    & \tabsize $[]$                     &   \tabsize $[]$                     &   \tabsize $[]$                     &   \tabsize $^*(10,58)\Ms$  \\ 
{\bf 10:1}  & \tabsize $^*(19,39)\Ms$  &   \tabsize $(17,48)\Ms$       &  \tabsize $(14,58)\Ms$      &   \tabsize $(11,74)\Ms$    \\ 
\multicolumn{5}{c}{}
\end{tabular} \newline
{\bf Adv. LIGO}
\begin{tabular}{c|cccc}
  &   ~~~\, $>3\%$ ~~~\,   &  ~~~\,  $>2\%$   ~~~\, & ~~~\, $>1\%$  ~~~\,  &  ~~~\, $>0.45\%$  ~~~\,  \\ \hline
{\bf 1:1}    &   \tabsize $(16,31)\Ms$    &   \tabsize $(14,37)\Ms$    &   \tabsize $(12,50)\Ms$      &   \tabsize $^*(10,500)\Ms$  \\ 
{\bf 2:1}    &   \tabsize $[]$                   &   \tabsize $[]$                   &   \tabsize $(13,14)\Ms$      &   \tabsize $^*(11,118)\Ms$  \\ 
{\bf 4:1}    &   \tabsize $[]$                   &   \tabsize $[]$                   &   \tabsize $^*(14,41)\Ms$  &   \tabsize $^*(12,366)\Ms$   \\ 
{\bf 10:1}  &   \tabsize $(18,114)\Ms$  &   \tabsize $(17,144)\Ms$  &   \tabsize $(14,196)\Ms$    &   \tabsize $(3,359)\Ms$      \\ 
\multicolumn{5}{c}{}
\end{tabular} \newline
{\bf Adv. Virgo}
\begin{tabular}{c|cccc}
  &   ~~~\, $>3\%$ ~~~\,   &  ~~~\,  $>2\%$   ~~~\, & ~~~\, $>1\%$  ~~~\,  &  ~~~\, $>0.45\%$  ~~~\,  \\ \hline
{\bf 1:1}    &   \tabsize $(11,44)\Ms$    &   \tabsize $(9,51)\Ms$      &   \tabsize $(6,65)\Ms$      &   \tabsize $^*(5,500)\Ms$  \\ 
{\bf 2:1}    &   \tabsize $[]$                   &   \tabsize $(10,15)\Ms$    &   \tabsize $(7,18)\Ms$      &   \tabsize $^*(6,118)\Ms$  \\ 
{\bf 4:1}    &   \tabsize $(16,17)\Ms$    &   \tabsize $(12,20)\Ms$    &   \tabsize $(8,44)\Ms$      &   \tabsize $(6,193)\Ms$    \\ 
{\bf 10:1}  &   \tabsize $(29,132)\Ms$  &   \tabsize $(18,155)\Ms$  &   \tabsize $(3,192)\Ms$    &   \tabsize $(3,296)\Ms$    \\ 
\end{tabular} \newline
\end{center}
&
\begin{center}
{\bf Initial LIGO}
\begin{tabular}{c|cccc}
  &  ~~\;  $>0.04$  ~~\;  &  ~~\;  $>0.02$  ~~\;  &  ~~\;  $>0.01$  ~~\;  &  ~~\;  $>00025$  ~~\;  \\ \hline
{\bf 1:1}   & \tabsize $(13,43)\Ms$   &  \tabsize  $^*(10,417)\Ms$   &   \tabsize  $^*(8,442)\Ms$    &  \tabsize  $(5,500)\Ms$    \\ 
{\bf 2:1}   & \tabsize $(14,31)\Ms$   &  \tabsize  $^*(11,407)\Ms$   &   \tabsize  $^*(9,422)\Ms$    &   \tabsize   $(6,456)\Ms$  \\ 
{\bf 4:1}   & \tabsize $(15,16)\Ms$   &  \tabsize  $(11,65)\Ms$      &    \tabsize  $^*(6,375)\Ms$      &  \tabsize  $(3,463)\Ms$   \\ 
{\bf 10:1} & \tabsize $(3,90)\Ms$     &  \tabsize  $(3,176)\Ms$      &    \tabsize  $(3,294)\Ms$         &  \tabsize  $(3,424)\Ms$    \\ 
\multicolumn{5}{c}{}
\end{tabular} \newline
{\bf Adv. LIGO}
\begin{tabular}{c|cccc}
  &  ~~\;  $>0.04$  ~~\;  &  ~~\;  $>0.02$  ~~\;  &  ~~\;  $>0.01$  ~~\;  &  ~~\;  $>00025$  ~~\;  \\ \hline
{\bf 1:1}      &   \tabsize $(14,92)\Ms$    &   \tabsize $^*(12,500)\Ms$   &   \tabsize $(10,500)\Ms$   &   \tabsize $(6,500)\Ms$    \\ 
{\bf 2:1}      &   \tabsize $(15,64)\Ms$    &   \tabsize $(13,94)\Ms$      &   \tabsize $(11,500)\Ms$    &   \tabsize $(8,500)\Ms$    \\ 
{\bf 4:1}      &   \tabsize $(15,54)\Ms$    &   \tabsize $(12,140)\Ms$    &   \tabsize $(5,368)\Ms$      &   \tabsize $(3,500)\Ms$    \\ 
{\bf 10:1}    &   \tabsize $(3,329)\Ms$    &   \tabsize $(3,500)\Ms$      &   \tabsize $(3,500)\Ms$      &   \tabsize $(3,500)\Ms$    \\ 
\multicolumn{5}{c}{}
\end{tabular} \newline
{\bf Adv. Virgo}
\begin{tabular}{c|cccc}
  &  ~~\;  $>0.04$  ~~\;  &  ~~\;  $>0.02$  ~~\;  &  ~~\;  $>0.01$  ~~\;  &  ~~\;  $>00025$  ~~\;  \\ \hline
{\bf 1:1}      &   \tabsize $(9,116)\Ms$    &   \tabsize $^*(6,500)\Ms$    &   \tabsize $(5,500)\Ms$      &   \tabsize $(4,500)\Ms$    \\ 
{\bf 2:1}      &   \tabsize $(10,84)\Ms$    &   \tabsize $(7,106)\Ms$     &   \tabsize $(6,500)\Ms$      &   \tabsize $(4,500)\Ms$    \\ 
{\bf 4:1}      &   \tabsize $(8,49)\Ms$      &   \tabsize $(4,175)\Ms$     &   \tabsize $(3,256)\Ms$      &   \tabsize $(3,500)\Ms$    \\ 
{\bf 10:1}    &   \tabsize $(3,287)\Ms$    &   \tabsize $(3,500)\Ms$     &   \tabsize $(3,500)\Ms$      &   \tabsize $(3,500)\Ms$    \\
\end{tabular} \newline
\end{center}
\\
\centering
(a) ineffectualness
&
\centering
(b) inaccuracy
\end{tabular}
\end{center}
\caption{Total mass ranges where the  ineffectualness (Eq.~(\ref{eq:def_effectualness})) and inaccuracy (Eq.~(\ref{eq:faith_cond_expanded})) of both PN(f) and PhenV2 models are greater than the quoted numbers in the top row of each table. The interpretation of these ranges are the following: \emph{(Left)} Mass ranges where both models have an ineffectualness larger than the quoted numbers; \emph{(Right)} Mass ranges where both models have an inaccuracy larger than
$1/\rho_{\rm eff}^2$, where
the effective SNR  $\rho_{\rm eff} = \{5, 7, 10, 20\}$ respectively. An asterisk in front of a mass range denotes that a small range in between the quoted $M$ values does not satisfy the quoted inequalities for the ineffectualness/inaccuracy.}
\label{tab:val_ranges}
\end{table*}


\begin{figure*}[t]
\begin{center}
\begin{tabular}{p{7pt}cccccc}
 & & & \multicolumn{2}{c}{PN(f) -- PhenV1} & & \\
 \vspace{-2.4cm} $\bar{\mathcal{E}}$:  &
\multicolumn{2}{c}{\includegraphics[width = 5.7cm]{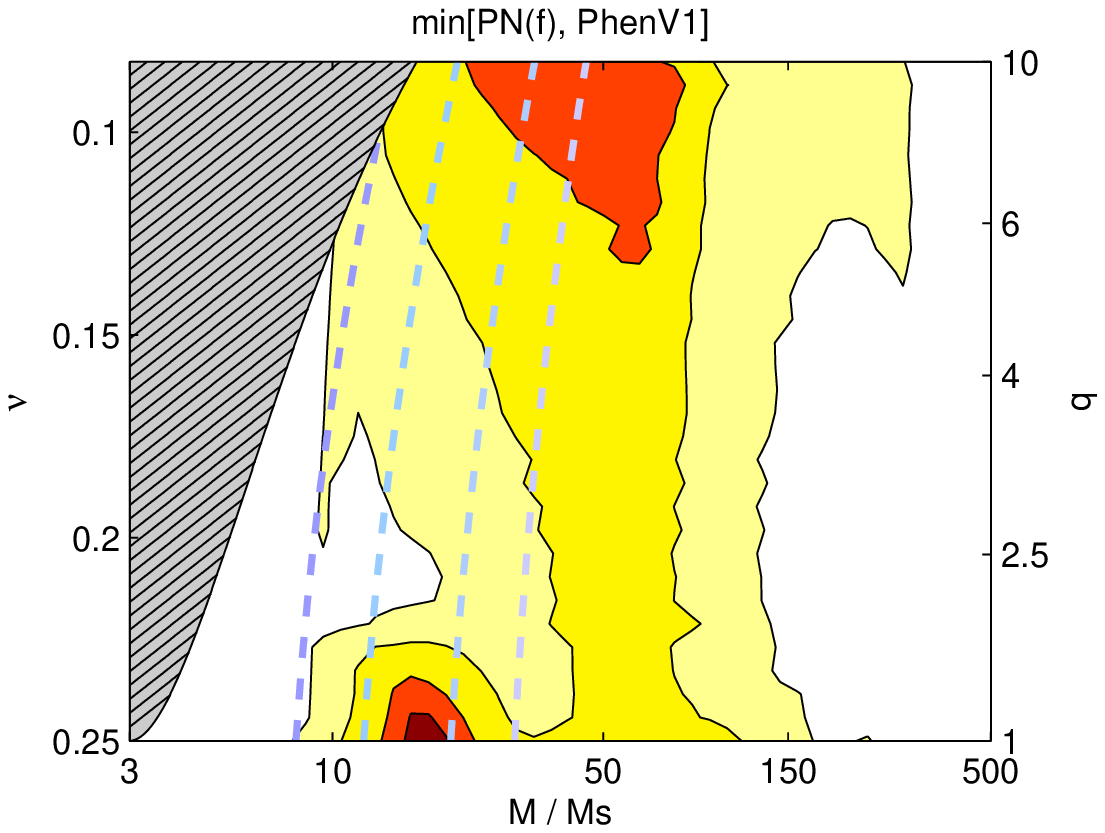}} & 
\multicolumn{2}{c}{\includegraphics[width = 5.7cm]{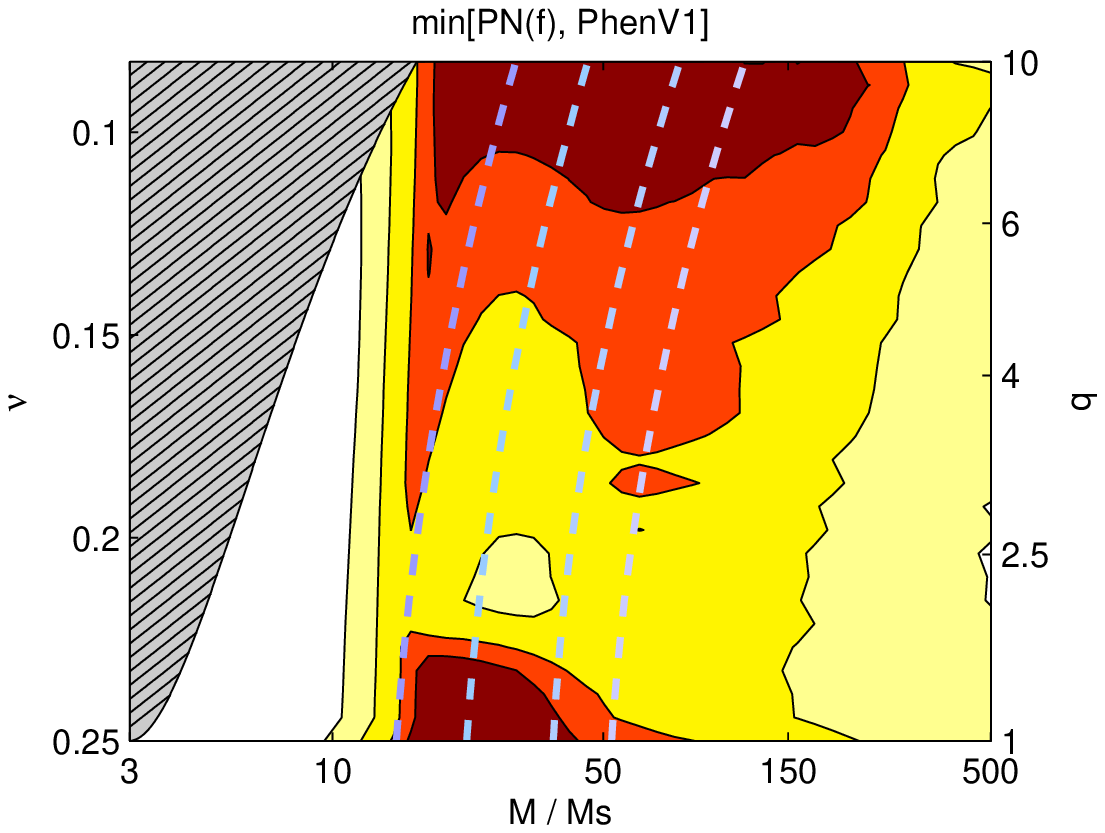}} & 
\multicolumn{2}{c}{\includegraphics[width = 5.7cm]{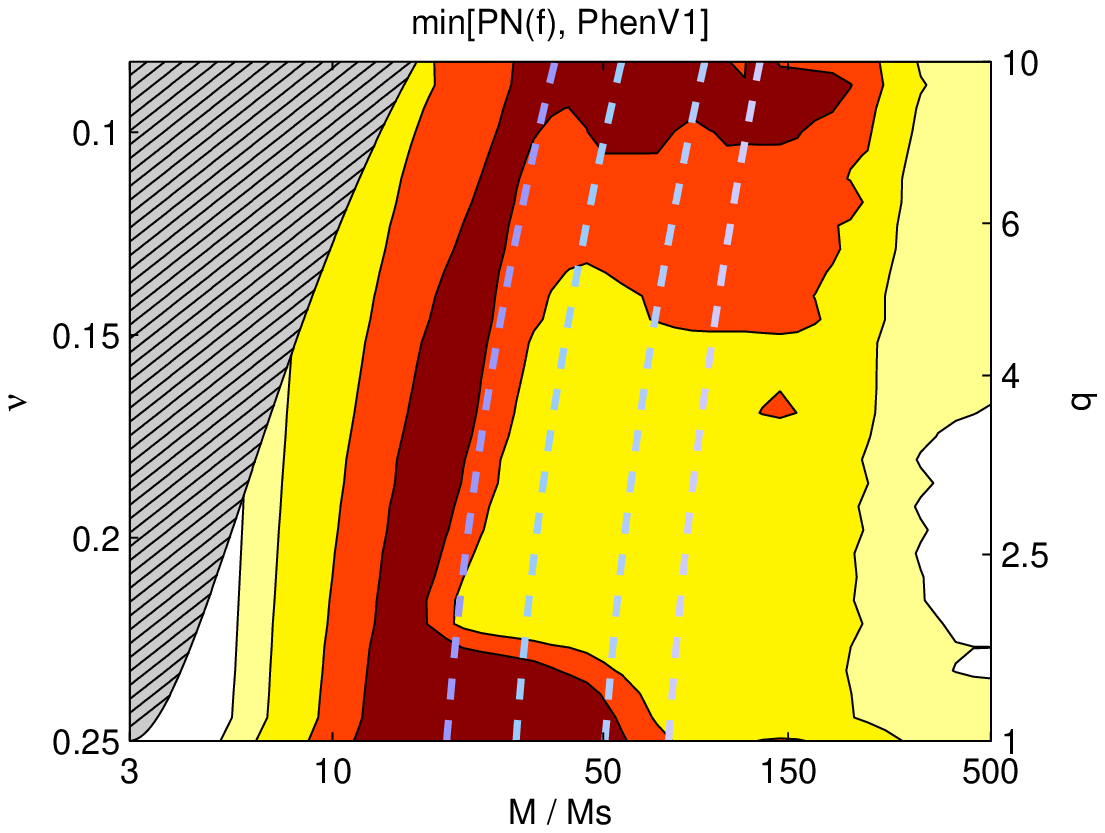}} \\
 &
\includegraphics[width = 2.8cm]{scan_param_space_eff_iLIGO_PN.eps} & 
\includegraphics[width = 2.8cm]{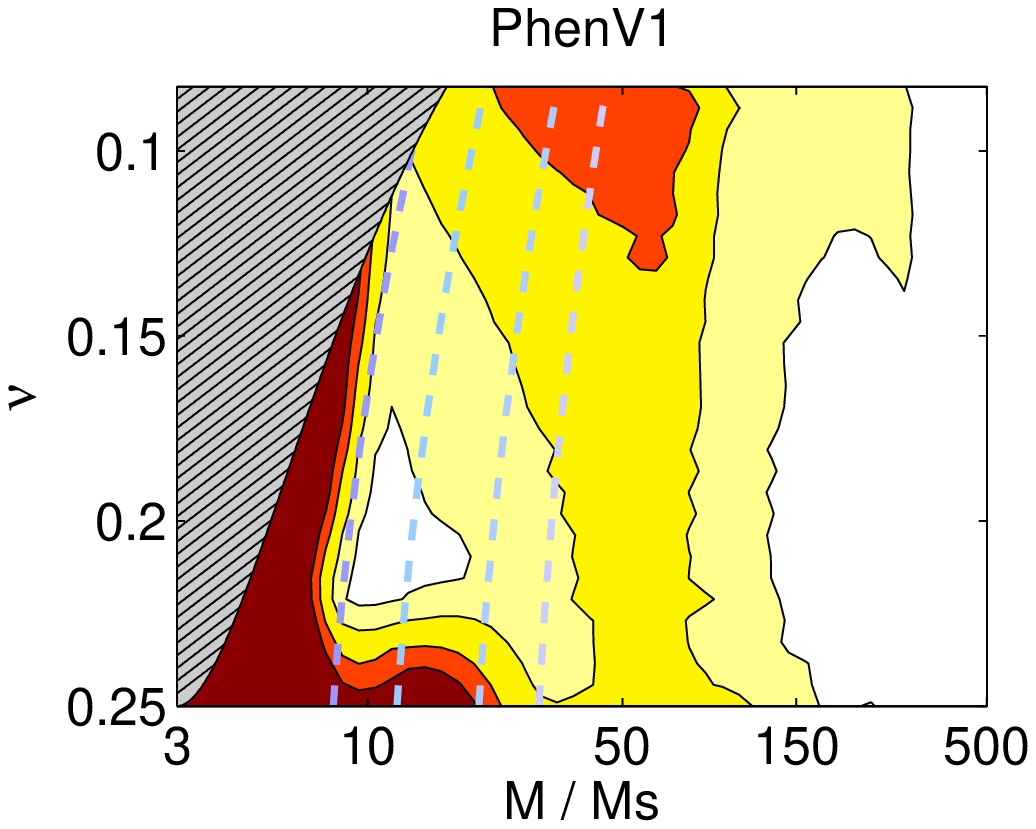} & 
\includegraphics[width = 2.8cm]{scan_param_space_eff_aLIGO_PN.eps} &
\includegraphics[width = 2.8cm]{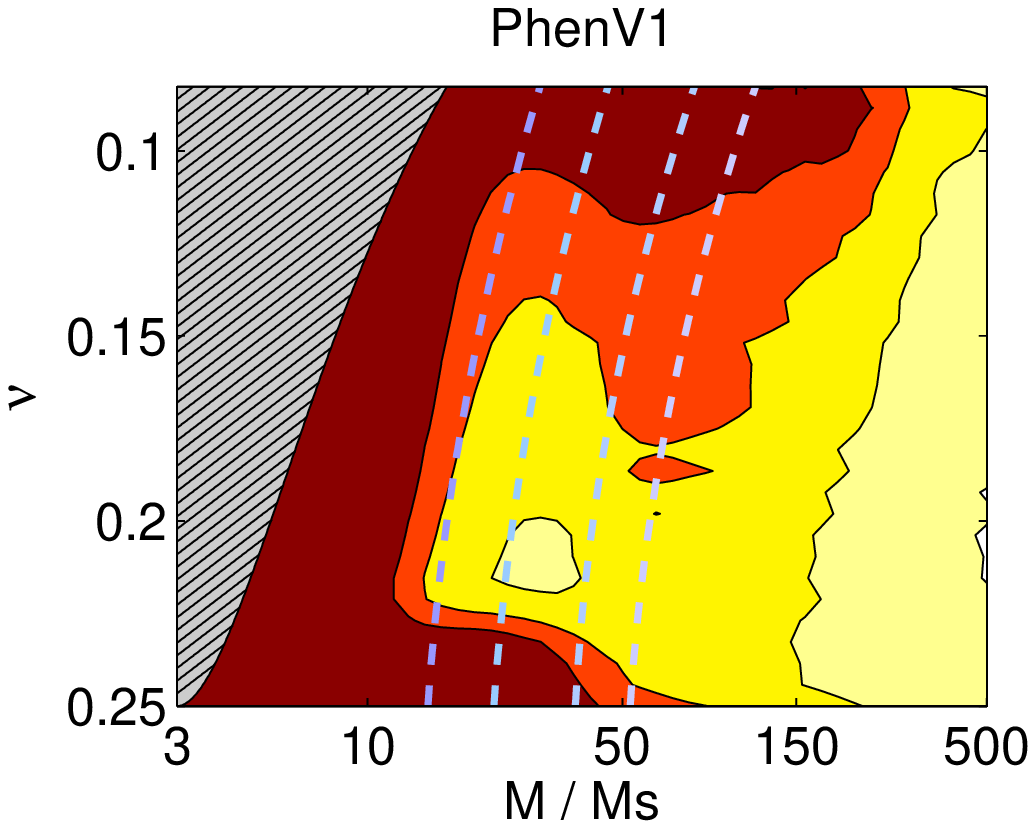} & 
\includegraphics[width = 2.8cm]{scan_param_space_eff_aVIRGO_PN.eps} & 
\includegraphics[width = 2.8cm]{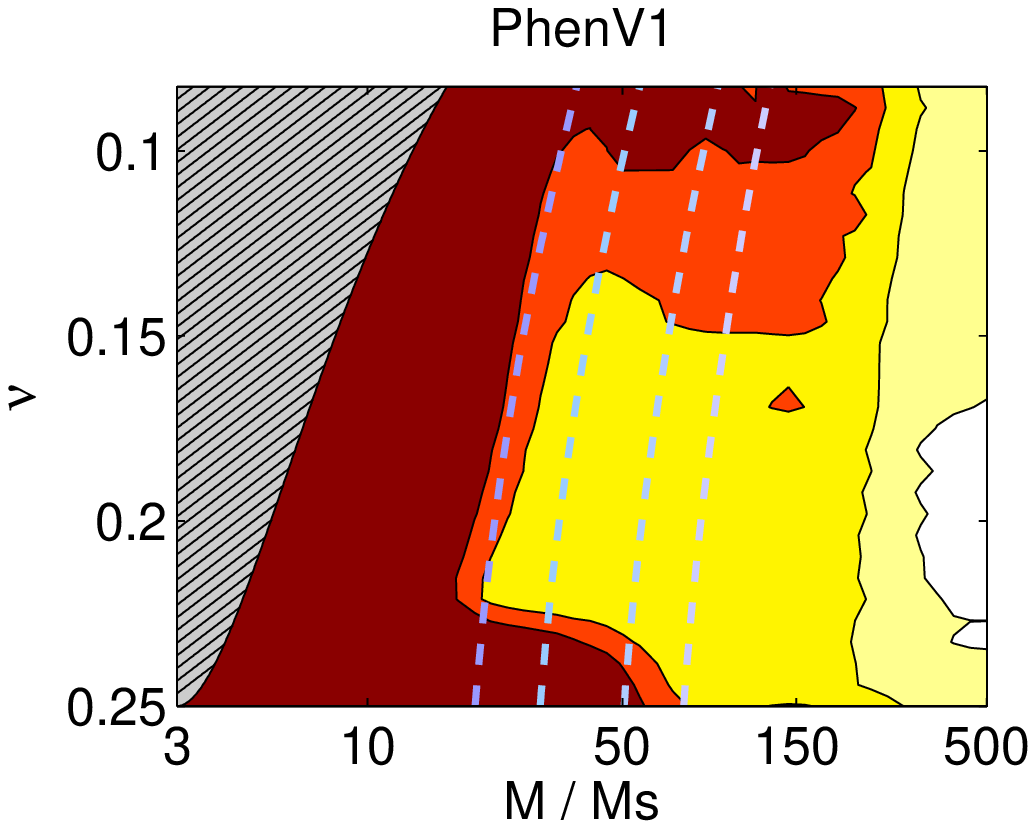} 
\\
 & \multicolumn{2}{c}{(a) iLIGO}
 & \multicolumn{2}{c}{(b) aLIGO}
 & \multicolumn{2}{c}{(c) aVirgo}
\end{tabular}
\end{center}
\caption{Same plots as top panels in Fig.~\ref{fig:scan_param_space} (ineffectualness),
 but considering PN(f) -- PhenV1 models instead. In this case, the accuracy results have not 
been plotted because neither PN(f) nor PhenV1 are accurate for $M>20\Ms$ at any mass ratio 
(between 1:1 and 10:1), even when considering an effective SNR of $5$.}
\label{fig:scan_param_space_V1}
\end{figure*}

\subsection{Effectualness. Detection}
\label{subsec:effectualness_results}

As recalled above, when one is only interested in detecting GW signals from compact coalescing 
objects but not in the extraction of accurate physical parameters from the observations,
 an important indicator to evaluate a given approximated model is the 
\emph{effectualness}, $\mathcal E$, Eq.~\eqref{eq:def_effectualness}, which represents
the fractional loss of SNR entailed by the use of a sub-optimal match filter. Actually, the
most important indicator is the  ``effective effectualness'' which takes into
account the discrete character of the template bank (see above). We have indicated in
Eq.~\eqref{eq:effectualness} the range of thresholds that one usually considers for
the effectualness. Let us say again that the often quoted effectualness level of $97 \%$ 
(corresponding to an ineffectualness $\bar{\mathcal E}=1- \mathcal E \leq 3 \%$)
is a very
minimal requirement that does not seriously take into account the constraints
coming from the discrete nature of the template bank.  It would seem more appropriate
to require an ineffectualness near   $\bar{\mathcal E} \leq 1 \%$, which, combined with
an equivalent minimal match $\epsilon_{MM} = 1 \%$, guarantees that one does not
lose more than $6 \%$ of the potential events.

To have a first view of the influence of  choosing an effectualness (and accuracy) threshold
on the validity range of various models, the
top panels of Fig.~\ref{fig:overlaps} plot as dark gray (blue online) lines the ineffectualness  
$\bar{\mathcal E}$ as a function of the total mass,  for a fixed mass ratio of  4:1, and using 
the Adv. LIGO noise PSD.  The three top panels respectively correspond to the three closed-form models $h_m$
introduced above: PN(f), PhenV1 and PhenV2. [We recall that EOB$_{\rm NR}$ is used as fiducial 
``exact'' target $h_x$.]

These top panels show that as one imposes a stricter threshold on the effectualness
(\ie a smaller level for the ineffectualness  $\bar{\mathcal E}=1- \mathcal E$) the
validity range of the various models change in the direction expected from the
preliminary discussion above.  For the PN(f) model (left top panel), the validity range at some ineffectualness threshold $\bar{\mathcal E}$ is an interval of the type 
$M< M_{\rm max}\superPN(\bar{\mathcal E})$,
where the maximum allowed total mass $ M_{\rm max}\superPN(\bar{\mathcal E})$ \emph{decreases}
as one decreases $\bar{\mathcal E}$. For instance, at the level $\bar{\mathcal E}= 0.03$,
indicated by the upper dashed line in the left top panel (blue online), one has 
$M_{\rm max}\superPN \approx 17 \Ms$, while the stricter level $\bar{\mathcal E}= 0.01$
leads to $ M_{\rm max}\superPN \approx 14 \Ms$. This behavior is consistent with the
fact that the PN model is only adequate during the inspiral, up to some maximum
dimensionless frequency $\hat{f} \equiv \fM \leq   \hat{f}_0 (\nu)$ (see 
Fig.~\ref{fig:PN-SPA_range}). Then, given the fixed detector sensitivity window
in Hz,  this translates into an upper limit on the total mass: $M \lesssim \hat{f}_0 (\nu)/f_{\rm detector}$. On the other hand, the phenomenological models considered in this paper were built 
with the specific aim of  reproducing the late inspiral, merger and ringdown phases of the evolution found by NR simulations, and therefore, they should be most effectual 
for the detection of high mass systems. And, indeed, one finds that, for the phenomenological
models (right top panels) the validity range at some ineffectualness threshold $\bar{\mathcal E}$ is an interval of the type 
$M> M_{\rm min}\superPhen(\bar{\mathcal E})$, where the minimum allowed total mass
$M_{\rm min}\superPhen(\bar{\mathcal E})$ \emph{increases} as  one decreases $\bar{\mathcal E}$. 
 For instance, at the level $\bar{\mathcal E}= 0.03$, and for PhenV2, as
indicated by the upper dashed line in the rightmost top panel (blue online), one has 
$ M_{\rm min}\superVtwo \approx 12 \Ms$, 
while the stricter level $\bar{\mathcal E}= 0.01$
leads to a much larger value: $M_{\rm min}\superVtwo \approx 41 \Ms$.
[For PhenV1, these minimum masses are $M_{\rm min}\superVone \approx 16 \Ms$
for $\bar{\mathcal E}= 0.03$; and $M_{\rm min}\superVone \approx 246 \Ms$
for $\bar{\mathcal E}= 0.01$].
This example (with $q=4$ and Adv. LIGO) shows that: (i) at the minimal effectualness 
level $\bar{\mathcal E}= 0.03$ there is an \emph{overlap} between the `low-mass' interval
where PN(f) is effectual, and the `high-mass' interval where PhenV2 is effectual, so that
the union of the two template banks is able to detect the full mass range; while, by
constrast, (ii) at the stricter  effectualness 
level $\bar{\mathcal E}= 0.01$, there exists a \emph{mass gap}, between
$ M_{\rm max}\superPN \approx 14 \Ms$ and $ M_{\rm min}\superVtwo \approx 41 \Ms$,
in which neither PN(f) nor PhenV2 (nor Phen V1) is effectual enough.
We shall focus on these mass gaps below, after having commented upon
the other panels of   Fig.~\ref{fig:overlaps}.

The bottom panels of Fig.~\ref{fig:overlaps} complete the discussion of the effectualness 
calculation by showing the fractional biases between the actual $M$ and $\nu$ parameters 
and the ones that maximize the overlap between the model and the target waveforms in the 
maximization, Eq.~(\ref{eq:maximization_effectualness}). In our analysis, we have restricted 
the parameters to have a physical meaning, i.e. $\nu \leq 0.25$. However,  the results 
obtained by allowing the parameters to be unphysical remain almost the same: only a small 
improvement of the effectualness near the equal-mass region is observed, because this is where the restriction becomes 
relevant. By looking at our particular results in Fig.~\ref{fig:overlaps} for $\nu=0.16$ and 
the Adv. LIGO noise PSD, we observe that the systematic biases in estimating the parameters 
are smaller than $1\%$ for $M<25\Ms$ when using PN waveforms; while they are smaller 
than $5\%$ for $M$ ($10\%$ for $\nu$) for $M\in [8,200]\Ms$ when using 
phenomenological models. Anyway, these fractional biases are merely 
informative since one would have to compare them with typical statistical errors from a proper 
parameter estimation study in order to truly interpret the consequences of having an effectual, 
but not accurate, model.  Let us mention that several works in the literature have studied the 
parameter estimation of BBH coalescences with ground-based detectors, using either PN 
templates~\cite{Cutler:1994ys, Poisson:1995ef, Sintes:2000, AISS_2005, VanDenBroeck:2006ar, 
Rover:2006bb, vanderSluys:2008qx, Veitch:2009hd} or phenomenological ones~\cite{Ajith:2009fz}.

Let us now study in detail the \emph{mass gaps}  where none of the closed-form models
are effectual enough. Since the non-precessing phenomenological waveforms 
(PhenV2)~\cite{PhenV2} are always more accurate and produce smaller mismatches with EOB 
than the original non-spinning ones (PhenV1)~\cite{PhenV1}, we shall study, for various
effectualness levels, the region of parameter
space where the union of PN(f) and  PhenV2 models fails to be effectual.  As we saw above,
this region is defined, for each given ineffectualness level  $\bar{\mathcal E}$,
and for each given value of $\nu$, by the mass gap between $ M_{\rm max}\superPN$ and
$ M_{\rm min}\superVtwo$, \ie by the interval 
 $ M_{\rm max}\superPN(\bar{\mathcal E}, \nu) \leq M \leq  M_{\rm min}\superVtwo(\bar{\mathcal E}, \nu)$, when it is non empty.
These mass gaps are listed in Table~I(a) as three sub-tables for the different
detectors: initial LIGO, advanced LIGO and advanced Virgo; and in there
the intervals $ [M_{\rm max}\superPN(\bar{\mathcal E}, \nu) ,  M_{\rm min}\superVtwo(\bar{\mathcal E}, \nu)]$ are given
for four different mass ratios ($q=1, 2, 4, 10$) and for four different effectualness levels:
$3\%, 2\%, 1\% , 0.45\%$. 

In Table~I(a), one can observe how these ``ineffectualness intervals''
$ [M_{\rm max}\superPN(\bar{\mathcal E}, \nu) ,  M_{\rm min}\superVtwo(\bar{\mathcal E}, \nu)]$
 become wider as we go to more restrictive effectualness bounds, and also as we 
consider more sensitive detectors.  Concerning their dependence upon $\nu$, we see that 
mass ratios between 2:1 and 4:1 generally lead to the narrowest intervals, while the other 
mass ratios (especially  $q=10$)  lead to wider intervals, mainly because of the increase of
their right end:  $M_{\rm min}\superVtwo(\bar{\mathcal E}, \nu)$.
 Typically, the lack of effectualness of the closed-form models takes place for intermediate mass 
systems, from a total mass of $10\Ms$ to $20-100\Ms$, depending of the mass ratio. For instance, 
considering the Adv. Virgo noise PSD and allowing a maximum loss of SNR of $2\%$, the analytical 
models would not be valid to search for equal-mass systems between $(4.5+4.5)\Ms$ and $(26+26)\Ms$; 
for 2:1 systems from $(3.5+7)\Ms$ to $(5+10)\Ms$; for 4:1 from $(2.4+9.6)\Ms$ to $(4+16)\Ms$ and 
for the most extreme mass ratio 10:1 from $(1.6+16.4)\Ms$ to $(14+141)\Ms$.

Finally, in order to have a better view of the regions
of the $(M, \nu)$ parameter space where the union of PN(f) and PhenV2 fails to be
``$\bar{\mathcal E}$-effectual'' (\ie where the \emph{infimum} of the ineffectualnesses
of  PN(f) and PhenV2 is larger than $\bar{\mathcal E}$),
 we have determined the ineffectualness intervals 
$ [M_{\rm max}\superPN(\bar{\mathcal E}, \nu) ,  M_{\rm min}\superVtwo(\bar{\mathcal E}, \nu)]$
on a finer grid resolution in the whole parameter space. 
The contour  plots in the large top panels of Fig.~\ref{fig:scan_param_space} 
and Fig.~\ref{fig:scan_param_space_V1}
represent the result of this ineffectualness calculation 
when one considers the two closed-form models simultaneously.
By contrast, the small sub-panels located below the large upper panels 
represent the separate values of the effectualnesses of the two
closed-form models: PN(f) on the left and PhenV2 on the right. 
The four contour lines represent the same 
effectualness thresholds used in Table~\ref{tab:val_ranges}
(\ie $3\%, 2\%, 1\% , 0.45\%$), but the higher resolution in $\nu$ allows us now to 
observe the continuous evolution of the validity ranges as the mass ratio varies. 
As before, normally the left-most bounds are determined by the total mass values 
$M_{\rm max}\superPN(\bar{\mathcal E}, \nu)$
above which the PN(f) model is not valid anymore, whereas the right-most limits are 
given by the minimum mass values $M_{\rm min}\superVtwo(\bar{\mathcal E}, \nu)$ 
where phenomenological models are effectual. Choosing a certain ineffectualness level
$\bar{\mathcal E}$, the corresponding regions (from the darkest one in which 
$\bar{\mathcal E}> 3\%$ to the lightest  [but not white] one where $\bar{\mathcal E} > 0.45\%$) on 
the contour plots represent the domain of parameter space where none of these closed-form 
models could be used in an actual search, even for the mere purpose of \emph{detection}, 
 without requesting  a faithful recovery of the physical parameters.
Let us start by noting that, in the case of initial LIGO, the \emph{simultaneous}
use of PN and PhenV2 waveforms (\ie, in fact, the use, for each particular $M, \nu$ of the best
of the two, namely the one having the smallest ineffectualness for these values of $M$ and $\nu$)
allows one to cover the full parameter space with an ineffectualness $< 3\%$, \ie an effectualness
larger than $97\%$. At higher effectualness levels, like $98\%$ or $99\%$, one is progressively
losing the complete coverage of the parameter space. It is, however, interesting to see, on the smaller sub-panels
below the top left panel, the \emph{separate performances} of PN and PhenV2 waveforms: PN has a good effectualness
only in the small-mass corner on the left. On the other hand, PhenV2 (for initial detectors) is $99 \%$ 
effectual in a large part of parameter space, though it fails to be $98 \%$ effectual in three sub-zones:
(1) in the small-mass corner, (2) in an ``island'' centered around $M = 10 \Ms$, $q=1$, and (3) in another
island located above $q \gtrsim 9$. This good effectualness performance of the closed-form models 
is, however, drastically decreased in the case of advanced detectors.
Note, for instance, how, in the case of advanced LIGO, the PhenV2 waveform fails to be $97 \%$
effectual in an extended ``C-shaped'' region which covers many
physically interesting systems: notably the range $ 3 \Ms < M < 40 \Ms, q \sim 1$, a domain 
located in the high q region, as well as essentially all systems with $ M \lesssim 10 \Ms$. However, the
simultaneous use of PN and PhenV2 waveforms allows one to substantially reduce the extension of this
ineffectualness domain. As seen on the middle top panel, the union of the two waveforms fails to be 
$ 98 \%$ effectual only in two ``islands'' located either around $ M \sim 20 \Ms, q \sim 1 $, 
or around $ M \sim 40 \Ms,  q > 7$. Note also, how, in the case of the advanced Virgo detector,
requiring an ineffectualness $ < 2\%$ (\ie an effectualness $> 98 \%$) excludes the
joint use of closed-form waveforms in an extended  ``ridge''  which separates a 
low-mass region on its left from a high-mass region on its right.

Before discussing the ``inaccuracy analogs'' of the ineffectualness regions plotted
in the top panels of Fig.~\ref{fig:scan_param_space}, let us mention the meaning of our
results for the most recent LIGO-Virgo searches for low-mass compact binary coalescences
(CBC) using PN templates. In particular,~\cite{CBC_search_1, CBC_search_2, CBC_search_4}
have performed searches for $M\in[2,35]\Ms$ and~\cite{CBC_search_3} for 
$M\in[2,40]\Ms$.  Upper limits for the particular systems $(1.35+1.35)\Ms$, $(5+5)\Ms$
and $(5+1.35)\Ms$ have also been set in \cite{CBC_search_2, CBC_search_4}. Our results show that the PN templates used in all the searches done so far (with the initial LIGO noise
PSD) had an ineffectualness $< 1\%$ for systems with $M\lesssim 10\Ms$, and that  this
performance could be extended to most of the analyzed parameter space  by also including
phenomenological waveforms. However, in a physically significant part of the parameter
space (namely systems with a total mass $M>10\Ms$ and mass ratios, either 
$q \lesssim 2$ or $q\gtrsim 5$), the ineffectualness of the PN templates exceed
$\bar{\mathcal{E}} = 1\%$, and even $\bar{\mathcal{E}} = 2\%$.  Keeping in mind that the
discretized template bank used in the searches had a maximal mismatch of
$\epsilon_{MM}=3 \%$, this means that, in a significant part of parameter space, the
searches could be missing $ 1- (1- \bar{\mathcal{E}} - \epsilon_{MM})^3 \approx 3 
(\bar{\mathcal{E}} + \epsilon_{MM})$, \ie, $12 \%$, or even $15 \%$ of the potential events. 
The probability to miss potential events because of the ineffectualness of closed-form
waveform models will become much more significant in searches performed with advanced
detectors (see top panels of Fig.~\ref{fig:scan_param_space}). This indicates the need to
use more effectual templates, and to construct template banks having a smaller maximal
mismatch.

\subsection{Accuracy. Measurement}
\label{subsec:accuracy_measurement}

As recalled above, if one wishes to use some waveform model not only for detecting the
presence of a GW signal, but also for \emph{measuring} the physical information contained
in the GW signal, one needs this waveform model to satisfy a stricter condition than
the one of effectualness studied in the previous section. Namely, the waveform model $h_m$
must be such that its \emph{inaccuracy} $\mathcal{I}[h_m;h_x] $, Eq.~\eqref{eq:inaccuracy1},
is smaller than $1/\rho_{\rm eff}^2$, where $\rho_{\rm eff} = \rho/\epsilon$ denotes
what we called the effective SNR (which is larger than the real SNR $\rho$ by a
safety factor $1/\epsilon > 1$), see Eq.~\eqref{eq:faith_cond_expanded}.
Thus, the accuracy condition for each model shall be well defined after choosing:
(i) what is the target signal $h_x$ that one is trying to measure; and (ii)
a certain  level for the effective SNR, $\rho_{\rm eff} = \rho/\epsilon$.
As before, we shall choose as fiducial target signal the 
NR-calibrated EOB waveform, EOB$_{\rm NR}$, whereas the effective SNR levels that
we shall consider are $\rho_{\rm eff} = 5, 7, 10, 20$. Note beforehand that most of
these levels are quite lax, and that only the last one is likely to be a really useful accuracy
requirement. [Indeed, we expect typical first observations to have $\rho \sim 10$, and,
as discussed above, adding a safety factor $\epsilon \sim 1/2$ seems to be a minimal
requirement.]

Following the same strategy as in the previous Section, we start by studying how
the inaccuracy evolves with the total mass of the system for a particular mass ratio,
4:1, and a particular noise PSD, namely Advanced LIGO's. The results
are plotted as light gray (red online) lines in Fig.~\ref{fig:overlaps} for the three 
closed-form models we have considered  (using EOB$_{\rm NR}$ waveforms as target).
As  expected from our effectualness results (remembering the fact mentioned above that
the inaccuracy is a non-minimized version of
-- twice -- the ineffectualness), we find that the PN(f)
 model is only accurate for low mass systems, while  the opposite happens for 
the phenomenological models. For instance, if we assume an effective SNR of $10$, 
we obtain that the PN(f) model is only \emph{accurate} for describing systems 
lighter than $5\Ms$, whereas PhenV1 is not accurate for any system we have 
analyzed within $M\in(3,500)\Ms$ and PhenV2  only provides an accurate 
representation for systems heavier than $368\Ms$. In this case, we are 
finding a \emph{very wide inaccuracy interval}, where neither PN(f), 
nor any of the phenomenological models provide waveforms accurate 
enough to extract faithful information from the GW data.

We have determined the \emph{inaccuracy intervals} of the three analytical 
models\footnote{As  in Sec.~\ref{subsec:effectualness_results}, their left-most 
boundaries are determined by the PN(f) model and the right-most ones 
by PhenV2.} for different mass ratio cases,  using also different detector 
noise PSDs and setting different SNR levels; the results are summarized 
in Table~\ref{tab:val_ranges}(b). 
Comparing the right part of this Table (``inaccuracy'') to its
left part (``ineffectualness'') we can see how stricter the accuracy condition is, compared
to the effectualness one. Typically we are now finding that PN(f) waveforms are only accurate
up to a few $\Ms$'s,  leaving then a wide  gap of at least $100\Ms$ 
until phenomenological models provide an accurate representation of the 
observed waveform.  Note in particular
that we find in nearly all cases that the expected paradigmatic BBHs with 
$m_1 \sim m_2 \sim 10-15 \Ms $  cannot be accurately described by any of the
closed-form waveforms. In other words, we find that, in  a large
fraction of the parameter space, no Fourier-domain closed-form models
can be used for measurement purposes.

In order to better visualize  the inaccuracy gaps that we are finding in
parameter space, we represent in Fig.~\ref{fig:scan_param_space} (bottom panels)
the contour plots corresponding to the  inaccuracy levels used in 
Table~\ref{tab:val_ranges}, namely 
$0.04, 0.0204, 0.01, 0.0025$, corresponding
to $\rho_{\rm eff} = 5, 7, 10, 20$. This means that the
darkest zone  is the  region where the inaccuracy\footnote{Note that we plot
as inaccuracy domains at some level $\mathcal I$ the
region of parameter space where \emph{both} PN(f) and PhenV2 have
an inaccuracy larger than $\mathcal I$.} is larger than $ 1/5^2=0.04$,
the next, less dark, zone is the region
where $\mathcal{I} >  1/7^2=0.0204 $ (but $\mathcal{I} < 0.04$), etc., until the lightest
[but non white]  zone where $\mathcal{I} >  1/(20)^2=0.0025 $ (but $\mathcal{I} < 0.01$).
Note how the darkest zones in the bottom figures, where the inaccuracy is larger than
the very lax level $1/5^2=0.04$,
span (even for initial detectors!) a very wide region of  parameter space.
For advanced detectors, this region of ``large inaccuracy'' can be schematized as:
(i) if $1 \leq q \leq 4$, all the systems with $10\Ms < M < 75\Ms$  belong to
this large-inaccuracy zone, while  (ii) if $q \geq 4$, all systems, 
whatever be their mass, belong to this large-inaccuracy 
zone\footnote{Let us recall that phenomenological models were created 
using NR data only for $q\leq4$.}!

The main conclusion of this study is that closed-form models can be used, for
measurement purposes, only in two {\it small islands} of parameter space:
(i) the vertically elongated  white islands on the left of the bottom figures, 
corresponding to small-mass systems (say with $ M < 5 \Ms$), where PN(f) gives 
an accurate waveform; and
(ii) the small whitish islands on the right of the bottom figures, corresponding to 
some special, large-mass systems (with $ M \sim  200 \Ms$ and $ q\sim 2$),  which
can be accurately described by PhenV2(f). In most of the rest of parameter space,
the inaccuracy of the closed-form models is larger than the minimal level
required for extracting faithful physical information from the GW observations.

\subsection{Implications of the non-validity regions}
\label{subsec:validity_regions}

Let us consider in more detail the meaning and implications of the presence
of extended regions in parameter space where closed-form models are either
ineffectual or inaccurate (dark, and darkish, regions in  Figs.~\ref{fig:scan_param_space} 
and~\ref{fig:scan_param_space_V1}).

To have a better grasp of the physics involved in the systems behind the
abstract $(M, \nu)$ parameter-space representation, we have overprinted 
on Figs.~\ref{fig:scan_param_space} and \ref{fig:scan_param_space_V1} 
 \emph{four dashed lines} which gauge the number of orbital cycles (before merger) contributing
$99\%$ of the SNR. [The consideration of a fraction of the SNR of order $99\%$
is natural in our discussion where we are ultimately interested in effectualnesses
of that order or better.]
 We recall that the SNR$^2$ is given by the 
logarithmic frequency integral of the squared ratio $r(f)\equiv (h_s(f)/h_n(f))^2$
between the (dimensionless) effective GW signal amplitude $h_s(f)$ and the 
(dimensionless) effective GW noise $h_n(f) = (f S_n(f))^{1/2}$. The integrand
$r(f)$ is maximum at some frequency $f_m$ and decreases on both sides of $f_m$.
For each given detector, each given fraction of SNR  (say $0.99$), and each given BBH system,
we can then find two frequencies, say $f_1, f_2$,
with $f_1< f_m<f_2$, and such that 
$\int_{f_1}^{f_2} r(f) d \ln f  = (0.99)^2 \int_{0}^{\infty} r(f) d \ln f $. Then, we can count
the number of orbital cycles, say $N_{0.99}$, (defined by the underlying EOB dynamics) 
spent by the considered BBH system 
between the GW frequencies $f_1$ and $f_2$. [When $f_2$ is reached after the merger, one 
counts the orbital cycles between $f_1$ and merger.]  Actually, the choice of the two
frequencies is not unique as we can move $f_1$ and $f_2$ so as to keep the
integral $\int_{f_1}^{f_2} r(f) d \ln f $ fixed. But we (numerically)  looked for the values
of $f_1$ and $f_2$ that \emph{minimize} the number of orbital cycles spent between them.
The result is unique, and defines a number of orbital cycles $N_{0.99}$ which is a function
of the parameters of the system (\ie, in our case, of $M$ and $\nu$) and of the choice of detector noise curve.
[In most cases, $N_{0.99}$ 
is dominated by the number of orbital cycles spent in the inspiral, and is approximately given by
the first term (with $f_0  \to  f_1$) of the Newtonian estimate Eq.~\eqref{eq:Norbits},
namely $N_{0.99} \sim (\pi Mf_1)^{-5/3}/(64 \pi \nu)$. But this formula does not help
to know the precise value of $N_{0.99}$ because the value of the minimizing lower
frequency $f_1$ is itself a function of $M, \nu$ and the detector PSD, which depends
very much on the relative location of the peak of the detector sensitivity
curve $1/(h_n(f))^2$, and the merger.]
Going from left to right,
the four dashed lines in  Figs.~\ref{fig:scan_param_space} and~\ref{fig:scan_param_space_V1} 
represent the locus of the systems for which $N_{0.99}(M,\nu)$ is constant, and successively
takes the values $\{ 100,50,20,10 \}$.
Clearly, for a given $\nu$, $N_{0.99}(M,\nu)$ will increase as the total mass $M$ decreases.
This means in particular that any system lying on the left of the left-most dashed line
will have $N_{0.99} > 100$, and similarly for the other lines.  

One should view the level lines $N_{0.99} = {\rm const.}$ as a (partial) ``intrinsic coordinate system'' in
parameter space. That is to say:  a level line $N_{0.99} ={\rm const.}$ has a more direct meaning,
from the data analysis point of view, than a coordinate line $M={\rm const}.$  
These lines allows one also to make a connection with the capabilities of NR simulations. 
Indeed, the current longest  NR simulations have at most $N \sim 20$ orbital cycles before merger.  
Anticipating faster NR simulations,  we can hope that they will  be able to explore in the 
foreseeable future $ N\sim 40$ orbital cycles, and possibly $N \sim 50$. On the other hand, it will 
probably be very difficult for NR to simulate $\gtrsim 100$ orbital cycles. We shall discuss below 
the issue of the minimum number of orbital cycles from NR simulations which would be needed to inform us, 
in the most effective way, about the dynamics and radiation of BBHs.

Having in mind the significance of the dashed lines in the figures, we can use them to
highlight  some interesting physical facts that appear on the inaccuracy plots, \ie 
 the lower plots of Fig.~\ref{fig:scan_param_space}. Let us first note that these
plots exhibit the existence of two ``islands of accuracy'' (\ie whitish zones), within 
an ``ocean'' of inacceptably large inaccuracy (darker zones). 
[For definiteness, we shall define this ocean of inaccuracy as the union of the two darkest zones.]
The (vertically elongated) island on the left of each plot lies in the small-mass region 
and corresponds to BBH systems for which  the inspiral PN(f) waveform
provides, by itself, an accurate description of the full EOB$_{\rm NR}$  waveform.
The other whitish island on the right of each plot (roughly centered around $ \sim 200 \Ms$,
$q \sim 2$) corresponds to BBH systems for which the  PhenV2(f) waveform accurately
agrees with the  EOB$_{\rm NR}$  waveform.

\begin{itemize}

\item  The first fact is that the left islands of accuracy (which we can call the
 islands of accuracy of pure PN waveforms) possess as \emph{right boundaries}
(\ie their ``shores'' in the dark ocean of inaccuracy)  some lines
which are approximately parallel to the dashed, $N_{0.99}={\rm const}.$ lines, with large values
of the constant. For instance, for initial LIGO, the right boundary of the left (PN) island
is approximately located along the $N_{0.99}=70$, while for advanced LIGO it is
approximately along the $N_{0.99}=100$ line. For advanced Virgo the value of $N_{0.99}$
on the boundary is significantly larger than $100$.

\item The second fact, is that the right islands of accuracy (i.e. the islands of accuracy
of phenomenological waveforms) do not extend, on their left side, beyond 
$N_{0.99}={\rm const}.$ lines, with a rather small value of the constant, 
say $N_{0.99}\sim 5$  or so. Notice also that the \emph{accuracy zone} of the 
phenomenological waveforms seems to be restricted to $q\leq4$.
This probably reflects the fact that the construction of phenomenological 
models did not correctly incorporate the known PN inspiral phasing, and it
is consistent with the parameter range of NR waveforms used to
build these models~\cite{PhenV1,PhenV2}.

\end{itemize}

The first fact is easily understood. Indeed, the PN(f) model does not include any merger signal.
Therefore it can have a small inaccuracy with respect to the target signal 
(which include the merger) only for systems whose SNR  is nearly completely
given by the inspiral signal. The second fact is a priori more surprising, because the motivation
for the construction of phenomenological signals was to approximately describe a waveform
comprising all the needed components of the full signal, namely a PN-type inspiral,
a NR-described merger, and a leading quasi-normal-mode-type ring-down.
However, we can see in Fig.~\ref{fig:waveforms} that the Phen waveforms exhibit significant
disagreements, both in amplitude and in phase, with the target (EOB$_{\rm NR}$) signal.
More quantitatively, if we remember, from the discussion in the Sec.~IID above, 
that an inaccuracy smaller than $1/7^2$ implies that the noise-weighted quadratic sum of the fractional
amplitude, and phase, differences must be smaller than $1/7\approx 0.14$, we understand 
upon looking carefully at  Fig.~\ref{fig:waveforms}
that the Phen models are too coarse approximations to be  $98 \%$ accurate during 
most evolutions that will have a significant mixture of inspiral and merger.
The conclusions we can draw from these remarks will be discussed in the next section.


The lack of validity of these analytical models in certain regions of the
parameter space, will require the use of more accurate (and also more time consuming)
models when performing a search, such as, either full NR simulations, or a new and more
complex phenomenological model, or EOB waveforms. In particular, the limited accuracy of
PN(f) that we are observing, specially at more extreme mass ratios, will require filling a
large gap of the BBH evolution between where PN-SPA fails and the merger (it may
represent several hundreds of orbital cycles, as we have seen in
Sec.~\ref{sec:PN-SPA_range}), that will probably be computationally 
unachievable for full NR simulations, but much more reasonable for semi-analytical 
methods such as the EOB approach.

\section{Conclusions}
\label{sec:conclusions}

Our knowledge of the dynamics of coalescing BBH systems and their gravitational emission has experienced a 
huge progress during the last $5-10$ years, both with the improvement of analytical methods, namely PN expansions 
and EOB approaches, and the breakthroughs that occurred in NR. A crucial question at this stage is to know whether 
the recently accumulated PN+EOB+NR knowledge is sufficient for the GW data analysis needs of the present
and upcoming GW detectors.  We can classify the currently available model waveforms 
into several categories: 

\begin{itemize}

\item  closed-form, frequency-domain models which are only accurate in a limited range of the evolution,  
\eg PN(f), ring-down models  and (NR-fitted) phenomenological models (which
try to extend their validity range by combining information coming from PN(f), ringdown(f) and some NR simulations); 

\item EOB models, which cover the whole evolution, and which are defined in the time-domain 
by integrating ODEs; they can easily incorporate information coming either from PN and from NR;

\item full NR simulations, which are necessary for describing  the nonperturbative physics
around merger, but can only cover a limited number of orbital cycles ($\lesssim 20$)
because they are very time consuming in terms of computational costs; 

\item time-domain PN models (either using closed-form expressions or integrating ODEs),
which are limited to the inspiral part of the evolution;

\item various types of \emph{hybrid} models, joining together (either in the time-domain,
or in the frequency-domain) early PN-type waveforms to later NR waveforms, so as to
cover the whole evolution.

\end{itemize}

The time needed to generate a waveform in one of these categories varies by many
orders of magnitude between, say, a closed-form frequency-domain waveform 
(a fraction of a second on a single CPU); an EOB waveform (several seconds, also, on a single CPU); or, a full NR simulation
($\sim$ a month on a computer cluster with several hundreds of CPUs).
From the data analysis point of view, it would be quite useful to be able to generate
waveforms as fast as possible [actually, the real issue is to be able to \emph{access} 
a dense bank of them very fast].  Though some trade off between speed and
accuracy can be allowed for, there are minimal accuracy requirements that 
waveforms have to satisfy. This is why our present paper has focused on a
detailed analysis of the accuracy of the fastest existing waveforms, namely
the closed-form, frequency-domain ones.

In this paper, we have performed an extensive study, in the whole parameter space,
 of the validity of the closed-form, frequency-domain   models  (for non-spinning BBHs): 
\ie PN(f) and phenomenological models. We have studied both their
\emph{effectualness} (detection) and their \emph{accuracy} (measurement)
using as  target model the currently most accurate NR-calibrated 
EOB waveform. We have considered both initial GW detectors and advanced ones.

Our conclusions concerning closed-form, frequency-domain waveform models are:

\begin{itemize}

\item When considering \emph{initial interferometers},
these closed-form models are \emph{just effectual enough} to be used for \emph{detection}.
Though their effectualness (\ref{eq:def_effectualness}) is, indeed, larger than $97 \%$ 
in nearly all the parameter space\footnote{Apart from a small island  near $M\sim 20 \Ms$ 
and $\nu=0.25$ in the PhenV1 plot, Fig.~\ref{fig:scan_param_space_V1}}, 
it is \emph{smaller than}  $98 \%$ in two separate islands
in parameter space, and \emph{smaller than}  $99 \%$ in sizable regions of parameter space.
See lef--top panel of Fig.~\ref{fig:scan_param_space} and left panel of
Fig.~\ref{fig:scan_param_space_V1}. In particular, notice (on the smaller sub-panels) 
that most of the effectualness is provided by the phenomenological model,
except for the very low mass end, and also around the crucial $M\sim 10M_\odot$, $q\sim1$ 
region where PhenV2 fails to be $97\%$ effectual.

\item When considering \emph{initial interferometers}, they
are \emph{not accurate enough} to be used for \emph{measurement}.
Their inaccuracy, Eq.~\eqref{eq:inaccuracy1}, is, indeed, larger than the threshold corresponding to 
an effective SNR of $5$ (which is an very low threshold) in a huge domain of parameter space.
See left--bottom panel of Fig.~\ref{fig:scan_param_space} where the darkest zone is the region
where the inaccuracy is larger than $1/5^2$. The union of the darkest and
second darkest zone correspond to an inaccuracy larger than $1/7^2$, which is still a very low threshold.

\item  When considering \emph{advanced interferometers}, they are \emph{neither effectual,
nor accurate} in large domains of parameter space.  More precisely, the domains
of ineffectualness are the extended islands  given by the dark zones of  the two right, 
upper panels in Fig.~\ref{fig:scan_param_space} (the darkest zone failing to be $97 \%$ effectual, and the
second darkest one failing to be $98 \%$ effectual).  Concerning the domains of inaccuracy
they have become so prominent and extended in the two right, lower panels in Fig.~\ref{fig:scan_param_space} 
(with respect to the domains of ineffectualness in the plots above them)\footnote{In other words,
we can say that the (dark) islands of ineffectualness of the upper panels have ``percolated''
among themselves,
and left only a reverse landscape where the accuracy zones exist only as separate islands.}
that it is better
to visualize them as an ocean of inaccuracy containing only two well-separated
islands of accuracy (namely the whitish zones in the two right, bottom panels).
In the left island of accuracy, PN(f) provides an accurate model for small-mass systems, 
while, in the right island of accuracy, PhenV2 provides an accurate model for systems 
with total mass $M \sim 200 \Ms$, and mass ratio $q \sim 2$.

\end{itemize}

Our conclusion is therefore that the current, existing closed-form, frequency-domain
models fail to be accurate enough to be used for measurement purposes, even with
initial GW detectors. On the other hand, if one uses \emph{jointly} the
two closed-form models with initial GW detectors sensitivities,
they provide a $97\%$-effectual coverage of the full
parameter space, and a $99\%$-effectual one  in 
an important fraction of parameter space. These effectualness levels are significantly reduced when considering advanced detectors
noise PSDs or taking the waveform models \emph{separately}.
These results call for the development of more accurate versions of these
phenomenological models. In particular, we think that an important defect of the
V1 and V2 versions of the phenomenological models is that they do not correctly
incorporate the analytically known (PN) phasing behavior during the (early) inspiral.
One can indeed see on Fig.~\ref{fig:waveforms} how this leads to large dephasings
during the inspiraling phase. In turn, these dephasings cause the ineffectualness and
inaccuracy properties. [See, below, additional results concerning the recently
constructed V3 version of the phenomenological models.]

In terms of the (partial) intrinsic coordinate system defined by the level lines
$N_{0.99}={\rm const}.$, we can summarize the performance of PhenV2 model
saying that it can provide an \emph{accurate} representation of the last $N_{0.99} \sim 5$
orbital cycles before merger for $q<3$ systems, and an \emph{effectual} representation of signals
with no more than $N_{0.99} \approx 200$ visible orbital cycles within the mass ratio range $1.5<q<8$.
[PhenV1 model is not accurate enough anywhere and its effectualness is restricted to signals
with $N_{0.99} \lesssim 100$ visible orbital cycles in the same mass ratio range as PhenV2].
Of course, as we consider more sensitive detectors, these upper limits on $N_{0.99}$ translate
into a restriction of  a larger portion of the parameter space.

Also, a side result of our study is the determination of the maximum frequency, as
as function of the mass ratio, at which PN(f) waveforms can be `joined' onto
EOB$_{\rm NR}$ (in the frequency domain) without undue loss of accuracy.
Our results are summarized in Fig.~\ref{fig:PN-SPA_range} and show wide validity
ranges for PN(f) up to less than $6$ orbits before merger for 1:1 and 2:1 systems,
but much more worrying results when considering $q>4$ systems. In particular, we
find gaps of $N \gtrsim 20$ orbital cycles for $\rho_{\rm eff} = 10$ [$N\gtrsim 100$
orbital cycles for $\rho_{\rm eff} = 20$] between the low-end of the PN(f) validity range
and the merger; which can hardly be covered with (computationally intensive) NR
simulations.

We are fully aware that the  analysis that led us to all these conclusions depends
on our use of the EOB$_{\rm NR}$ waveform as a reference model for defining
the inaccuracy functional Eq.~\eqref{eq:inaccuracy1}, Eq.~\eqref{eq:inaccuracy2}.
Strictly speaking the dark zones in Fig.~\ref{fig:scan_param_space} only represent
the regions of parameter space where the fractional squared Wiener distance
between the closed-form models and EOB$_{\rm NR}$ is larger than $1/7^2 \approx 0.02$.
However, we think that there is a good rationale for considering that our results
tell us something about the exact inaccuracy of the closed-form models, \ie their
fractional squared Wiener distance away from the (unknown) exact waveform.
Indeed, one should keep in mind that the inaccuracy level corresponding to the
two darkest zones corresponds to fractional amplitude, and phase inaccuracies (in radians),
which are only $\sim 1/7 \approx 0.14$. By comparison, the NR-calibrated version
of the EOB formalism that we used here \cite{Damour:2009kr} 
has been proven to agree with some of the
current most accurate NR simulations with significantly smaller amplitude and phase inaccuracies.
More precisely, in the $q=1$ case, the  phase disagreement $\Delta \phi$ between EOB and the
Caltech-Cornell data~\cite{Boyle:2007ft, Scheel:2008rj} remained smaller than $\pm 0.02$ radians
during  the entire $16$ orbital cycles of inspiral and plunge, which is comparable with the
estimated NR errors.  At the merger and during the ringdown $\Delta \phi$ took somewhat
larger values ($\pm 0.1$ radians), but it oscillated around zero so that, on average, it stayed
very well in phase with the NR waveform. As for the fractional amplitude
disagreement $|\Delta A|/A$ it  remained smaller than about $ 0.003$
during the entire inspiral, and only momentarily exhibited a larger value ($0.025$)
just after the merger. In the $q=2$ and $q=4$ cases, the agreement between EOB and the Jena data
(published in~\cite{Damour:2008te}) was excellent, and within the numerical errors during the
entire span of the simulation (which comprised $17$ GW cycles); see Fig.~10 in~\cite{Damour:2009kr}.  
However, the numerical errors in the latter $q=2$ case were significantly larger than in the $q=1$ case, 
but smaller than the   $\sim 1/7 \approx 0.14$ level used in our inaccuracy study (remaining
 probably smaller than $\pm 0.1$ radians). Let us also mention that an earlier
version of the EOB formalism has also been succesfully calibrated (within NR errors) to accurate 
NR simulations for several mass ratios $q=1, 2 $ and $3$ \cite{Buonanno:2009qa} (however,
the $8$-orbit $q=2, 3$ simulations used in~\cite{Buonanno:2009qa} cover only part
of the late inspiral, without the merger). For $q>4$, no EOB--NR calibrations
have been published yet, due to lack of accurate and long enough NR waveforms available to us
in this mass-ratio regime; however, let us note that recent work~\cite{Bernuzzi:2010ty,BN2010b} 
has accurately confirmed the validity of the EOB waveform model in the extreme mass ratio limit.

We have also explored how the results would change in the hypothetical case where the
low frequency cut-off of the advanced detectors would be increased from $10\Hz$ to $20\Hz$, since
improving the low frequency region of the ground-based detectors sensitivity is always the most
challenging task from the experimental point of view. All our results remain almost unaltered,
except for a tiny improvement in the effectualness of phenomenological models due to
a reduction of the number of visible orbital cycles. In any case, all the conclusions described 
in this article would remain the same.


\begin{figure*}[t]
\begin{center}
\begin{tabular}{p{7pt}ccc}
 & & PhenV3 & \\
 \vspace{-2.4cm} $\bar{\mathcal{E}}$:  &
\includegraphics[width = 5.7cm]{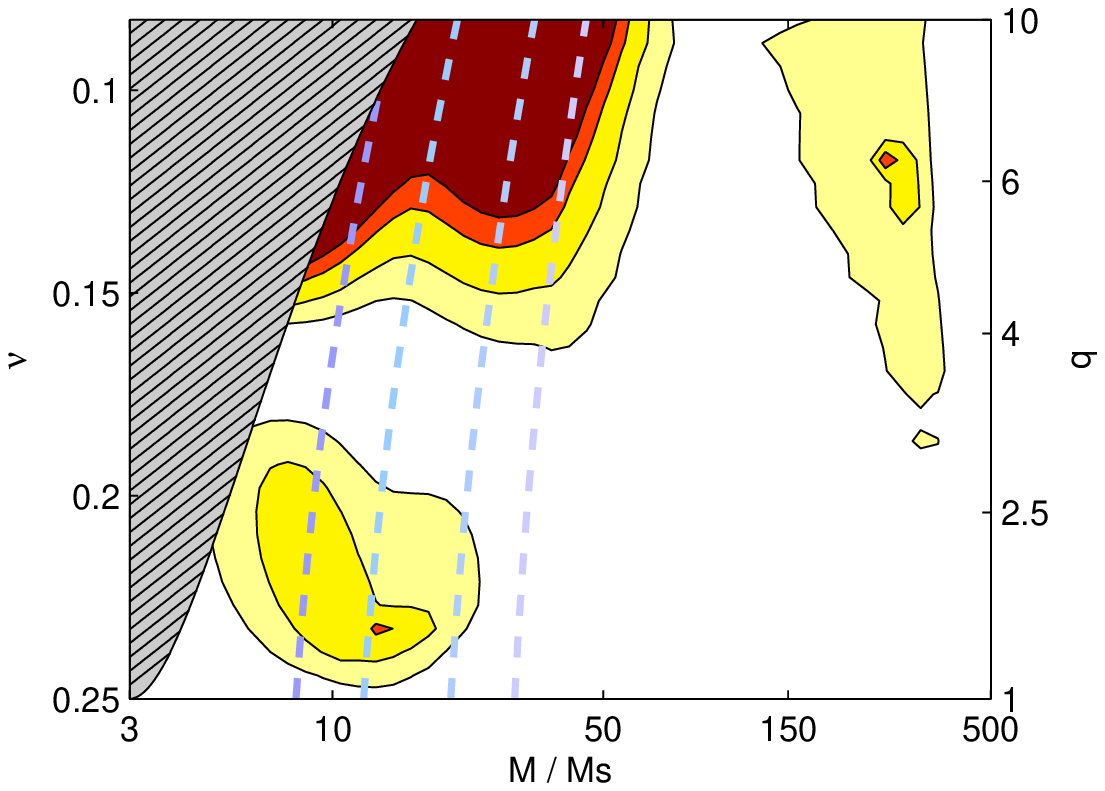} & 
\includegraphics[width = 5.7cm]{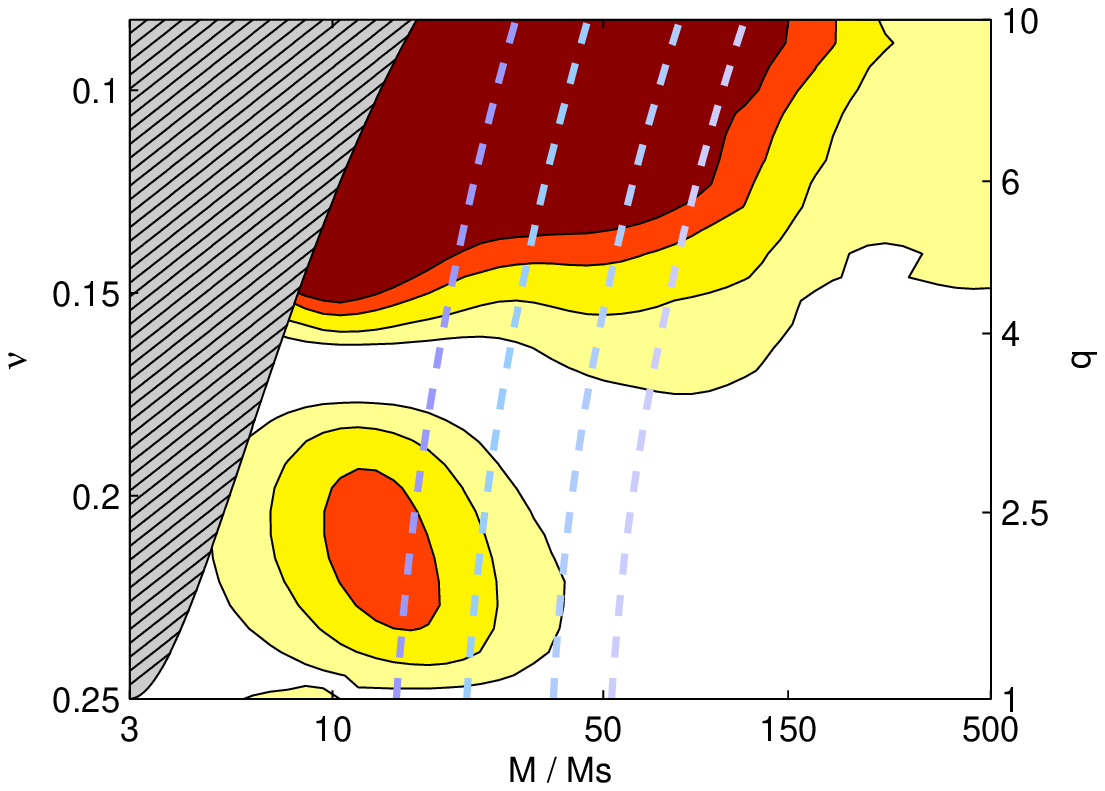} & 
\includegraphics[width = 5.7cm]{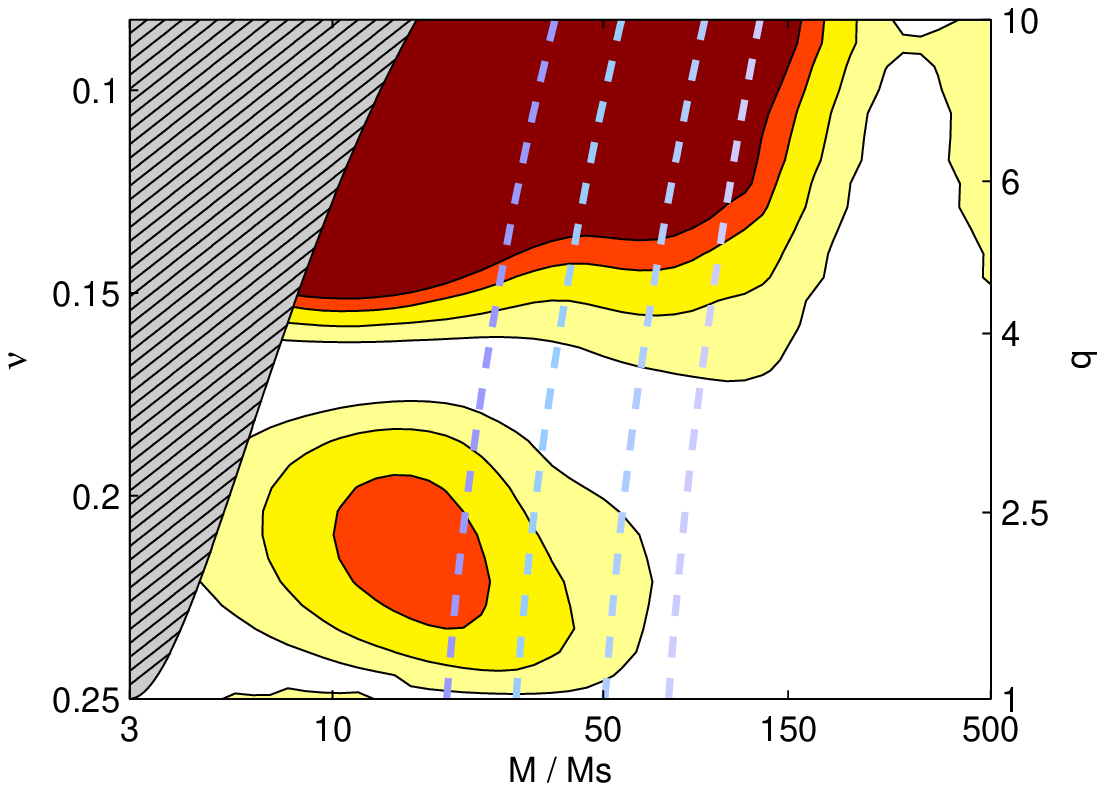} \\[1cm]
 \vspace{-2.4cm} $\mathcal{I}$:  &
\includegraphics[width = 5.7cm]{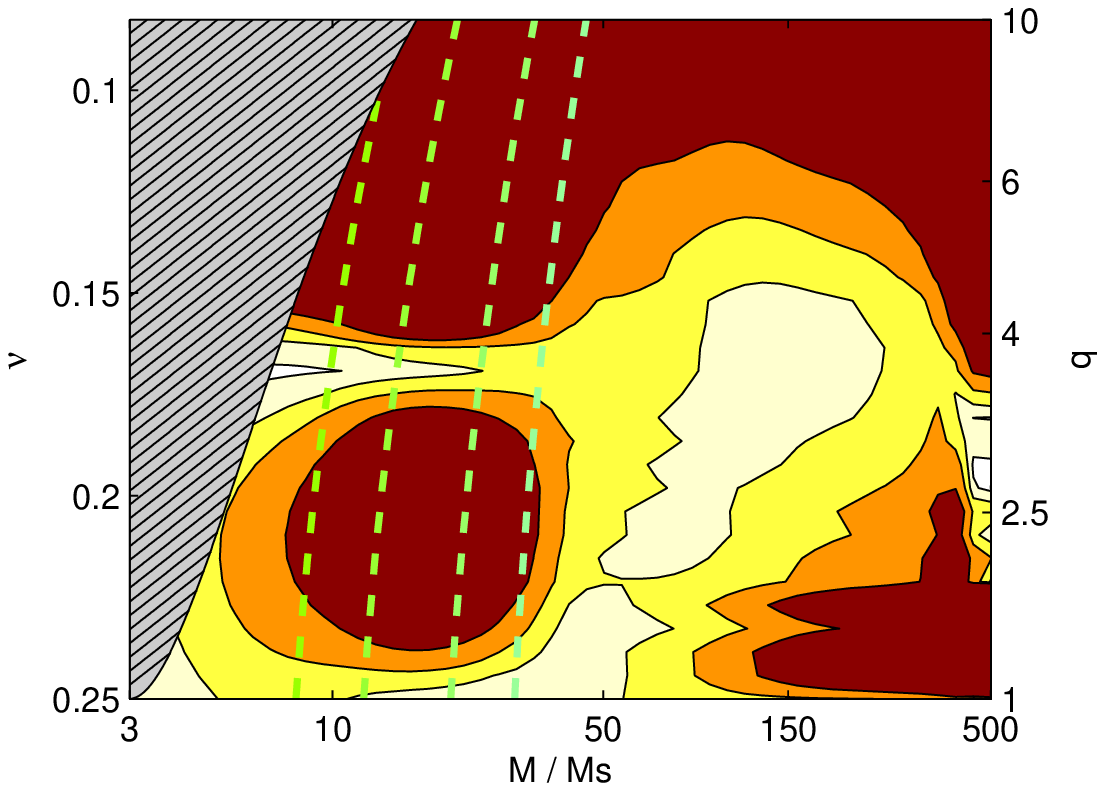} & 
\includegraphics[width = 5.7cm]{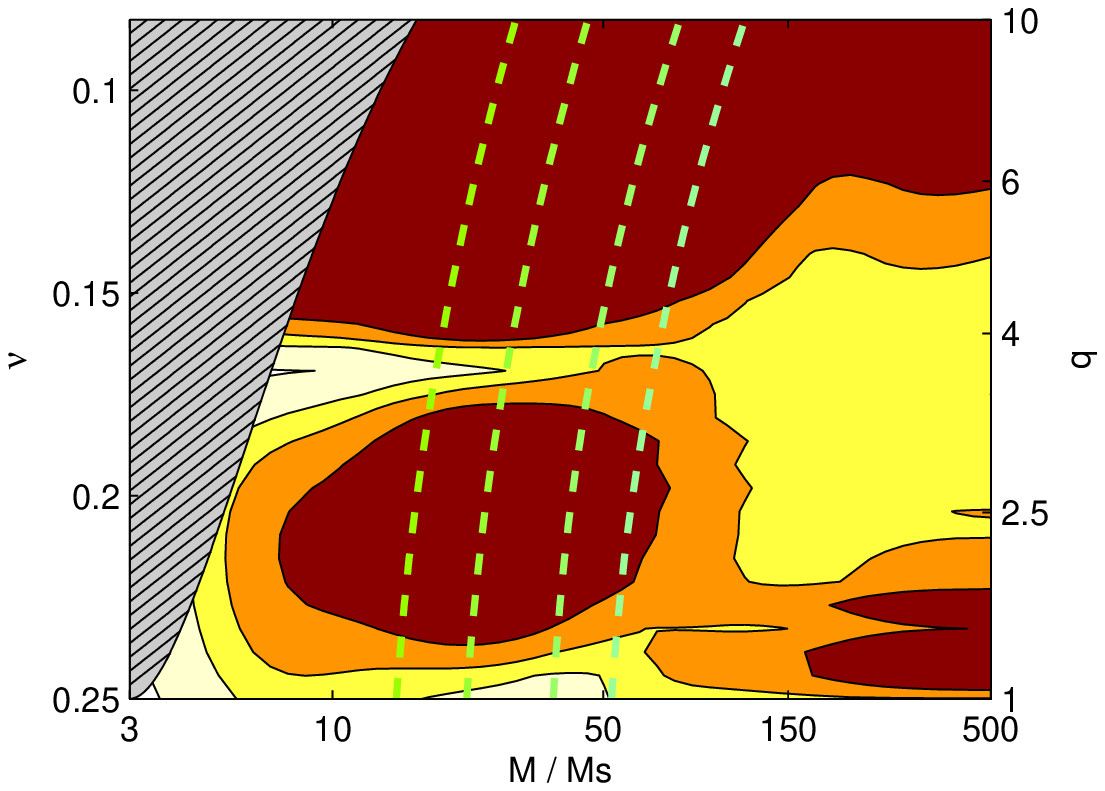} & 
\includegraphics[width = 5.7cm]{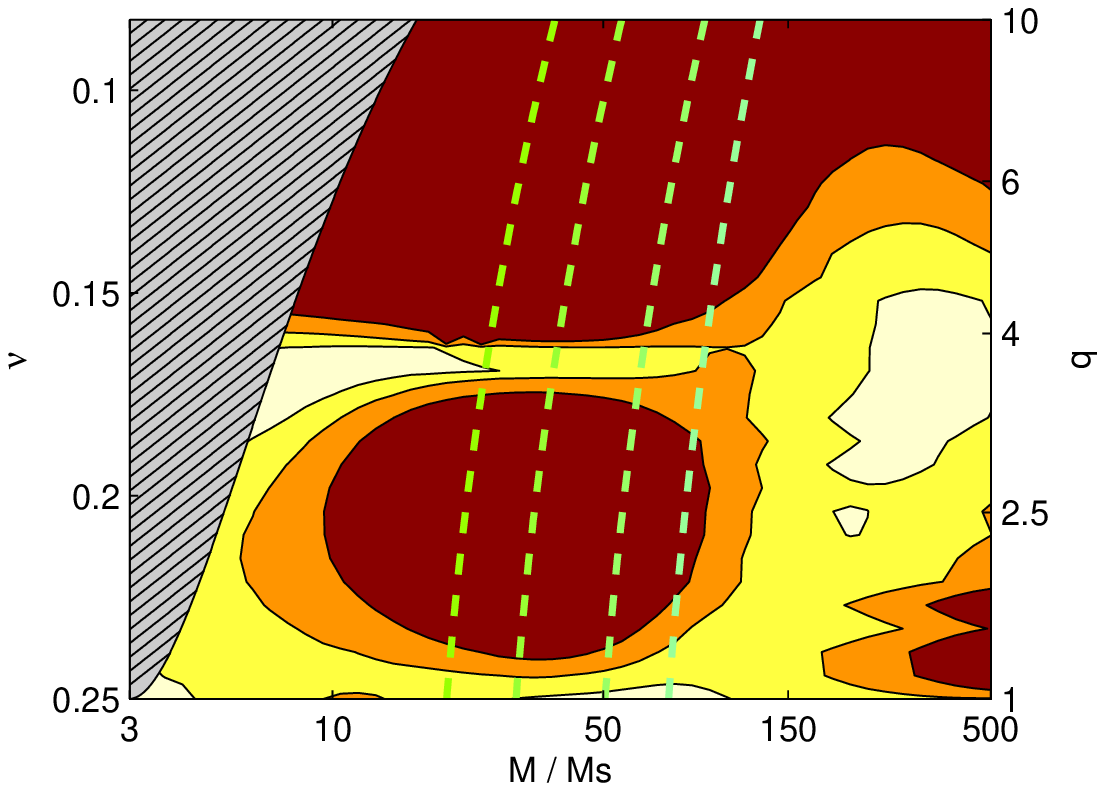} \\
 & (a) iLIGO
 & (b) aLIGO
 & (c) aVirgo
\end{tabular}
\end{center}
\caption{Same plots as in Fig.~\ref{fig:scan_param_space}, but considering the recently published
PhenV3 \cite{Santamaria:2010yb} model. Since it already includes all the PN contributions in the low
frequency region, here we consider the phenomenological model alone. We recall that the top
panels plot the regions of the parameter space where the ineffectualness is larger
than $\{ 3\%, 2\%, 1\%, 0.45\% \}$ going from the darkest to the lightest areas; whereas the
bottom panels show the regions where the inaccuracy is larger than $1/\rho_{\rm eff}^2$,
where the effective SNR is $\rho_{\rm eff} =\{5, 7, 10, 20\}$, when going from the darkest to
the lightest region.}
\label{fig:scan_param_space_V3}
\end{figure*}


By the time this article was ready for submission, a new phenomenological
waveform model for non-precessing spinning BBH systems was published by
Santamaria \etal \cite{Santamaria:2010yb}. This model (that we shall call ``PhenV3'')
represents an improvement over the previous PhenV2 model. This new PhenV3 model
aims at describing the whole BBH coalescing process by joining a state-of-the-art PN
description of the early inspiral (which was lacking in V1 and V2) to separate
phenomenological representations of a ``pre-merger signal", and a ``ring-down" one.
In view of the improved nature of this PhenV3 model, we have decided to complete our
analysis by studying its effectualness and accuracy performances (i.e. the analogs of
Figs.~\ref{fig:scan_param_space} and \ref{fig:scan_param_space_V1} above). Our results
for the PhenV3 waveform\footnote{The authors of \cite{Santamaria:2010yb} kindly 
let us know about the existence of a typo in their Eq.~(5.12): the `$\sigma$' parameter
of the Lorentzian function should read $\delta_2 f_{\mathrm{\textsc{rd}}} / Q$ instead of 
$\delta_2 Q$. Here, we have used the corrected expression when generating our results.}
are displayed in Fig.~\ref{fig:scan_param_space_V3}. As this
new model now incorporates a 3.5 PN-accurate description of the early inspiral, we
directly plot in Fig.~\ref{fig:scan_param_space_V3} the ineffectualness and inaccuracy
(with respect to EOB$_{\rm NR}$) of PhenV3(f) alone, without computing the infimum
with respect to PN(f). If we focus our attention on the $q \leq 4$ region (where the
phenomenological parameters were fitted), we observe that the PhenV3 model is
$>98\%$ effectual for initial detectors and $>97\%$ effectual when considering
advanced detectors noise PSDs. The effectualness is higher for massive systems, 
$M>50\Ms$, (where only the last few cycles of the evolution are observable) and reaches
minimum values in the [astrophysically relevant] region around $M\simeq 10\Ms$
(where there are still many observable cycles but PN theory has already started to fail).
Notice that these results are consistent with the overlaps plotted in Fig.~10 
of~\cite{Santamaria:2010yb}, despite the fact that here we do not maximize the
effectualness over any spin parameter, as we have restricted ourselves to non-spinning
systems. The fact that the ineffectualness (as well as the inaccuracy) is maximum around
$q \simeq 2$ and $ M \sim 20 \Ms$, is compatible with the fact that, when plotting the
$q=2$ analog of Fig.~\ref{fig:waveforms} for the PhenV3/EOB comparison, we observed
the existence of significant dephasings, of order $\sim 0.5$ rad, around 
$\fM \sim 8 \times 10^{-3}$. Similarly to what was observed with PhenV1 and
PhenV2 models, the PhenV3 effectualness becomes much worse for systems with $q>4$.
In addition, it is important to note that the accuracy of the PhenV3 model is only
slightly improved with respect to the PhenV2 one. Actually,
Fig.~\ref{fig:scan_param_space_V3} shows that the PhenV3 model cannot be used for
measurement purposes (accuracy) in most of the parameter space. This shows that the
conclusions drawn above from a study of the V1 and V2 models hold also for the
PhenV3 model.

Also, while writing up our results for publication, another paper related to our work
appeared on the archives. Indeed, the article \cite{Hannam:2010}, presented another
study similar in spirit to the one of our  Sec.~V.  However, the context, and details of
their study are quite different from ours:  they are using as target waveform a (spinning)
PN$\cup_{f_{\rm start}}$NR  hybrid waveform (where the starting frequency $f_{\rm start}$
of their NR waveform covers 6-10 orbital cycles before merger) and they compute the
\emph{faithfulness}  between this target and a hybrid of the form PN$\cup_{f_{ 0}}$NR,
where the junction frequency $f_0$ is \emph{larger}  than $f_{\rm start}$.  In other words,
they are monitoring the loss of effectualness in  PN$\cup$NR hybrids as one includes less
and less NR orbital cycles in the hybrid (starting from 6-10 NR orbital cycles).

\section{Future work}
\label{sec:future_work}

In view of our results, we think that, especially for  future searches in advanced detectors data,
more effectual and more accurate model waveforms will need to be considered in order not to
miss a significant fraction of the hard-won potential detections and, to extract reliable 
physical data out of the observations. Though some improved (more effectual) phenomenological
waveforms might be useful for detection purposes, we think that the NR-calibrated EOB
waveforms are our current ``best bet''  models for building waveforms that are \emph{both}
effectual and accurate. More effort should be put into improving the EOB waveforms as well as
into improving their implementation in data-analysis platforms.  Here is a list of suggestions
towards this aim.

\begin{itemize}

\item More work is needed to refine the results of Sec.~V, displayed in 
our Fig.~\ref{fig:PN-SPA_range}  concerning the
maximum junction frequency $M f_0$  at which one can smoothly join 
(in the frequency-domain) an early PN(f) waveform to a later EOB(f) one without
losing more than some given accuracy.  Our results suggest that it
is ill-advised to continue the common practice of defining  (either in the
time-domain, or the frequency-domain) hybrid PN$\cup_{f_0}$NR waveforms
by joining an early PN waveform to a later NR one, at a frequency $f_0$ near the
starting frequency of the NR waveform.  Such a procedure is probably justified
(without losing too much accuracy) only for NR waveforms containing 
more orbital cycles before merger than the numbers displayed in the last columns of
the Table contained in  Fig.~\ref{fig:PN-SPA_range}.  This would make it practically 
impossible to build accurate PN$\cup$NR waveforms for mass ratios $q > 4$.

\item We think, however, that most of the accuracy loss behind Fig.~\ref{fig:PN-SPA_range}
is due to the poor quality of (any of) the \emph{non-resummed PN expansions},
and that the better quality of the \emph{resummed} expressions\footnote{The effectiveness
of the resummation methods used in the most recent version of EOB is exemplified, for instance,
 in Fig.~1 of \cite{DIN} for the $\nu \ll 1$ case, and in Fig.~4 of 
\cite{Damour:2009kr} for the $\nu=1/4$ case.} used in the EOB formalism will 
allow one (if one insists on doing so) to join an early EOB waveform to a later 
NR waveform around a frequency $f_0$ corresponding to significantly less orbital 
cycles before merger than the numbers displayed in the Table contained 
in Fig.~\ref{fig:PN-SPA_range}.  We leave to a future work \cite{DNT2} a 
discussion of the minimum number of NR orbital cycles needed to improve the 
NR-calibation of EOB to the level of accuracy needed for the analysis of the future GW data. 

\item  Independently of the quantitative issue of this minimum number of NR orbital cycles, we
think that it is more efficient to use accurate NR data to directly \emph{inform}
the EOB formalism by calibrating some unknown higher-order theoretical parameters,
and then to use the resulting NR-calibrated EOB waveforms as best-bet models
(rather than passing through the intermediate step of building hybrid EOB$\cup$NR
waveforms).  These EOB$_{\rm NR}$ waveforms could then be used (if wished) as
basic material for defining more accurate \emph{closed-form} frequency-domain
phenomenological-type waveforms.

\item However, we think that it should be possible to set up a (multi-point)
\emph{interpolation}  technique using a relatively sparse bank of (pre-computed)
EOB$_{\rm NR}$(f) waveforms to compute, \emph{in a numerically fast way},  any wished
EOB$_{\rm NR}$(f) waveform, with the \emph{accuracy} required for measurement purposes.
A detailed analysis is needed to estimate the CPU time needed, at some
required inaccuracy level $\mathcal I$, by such an interpolation technique, 
which would then obviate the need for having at hand a closed-form, frequency-domain
model.  [However, improved closed-form models might still be useful for detection purposes.]

\end{itemize}

Let us note finally that several ongoing works might soon allow one to define
more accurate  analytical descriptions of  BBH systems. First, 
 a recent joint effort between several numerical relativity and analytical 
relativity groups~\cite{NRAR_col} has begun to make use of TeraGrid resources 
in order to produce longer and more accurate simulations of comparable-mass BBH systems. 
Second,  the case of extreme mass-ratio BBH mergers is amenable to separate
Regge-Wheeler-Zerilli-type studies which inform us about the structure of
the waveform in the $\nu \ll 1$ limit~\cite{Poisson:1995vs, Yunes:2008tw, Yunes:2009ef, Lousto:2010qx, Bernuzzi:2010ty}. Third,
 progress in Gravitational Self Force (GSF) studies~\cite{Barack:2009ey,Barack:2010tm}  
has recently allowed one to get a new, accurate source of information about 
the \emph{strong-field behavior} of some of the basic ingredients of the EOB 
formalism~\cite{Damour:2009sm,Barack:2010ny}.
We note also that our current study has only considered the leading quadrupolar waveform
($\ell =2, m=\pm 2$),
and non-spinning systems. The EOB formalism provides (resummed) waveforms for all multipolarities
 $\ell, m$~\cite{DIN} so that it will be straightforward to extend our analysis
to the full waveform (at least for the comparison with the corresponding higher-multipolarity
PN waveform). The EOB formalism has  been defined for spinning BBH systems
\cite{Damour:2001tu,Damour:2008qf,Barausse:2009xi,Pan:2010hz,Pan:2009wj},
so that our study can also be extended to the case of spinning BBH waveforms.


\section*{Acknowledgements}

We thank P. Ajith, S. Husa, M. Hannam, L. Santamaria and A. Vecchio for helpful
comments on the manuscript. MT thanks IHES for hospitality while this work was 
carried out and is grateful for the support of the European Union FEDER funds 
and the Spanish Ministry of Science and Education (projects
FPA2007-60220 and CSD2009-00064).



\end{document}